\documentclass[11pt]{article}

\usepackage{arxiv}

\usepackage{amsmath,amsfonts,latexsym,fullpage,graphicx,vmargin, mathrsfs}
\usepackage{pstricks}
\usepackage{subcaption}
\usepackage{amsmath}
\usepackage{multirow}
\usepackage{comment}
\usepackage[inline]{enumitem}

\RequirePackage[OT1]{fontenc}

\usepackage{amsmath}
\usepackage{amsfonts}
\usepackage{caption}
\usepackage{subcaption}
\usepackage{multirow}
\usepackage{tikz}
\usetikzlibrary{graphs}
\usetikzlibrary{cd}
\usepackage{bm}
\usepackage{url}
\usepackage{placeins}

\DeclareMathOperator*{\argmax}{arg\,max}
\DeclareMathOperator*{\argmin}{arg\,min}

\RequirePackage{natbib}

\newtheorem{prop}{Proposition}
\newtheorem{defi}{Definition}
\newtheorem{rmk}{Remark}
\newtheorem{cor}{Corollary}

\newcommand{\virgolette}[1]{``#1''}
\newcommand{\Sp}{\textrm{Sp}}

\newcommand{\Xvar}{\mathcal{X}}
\newcommand{\Yvar}{\mathcal{Y}}

\newcommand{\Zvar}{\mathcal{Z}}

\usepackage[normalem]{ulem}

\firstpageno{1}

\begin{document}

\title{Projected Statistical Methods for Distributional Data on the Real Line with the Wasserstein Metric}

\author{Matteo Pegoraro\thanks{MOX -- Department of Mathematics, Politecnico di Milano}, Mario Beraha\thanks{Department of Mathematics, Politecnico di Milano}\thanks{Department of Computer Science, Universit\`{a} degli Studi di Bologna}}

%
\maketitle

\begin{abstract}
We present a novel class of \emph{projected} methods to
perform statistical analysis on a data set of probability distributions on 
the real line, with the 2-Wasserstein metric. We focus in particular on
Principal Component Analysis (PCA) and regression.
To define these models, we exploit a representation of the Wasserstein space closely related to its weak Riemannian structure, by mapping the data to a suitable linear space and using a metric projection operator to constrain the results in the Wasserstein space.
By carefully choosing the tangent point, we are able to derive fast empirical
methods, exploiting a constrained B-spline approximation. 
As a byproduct of our approach, we are also able to derive faster routines for previous work on PCA for distributions.
By means of simulation studies, we compare our approaches to previously
proposed methods, showing that our \emph{projected} PCA has similar performance for a fraction of the computational cost and that the \emph{projected} regression is extremely flexible even under misspecification.
Several theoretical properties of the models are investigated, and asymptotic
consistency is proven.
Two real world applications to Covid-19 mortality in the US and wind speed
forecasting are discussed.
\end{abstract}


\section{Introduction}

In many fields of machine learning and statistics, performing inference on a 
set of distributions is an ubiquitous but arduous task.
The Wasserstein distance provides a powerful tool to compare distributions,
as it requires very little assumptions on them and is at the same time
reasonably easy to compute numerically. In fact, many other
distances for distributions either require the existence of a probability
density function or are impossible to evaluate, cf.\
\cite{cuturi2013sinkhorn}, \cite{peyre2019computational}, \cite{panaretos}.

The Wasserstein distance recently gained popularity both in the statistics
and machine learning community.
See for instance \cite{bassetti2006minimum}, \cite{bernton2019parameter},
\cite{catalano2021annals} for statistical properties of the Wasserstein 
distance, \cite{cao2019multi}, \cite{cuturi2019differentiable} and
\cite{cuturi2014fast} for applications in the field of machine and deep learning,
\cite{bernton2019approximate} and \cite{srivastava2015wasp}
for applications in Bayesian computation.

In this work, we focus on the situation in which the single observation itself 
can be seen as a distribution, 
as in the analysis of images \citep{cuturi2014fast, geod_regression_chakra},
census data \citep{geod_vs_log}, econometric surveys \cite{potter2017advantages}
and process monitoring \citep{menafoglio}.
In particular, we consider observations to be distributions on the
real line.
There exist several possible ways to represent 
distributions, such as histograms, probability density functions (pdfs)
and cumulative density functions (cdfs), each 
characterized by different constraints.
For instance, histograms sum to one, pdfs
integrate to one, and the limits for cdfs are 0 and 1, moreover
all of these functions are nonnegative.
These constraints translate into complex geometrical structures that
characterize the underlying spaces these objects live in.

\subsection{Previous work on distributional data analysis}
\label{sec:prev_work}

One of the first works defining PCA for a data set of distributions is
\cite{kneip}, where the authors apply tools from functional data analysis (FDA) directly to a collection of probability density functions.
This approach, however, completely ignores the constrained nature 
of probability density functions, leading to poor interpretability of the results.

Based on theoretical results in \cite{egozcue2006hilbert}, who defines
a Hilbert structure on a space of probability density functions on a compact interval (called a Bayes space),
\cite{delicado} and \cite{menafoglio}, propose a more reasonable approach
to the problem of PCA for density functions.
In particular, in \cite{menafoglio}, the authors use the geometric properties of the Bayes space, coupled with a suitable transformation from the Bayes space to an
$L_2$ space, to perform PCA on a set of pdfs using FDA tools, and
then map back the results to the Bayes space.

Another, perhaps less widely used, approach focuses on borrowing
tools from symbolic data analysis (SDA) in the context
of histogram data \citep{histo_PCA_2007, diday, histogram_pca_2017}.
Moreover, in \cite{histogram_PCA_wass} some of these attempts are extended to 
generic distributional data using Wasserstein metrics.

Finally, \cite{geodesic} and \cite{geod_vs_log} propose two PCA formulations
based on the geometric structure of the Wasserstein space: a \emph{geodesic} PCA
and a \emph{log} PCA.
In a similar fashion, the recent preprints of \cite{muller} and \cite{zhang2020wasserstein}
propose linear regression and autoregressive models, respectively, for 
distributional data using the Wasserstein geometry.

We now highlight some key aspects of the aforementioned approaches.
\cite{menafoglio} assumes that
all the probability measures have the same support. 
This is hardly verified in practice, so that to apply their techniques one needs either to truncate the support of some of the probability density functions, or to extend others (for instance, by adding a small constant value and renormalizing), leading to numerical instability as discussed in Sections~\ref{sec:PCA_simulations} and \ref{sec:reg_simulations}.

The SDA-based methods in \cite{histo_PCA_2007, diday, histogram_pca_2017} and \cite{histogram_PCA_wass} share the poor interpretability of SDA.

The methods in \cite{geodesic}, \cite{geod_vs_log}, \cite{muller} and \cite{zhang2020wasserstein} 
are based on the weak Riemannian structure of the Wasserstein space, cf. Section~\ref{sec:weak_riemann}.
Such structure enables the authors to borrow ideas and terminologies from statistical frameworks defined on Riemannian manifolds
 \citep[see][]{extrinsic, intrinsic, pennec_manifolds, 
huck_geod, patra, geod_regression_flet, geod_regression_chakra}.
We can roughly distinguish those frameworks in two main approaches: the intrinsic/geodesic one and extrinsic/log one.

Briefly, intrinsic methods are defined using the metric structure 
of the Wasserstein space, working with geodesic curves and geodesic subsets, so that they faithfully respect the metric of the underlying space.
However, in general, intrinsic methods present many 
practical difficulties in that the optimization problems they lead to are 
usually nontrivial, as we discuss in Section~\ref{sec:new_geod}.
Instances of intrinsic methods for distributional data are the
\emph{geodesic} PCA in \cite{geodesic} and, under some rather restrictive assumptions,
 the linear models in \cite{muller} and the autoregressive models in \cite{zhang2020wasserstein}, see Sections~\ref{sec:vs_int} and \ref{sec:vs_extrinsic}.

On the other hand, extrinsic methods resort to the linear structure 
of suitably defined tangent spaces, by mapping data from the Wasserstein
space to the tangent (through the so-called \emph{log} map) and then mapping back the results to the Wasserstein space (through the \emph{exp} map).
Of course, this approach is less respectful of the underlying geometry
than the intrinsic one, but usually presents several numerical advantages.
An example of such extrinsic methods defined in the Wasserstein space is the \emph{log} PCA in \cite{geod_vs_log}.

The main issue with this \emph{log} PCA is that the image of the \emph{log} map inside the tangent of the Wasserstein space is not a linear space, but rather a convex cone embedded in a linear space
(see Section~\ref{sec:weak_riemann}).
Hence, while exploiting the linear structure of the tangent, it is possible that the projection of some points onto the principal components end up outside of the cone.
For these points, the \emph{exp} map from the tangent to the Wasserstein space used in \cite{geod_vs_log} is not a metric projection, which in general is not
available, so that the results in this setting are hardly interpretable.

\subsection{Our contribution and outline}

The contribution of this work is three folded. First, we propose alternative 
PCA and regression models for distributional data in the Wasserstein space.
We term these models \emph{projected}, in opposition to the \emph{log} PCA in \cite{geod_vs_log}.
Second, by exploiting a geometric characterization of Wasserstein space closely related to its weak Riemannian structure, we build a novel approximation of the Wasserstein space using monotone B-spline. This allows us to 
represent the space of probability measures as a convex polytope in $\mathbb{R}^J$.
Lastly, we obtain faster optimization routines for the \emph{geodesic} PCAs defined in \cite{geodesic}, exploiting the aforementioned B-spline representation.

Our \emph{projected} framework lies in between the \emph{log} one and the \emph{geodesic} one, 
since we use an analogous to the \emph{log} map to transform our data, as 
for extrinsic methods, but do not resort to the \emph{exp} map to return to the Wasserstein
space, using instead the metric projection operator. 
Thanks to this, our \emph{projected} methods are more respectful of the
underlying geometry than the \emph{log} ones, while at the same time retaining
the same reduced computational complexity.
Thus, the \emph{projected} methods expand the range of situations where \emph{extrinsic} methods are an effective and efficient alternative to
intrinsic tools: in our examples, the performance loss in general is marginal (see Section~\ref{sec:PCA_simulations}). 
 
By centering the analysis in appropriate points of the Wasserstein space, one can identify the space of probability measures (with finite second moment) with the space of square integrable monotonically non-decreasing functions on a compact set. We use a suitable quadratic B-spline expansion to get
a very handy representation of such functions.
Through such B-spline expansion, it is possible to approximate the metric projection
onto the Wasserstein space as a constrained quadratic optimization problem
over a convex polytope, that is a well-established problem, cf.\ \cite{POTRA2000281}.
This allows us to exploit the underlying linear structure of an $L_2$ space, so that all the machinery developed for functional data
analysis can be directly applied to this setting.
We address the issue of interpretability of the results, tackling a number of diverse applications and developing different ways to measure the loss of information caused by the \emph{extrinsic} nature of our methods.

We observe that the idea of representing nondecreasing functions through
B-splines for statistical purposes has been proposed also by \cite{das2017bayesian}, in the 
context of Bayesian quantile regression, where the authors
use B-splines with (random) monotonic coefficients as a generative model
for random quantile functions.
However, their focus is on defining a generative model, and not 
on developing a statistical setting exploiting the geometry given by the constrained representation.
Along this direction, they
do not restrict their attention to quadratic splines and consider
cubic ones.

As already mentioned, a further contribution of this work is the derivation of alternative
numerical optimization schemes for the \emph{geodesic} PCA in \cite{geodesic} and \cite{geod_vs_log}, based on the proposed quadratic B-spline expansion.

The remaining of the paper is organized as follows.
Section~\ref{sec:preliminaries} covers the basic concepts
of Wasserstein distance and the weak Riemannian structure of the
Wasserstein space, along with a brief discussion on a suitable way to exploit such structure for our purposes.
Section~\ref{sec:proj_methods} defines the \emph{projected} PCA
and \emph{projected} regression in a general setting.
In Section~\ref{sec:metric} we discuss the choice of the base point in which we center our analysis
and how to efficiently approximate the metric projection through B-splines;
in Section~\ref{sec:empirical} we present the numerical algorithms
needed to compute our \emph{projected} methods and an alternative
optimization routine for the \emph{geodesic} PCA in \cite{geod_vs_log}.
Section~\ref{sec:asympt} discusses the asymptotic properties of the
spline approximation and of the \emph{projected} models, establishing
consistency of the estimators under some assumptions.
Numerical illustrations on real and simulated data sets are shown in Sections~\ref{sec:PCA_simulations} and \ref{sec:reg_simulations}. In particular, we apply our projected methods to two real world problems:
we perform PCA on the US data on Covid-19 mortality by age and sex and 
perform a distribution regression to forecast the wind speed near
a wind farm.
Finally, the article concludes in Section~\ref{sec:discussion}.
The Appendix collects all the proofs of the theoretical results,
additional details on the simplicial PCA and regression, and further
simulations.
Code for reproducing the numerical results is available at
\url{https://github.com/mberaha/ProjectedWasserstein}.

\section{Preliminaries}\label{sec:preliminaries}

In the following, we will consider probability measures on the real line
$\mathbb{R}$ endowed with the usual Borel $\sigma$-field, we will skip
references to the $\sigma$-field whenever it is obvious.

Given a measure $\mu$ on $\mathbb{R}$ define its cumulative distribution function
$F_\mu(x) = \mu((-\infty, x])$ for $x \in \mathbb{R}$ and the associated 
quantile function $F^{-}_\mu(t) =  \inf \{ x \in \mathbb{R}: \ t \leq F_\mu(x) \}$.
When $F_\mu$ is continuous and strictly monotonically increasing, $F^{-}_\mu = \left(F_\mu\right)^{-1}$. 

\subsection{Wasserstein metric and Wasserstein spaces}
\label{sec:wass_metr}

We start by recalling the definition of the 2-Wasserstein 
distance between two
probability measures $\mu, \nu$ on $\mathbb{R}$:
\begin{equation}\label{eq:wass_kolmo}
    W_2^2(\mu, \nu) = \inf_{\gamma \in \Gamma(\mu, \nu)}
     \int_{\mathbb{R} \times \mathbb{R}} | x - y|^2 d\gamma(x, y),
\end{equation}
where $\Gamma(\mu, \nu)$ is the collection of all probability measures on 
$\mathbb{R} \times \mathbb{R}$ with marginals $\mu$ and $\nu$.
Closely related to the definition of Wasserstein distance lies the 
one of Optimal Transport (OT). 
In particular, \eqref{eq:wass_kolmo} identifies the Wasserstein
distance with the minimal total transportation cost between $\mu$ and $\nu$ in
the Kantorovich problem with quadratic cost \citep{ambrosio2008gradient}.

For our purposes, it is convenient to consider another formulation
of the OT problem, originally introduced in \cite{monge1781memoire}.
Given two measures $\mu, \nu$ as before, the optimal 
transport map from $\mu$ to $\nu$ is the solution of the problem
\begin{equation}\label{eq:ot_monge}
\inf_{T: T \# \mu = \nu} \int_\Omega |x - T(x)|^2 d\mu(x),
\end{equation}
where $\#$ denotes the pushfoward operator, that is for any measurable
set $B$ and measurable function 
\begin{equation}\label{eq:push}
f: \mathbb{R} \rightarrow  \mathbb{R}, \qquad  (f\#\mu)(B) = \mu(f^{-1}(B)).
\end{equation}
Note that any solution of \eqref{eq:ot_monge} induces one and 
only one solution of \eqref{eq:wass_kolmo}; moreover if the OT 
problem has a unique solution, then also the Wasserstein distance 
problem has only one solution. 
However not all Wasserstein distance problems can be solved through Monge's formulation \citep{ambrosio2008gradient}.

The unidimensional setting is a remarkable exception in that there 
exist explicit formulas for both problems.
In particular, the Wasserstein distance can be computed as
\begin{equation}\label{eq:wass1d}
     W_2^2(\mu, \nu) = \int_0^1 |F^{-}_\mu(s) - F^{-}_\nu(s)|^2 ds,
\end{equation}
and, if the measure $\mu$ has no atoms, then there exists
a unique solution to Monge's problem given by 
$T_{\mu}^\nu = F^{-}_{\nu} \circ F_\mu$.
For a proof of these results, see Chapter 6 of 
\cite{ambrosio2008gradient}.

It is clear that, in general, the Wasserstein distance between two 
probability
measures can be unbounded (for instance when in \eqref{eq:wass1d} $F^{-}_\mu$ is not square integrable on $[0, 1]$).
Nonetheless, when restricting the focus on the set
of probability measures
with finite second moment, then it holds that $W_2$ defines a 
metric \citep[see, for instance,
Chapter 7 of][]{villani2008optimal}.
Formally, let the Wasserstein space:
\[
    \mathcal{W}_2(\mathbb{R}) = \Big\{ \mu \in 
    \mathcal{P}(\mathbb{R}) \ : 
    \int_\mathbb{R} x^2d\mu < +\infty \Big\}
\]
then $(\mathcal{W}_2(\mathbb{R}), W_2)$ is a separable 
complete metric space.

\subsection{Weak Riemannian structure of the Wasserstein Space}
\label{sec:weak_riemann}

Thanks to the uniqueness of the transport maps, 
by fixing an absolutely continuous (a.c.) probability
 measure $\mu\in \mathcal{W}_2(\mathbb{R})$,
we can associate to any $\nu\in  \mathcal{W}_2(\mathbb{R})$  
the optimal transport map $T_\mu^\nu$.
Since $\int_{\mathbb{R}}|T_\mu^\nu(x)|^2d\mu = 
\int_{\mathbb{R}}x^2d\nu$ we can define the following map $\varphi_\mu:\mathcal{W}_2(\mathbb{R})
\rightarrow L_2^\mu(\mathbb{R})$ with the rule: $\varphi_\mu(\nu)=T_\mu^\nu$.

We note several immediate but interesting properties of the map $\varphi_\mu$. First, it is an isometry (and so a homeomorphism onto its image) since
\[
\int_{\mathbb{R}}|T_\mu^\nu(x)-T_\mu^\eta(x)|^2d\mu = 
\int_{[0,1]}|F_\nu^--F_\eta^-|^2ds = W^2_2(\nu,\eta).
\]
Second, the image of $\varphi_\mu$ is a closed convex cone in $L_2^\mu(\mathbb{R})$:
a set closed under addition and positive scalar multiplication. 
In fact, for any $\lambda \geq 0$, $\lambda T_\mu^\nu$ is still a transport map from 
$\mu$ to another measure whose quantile is $\lambda F_\mu^-$; and similarly 
$T_\mu^\nu+T_\mu^\eta = (F_\nu^-+F_\eta^-) \circ F_\mu$.
Being  $\mathcal{W}_2(\mathbb{R})$ complete, $\varphi_\mu(\mathcal{W}_2(\mathbb{R}))$ is closed in $L_2^\mu(\mathbb{R})$.
Third, $\varphi_\mu(\mu)=id_\mathbb{R}$ (where $id_C$ denotes the identity map of the set $C$).
Finally, as shown in \cite{panaretos}, $\varphi_\mu$ is not surjective and 
$\varphi_\mu(\mathcal{W}_2(\mathbb{R}))$ is the set of $\mu$-a.e. 
non decreasing functions in $L^\mu_2(\mathbb{R})$.

The inverse of the map of $\varphi_\mu$ is the measure pushforward (see Equation \ref{eq:push})
and it is defined on the whole $L_2^\mu(\mathbb{R})$: given $f\in L_2^\mu(\mathbb{R})$, 
then $\nu = f\#\mu$ is a measure in $\mathcal{W}_2(\mathbb{R})$.
In fact:
\[
\int |x|^2 d\nu = \int |f(x)|^2 d\mu = \|f\|^2_\mu
\]

A natural way to define a tangent structure for 
$\mathcal{W}_2(\mathbb{R})$ is therefore to take advantage of the 
cone structure given by $\varphi_\mu$. In fact for closed 
convex cones, there are already notions of tangent cones. 
Similarly to \cite{RockWets98}, Theorem 6.9, we can 
define:
\begin{equation}\label{eq:tan}
\text{Tan}_\mu(\mathcal{W}_2(\mathbb{R})):=
\text{Tan}_{id_\mathbb{R}}(L_2^\mu(\mathbb{R}))=
\overline{\{f\in L_2^\mu(\mathbb{R})|
\exists h>0 : id + hf\in 
\varphi_\mu(\mathcal{W}_2(\mathbb{R}))\}}^{ L_2^\mu(\mathbb{R})}
\end{equation}

We remark that Theorem 6.9 in \cite{RockWets98} is stated in 
$\mathbb{R}^n$, but it holds also more generally, for instance in 
an Hilbert space (see \cite{tangent_cone}, Chapter 4).

A geometric interpretation of \eqref{eq:tan} is the following.
The tangent space consists of all the vectors $f$ that move the base point inside the cone $\varphi_\mu(\mathcal{W}_2(\mathbb{R}))$, when considered up to a scale factor $h$.
Hence,  $f$ plays the role of direction of a tangent vector going out from the tangent point.
Furthermore, since for every $f\in \varphi_\mu(\mathcal{W}_2(\mathbb{R}))$ then $f+id \in \varphi_\mu(\mathcal{W}_2(\mathbb{R}))$ we have that $\varphi_\mu(\mathcal{W}_2(\mathbb{R}))$ is included in the tangent space.
As shown later in this Section, the inclusion is strict and
the tangent space is much larger than $\varphi_\mu(\mathcal{W}_2(\mathbb{R}))$.

Note that we can recover the definition of tangent space given by
\cite{ambrosio2008gradient} and \cite{panaretos}
by a simple \virgolette{change of variable}: calling $g=id+hf$ then substituting
$(g - id)/h$ in \eqref{eq:tan} gives the following definition of
tangent
\[
\text{Tan}_\mu(\mathcal{W}_2(\mathbb{R}))=
\overline{\{\lambda(f-id)|f\in 
\varphi_\mu(\mathcal{W}_2(\mathbb{R})); 
\lambda>0 \}}^{ L_2^\mu(\mathbb{R})},
\] 
which is the one given in \cite{ambrosio2008gradient} and \cite{panaretos}.
As shown in \cite{panaretos} the tangent cone
$\text{Tan}_\mu(\mathcal{W}_2(\mathbb{R}))$ is indeed a linear space.
For this reason we refer to it as tangent space, instead of cone. 

In analogy to Riemannian geometry, following \cite{ambrosio2008gradient} and \cite{panaretos}, we define the $\log_\mu$ and $\exp_\mu$ maps. Having fixed $\mu$ absolutely continuous:
\begin{equation}\label{eq:log_exp}
    \begin{aligned}
        \log_\mu: \mathcal{W}_2(\mathbb{R}) &\rightarrow \text{Tan}_\mu(\mathcal{W}_2(\mathbb{R})) \\
        \nu & \mapsto T_\mu^\nu - id
    \end{aligned}
    \qquad
    \begin{aligned}
     \exp_\mu:\text{Tan}_\mu(\mathcal{W}_2(\mathbb{R})) &\rightarrow \mathcal{W}_2(\mathbb{R}) \\
     f & \mapsto (id + f)\#\mu
    \end{aligned}
\end{equation}

We briefly highlight some properties of these maps; properties which immediately follows from the discussion above.

\begin{rmk}\label{rmk:curvature}
The map $\log_\mu$ is defined on the whole space $\mathcal{W}_2(\mathbb{R})$. Moreover, it is clearly an isometry: $
W_2(\eta,\nu)=\|\log_\mu(\eta)-\log_\mu(\nu)\|_{L^\mu_2(\mathbb{R})}
$ \citep{panaretos}.
This shows that there is no local-approximation issue when working in the tangent space, in contrast with the usual Riemannian manifold setting.
There, the tangent space usually provides good approximation only in a neighborhood of the tangent point. 
\end{rmk}

\begin{rmk}\label{rmk:log_exp_prop}
The map $\log_\mu$ is not surjective on $\text{Tan}_\mu$, indeed its image $\text{Im}(\log_\mu)$ is a closed convex 
subset of $L_2^\mu(\mathbb{R})$ given by all the maps $f$ such that $f+id\in \varphi_\mu(\mathcal{W}_2(\mathbb{R}))$, that is, $f+id$ is $\mu$-a.e. increasing.
The restriction of $\exp_\mu$ on $\text{Im}(\log_\mu)$, henceforth denoted by $\exp_{\mu|\log_\mu(\mathcal{W}_2(\mathbb{R}))}$,
is an isometric
homeomorphism and its inverse is $\log_\mu$.
In particular, we observe that $\log_\mu \circ \exp_\mu$ is not a metric projection in $L_2^\mu$. That is, in general $\log_\mu \circ \exp_\mu(f)\neq \argmin_{g\in\text{Im}(\log_\mu)} ||f-g||_{L_2^\mu}$ .
\end{rmk}

\subsection{Intrinsic and extrinsic methods in the Wasserstein space}

As mentioned in Section~\ref{sec:prev_work}, borrowing ideas from 
Riemannian geometry leads to discerning statistical methods on 
the Wasserstein space in the classes of \emph{intrinsic} and \emph{extrinsic}
methods.

The Weak Riemannian structure presented in Section \ref{sec:weak_riemann} provides a suitable environment for developing intrinsic methods. In fact, the geodesic structure of $\mathcal{W}_2(\mathbb{R})$ can be recovered through the linear structure of any $L_2^\mu(\mathbb{R})$ space through the isometry $\varphi_\mu$. Pointwise interpolation of the transport maps coincide with the geodesic between measures. In other words, given $\mu$ a.c., the geodesic between $\nu$ and $\eta$ is given by:
\begin{equation}\label{eq:wass_geodesic}
\gamma(t) = ((1-t)\cdot T_{\mu}^\nu + t\cdot T_{\mu}^\eta)\#\mu	
\end{equation}
Thus, such geodesic structure can be recovered in many different 
(but equivalent) ways, depending on $\mu$. 

On the other hand, Remark~\ref{rmk:curvature} motivates the 
development of extrinsic tools, since working in the 
image of  $\text{log}_\mu$ inside the tangent space 
$\text{Tan}_\mu$ is exactly like working in 
$\mathcal{W}_2(\mathbb{R})$.
This is not common in Riemannian manifold framework, 
since usually the 
tangent space provides a good approximation only near to the tangent point.
As a consequence, if in the general Riemannian manifold framework the choice 
of the tangent point $\mu$ is
crucial (since results for extrinsic methods might be significantly altered
for different choices of $\mu$)  when working with 
$\mathcal{W}_2(\mathbb{R})$ this is not the case.

To further motivate this key point, consider $\mu$ and $\nu$ a.c. measures; 
the maps 
$\allowbreak \log_\nu \circ
(\text{exp}_{\mu|\text{log}_\mu(\mathcal{W}_2(\mathbb{R}))})$  
and $\varphi_\nu\circ\varphi_\mu^{-1}$ are isometric 
homeomorphisms (as composition of isometries and homeomorphisms). 
In other words, they preserve distances and send border elements of 
$\text{log}_\mu(\mathcal{W}_2(\mathbb{R}))$ or 
$\varphi_\mu(\mathcal{W}_2(\mathbb{R}))$  into border elements of $\text{log}_\nu(\mathcal{W}_2(\mathbb{R}))$ and 
$\varphi_\nu(\mathcal{W}_2(\mathbb{R}))$, respectively,
and the same with internal points (and so in particular, they 
preserve distances from any point to the border).
In \cite{muller}, \cite{geodesic} and \cite{zhang2020wasserstein} $\mu$ is
chosen as the barycentric measure $\bar x$  of the observations $x_i\in \mathcal{W}_2(\mathbb{R})$.
The discussion above implies that considering the tangent space 
at the Wasserstein barycenter $\bar{x}$ and working on
$\text{log}_{\bar{x}}(x_i)= \text{log}_{\bar{x}}(x_i)-
\text{log}_{\bar{x}}(\bar{x})$
is exactly the same as considering 
the tangent space at any $\mu$ a.c. and working  on 
$\text{log}_{\mu}(x_i)-\text{log}_{\mu}(\bar{x})$ 
for our statistical purposes.
So the choice of the tangent space from the theoretical point of view is completely arbitrary.

Moreover, centering the analysis in the barycenter presents a drawback when studying asymptotic properties
of the models under consideration, since $\bar x$ changes as the sample size
grows.
In Section~\ref{sec:uniform} we propose to fix $\mu$ as the uniform measure
on $[0, 1]$.
This choice not only allows us to derive empirical methods that are extremely
simple to implement, cf.\ Section~\ref{sec:empirical}, but also allows us
to study asymptotic properties of the models in Section~\ref{sec:consistency}
without resorting to parallel transport, as done 
for instance in \cite{muller}.

\subsection{Tangent vs. $L_2^\mu$}\label{sec:tangent_vs_l2}

Lastly, we briefly discuss the major differences between using 
a tangent space representation of $\mathcal{W}_2(\mathbb{R})$ and 
using the representation given by some $\varphi_\mu$.

We recall that, for a fixed $\mu$ a.c., the two representations are indeed quite similar $\varphi_\mu(\nu) = T_\mu^\nu $, $\text{log}_\mu(\nu)= T_\mu^\nu-id$; a priori one may prefer the tangent representation, because it already expresses data as vectors coming out of a point. Therefore, for instance, it might result practically more convenient to center the analysis in the barycenter and work on vectors, taking away any \virgolette{data centering} issues. 
At the same time, also notational coherence with already existing methods might benefit from this choice.

However, especially when dealing with extrinsic techniques, we 
found slightly more practical to use the $\varphi_\mu$ representation in that it is more straightforward to represent $\varphi_\mu(\mathcal{W}_2(\mathbb{R}))$ compared to $\text{log}_\mu(\mathcal{W}_2(\mathbb{R}))$: the first one can in fact be represented directly as the cone of the $\mu$-a.e non-decreasing functions.

\section{Projected Models in the Wasserstein Space}
\label{sec:proj_methods}

In this section, exploiting the embeddings given by $\varphi_\mu$, we
define a class of \emph{projected} statistical methods to perform
extrinsic analysis for data in the Wasserstein space.

To give a general framework, we do not restrict our attention to a particular 
$\varphi_\mu$ yet, even though in Section~\ref{sec:metric} we argue that
a natural choice which allows an easier implementation of the empirical methods is letting $\mu$ be the uniform distribution on $[0, 1]$.
Hence, for the sake of notation, we consider a generic case of data
lying in a closed convex cone $X$ inside a separable Hilbert space $H$.
In our setting, $H$ would be $L_2^\mu(\mathbb{R})$ and $X=\varphi_\mu(\mathcal{W}_2(\mathbb{R}))$, for some $\mu\in\mathcal{W}_2(\mathbb{R})$ absolutely continuous.

\subsection{Principal component analysis}
\label{sec:proj_PCA}

We start by defining one of the main contributions of our work: the \emph{projected} PCA.
We recall that for an $H$-valued random variable $\Xvar$, 
PCA is a well established technique and amounts to finding the eigenfunctions
of the Karhunen-Lo{\'e}ve expansion of the covariance operator of $\Xvar$, see
\cite{ramsay}.
Observe that any $X$-valued random variable can be considered as an $H$-valued one
(by the inclusion map), so that a notion of PCA is already available. 

When defining principal components, a key notion is the one of dimension
of the principal component (PC). 
In this work, principal components will be closed convex subsets of $H$, 
and we will always define the dimension of a subset of $H$ as the dimension of the smallest affine subset of $H$ containing it.
For a generic closed convex set $C \subset H$, let $\Pi_C$ denote the metric projection onto $C$: $\Pi_C(x):= \argmin_{c\in C}||x-c||$
and, for a set of vectors $U$, denote with $Sp(U)$ its linear span.

In what follows, we denote by $x_0$ the \virgolette{center} of the PCA. For us, $x_0 = \mathbb{E}[\Xvar]$, or its empirical counterpart.
To have a well defined PCA, we always assume that $x_0$ belongs to the relative interior of the convex hull of the support of $\Xvar$, see Appendix~\ref{sec:proofs} for the definition of relative interior and further details. This is a rather technical hypothesis but it is not a restrictive one. For instance, it is always verified for empirical measures and when $X \subseteq \mathbb{R}^d$ and hence for our empirical methods, cf. Section~\ref{sec:emp_pca}.

\begin{defi}{(Projected PCA).}\label{def:proj_pca}
Given $\Xvar$ a random variable with values in $X \subset H$, let $U_k=\{w_1,...,w_k\}$ be
its first $k$ $H$-principal components centered in $x_0 = \mathbb{E}[\Xvar]$. 
A $(k,x_0)-$projected principal component of $\Xvar$
is the biggest closed convex subset $U_X^{x_0,k}$ of $X$ such that:
\begin{enumerate*}[label=(\roman*)]
    \item $x_0 \in U_X^{x_0,k}$,
    \item $dim(U_X^{x_0,k}) = k$, and
    \item $U_X^{x_0,k} \subseteq \Pi_X(\Sp(U_k))$.
\end{enumerate*}
\end{defi}

In other words, the projected principal component is obtained by approximating the 
span of the principal components found in $H$, with convex subsets in $X$.
Note that the principal components in $H$ might \virgolette{capture} some variability
which is not present when measuring distances inside $X$.
In fact the projection of a point belonging to $X$ onto a direction
$w_j$ might end up  being outside $X$, see Section~\ref{sec:vs_int}.
However, as we will show in Section~\ref{sec:PCA_simulations}, in our examples
the projected PCA behaves well and this issue does not seem to affect significantly
the performance.

\begin{rmk}
Convex sets are essential in our analysis since, thanks to \eqref{eq:wass_geodesic} convex sets in $X$ are precisely the subsets of $\mathcal{W}_2(\mathbb{R})$ which are geodesically complete: the geodesic connecting any pair of points in the subset, is contained in the subset. Geodesic subsets are a natural generalization of linear spaces.
\end{rmk}

\begin{rmk}\label{rmk:convex_proj}
The metric projection of a linear subspace onto a convex subset can end up being a nonconvex set. In addition to that, while loosing convexity, the dimension of the metric projection of a convex subset can be bigger of the dimension of the original subset. 
A simple example where both cases happen is
the projection of $y=-x$ onto $x,y\geq 0$ in $\mathbb{R}^2$. 
\end{rmk}

We observe that inside a projected principal component, we have a preferential orthonormal basis given by the principal components in $H$; for this reason we call $U_k=\{w_1,...,w_k\}$ \emph{principal directions}.

Although it might seem impractical to find the projected component, the following Lemma provides a more convenient alternative characterization.

\begin{lemma}\label{lemma:pca_easy}
Let $x_0$ and $U_X^{x_0,k}$ be as in Definition~\ref{def:proj_pca}, then $U_X^{x_0,k} = (x_0+\Sp(U_k)) \cap X$.
\end{lemma}

Natural alternatives to Definition~\ref{def:proj_pca} would be, for instance,
to let the projected principal directions (component) be the metric projection of $w_1, \ldots, w_k$ (the linear span of $\{w_1, \ldots, w_k\}$) onto $X$, respectively.
In the former case, the projection would not guarantee the orthogonality of the
projected directions, which is instead essential to properly explore the
variability.
Moreover, since the \virgolette{tip} of the projected unit vectors would likely
lie on the border of $X$, the projection of a new observations on a direction would still lie outside of $X$ as soon as the score associated
to that direction is larger than $1$.
The latter case, instead, presents the drawbacks pointed out in Remark~\ref{rmk:convex_proj}.

We argue that, despite its simplicity, Definition \ref{def:proj_pca}
is indeed very well suited for statistical analysis in the Wasserstein Space.
For instance, we are guaranteed that, as the dimension grows up, the $k$ projected components provide a monotonically better fit to the data. This is easily verified because $\Pi_X$ is a strictly non-expansive operator, being $X$ closed and convex (see \cite{metric_projection}), which implies the following Proposition.

\begin{prop}\label{prop:err_convergence}
With the same notation as Definition~\ref{def:proj_pca}, for any $x\in X$ we have: 
\[
    \| \Pi_{U_X^{x_0,k}}(x)-x \| \geq \|\Pi_{U_X^{x_0,k+1}}(x)-x\|\rightarrow 0 \text{ with } k\rightarrow +\infty.
\]
\end{prop}

Once a principal component is found, a classical task that one may want to
perform is to project a new \virgolette{observation} $x^{*} \in X$ onto $U_X^{x_0, k}$,
for instance for dimensionality reduction purposes.
In general, the metric projection on generic convex subsets might be arduous
to find, we will deal with this issue in Section~\ref{sec:metric}.
Nevertheless, we can use the following Proposition to reduce in advance the 
dimension of the parameters involved in the problem; turning it into a 
projection problem inside the principal projected component, which allows
for faster computations (see Equation~\ref{eq:proj_emp}).

\begin{prop}\label{prop:metric_proj}
    Let $x^* \in X$ and let $\Pi_k$ be the orthogonal projection on $Span(U_k)$. The projection of $x^*$ onto $U_X^{x_0, k}$ is given by
    \begin{equation}\label{eq:pca_proj}
    \argmin_{v'\in U_X^{x_0,k}} \|x^* - v' \| = \Pi_{Sp(U_k)\cap (X-x_0)}(\Pi_k(x^*-x_0))+x_0.
    \end{equation}
\end{prop}

Lastly, we observe that, since projected principal components are not linear subspaces, the scores of some points on a principal direction can vary as we increase the dimension of the principal component.

\subsection{Regression}\label{sec:proj_reg}

Broadly speaking, a regression model between two variables with values in two different spaces is given by an operator between such spaces, which for every input value of the independent variable, returns a predicted value for the dependent variable.
In the following, let us denote with $\Zvar$ the independent variable
and with $\Yvar$ the dependent one.
A regression model is usually understood as an operator $\Gamma$ specifying the conditional value of $\Yvar$ given $\Zvar$, that is, $\mathbb{E}[\Yvar | \Zvar] = \Gamma(\Zvar)$.

If the spaces where $\Zvar$ and $\Yvar$ take values possess a linear structure,
this linearity is usually exploited by means of a (kernel) linear operator, with possibly
an \virgolette{intercept} term.
To define our \emph{projected} regression model, we want to
exploit the cone structure of $X$ in a similar fashion.
In fact, such linear kernel operators combine good optimization properties and interpretability since their kernels can provide insights into the analysis, much like coefficients in multivariate linear regression.

We treat separately the cases where the $X$-valued variable is the independent
or the dependent one. The case when both variables are $X$-valued follows naturally.
To keep the notation light, in what follows we will not distinguish between \virgolette{proper} linear operators and linear operators with an added intercept term, which could as well be employed in all the incoming definitions to gain flexibility. 

Consider the case in which we have an independent $X$-valued random variable, and denote with $V$
the space where the dependent variable takes value.
Despite the fact that $X$ is not a linear space, with an abuse of notation, we call \virgolette{linear} an operator which respect sum and positive scalar multiplication for elements in $X$.
Such operators are in fact obtained by restricting on $X$ linear operators defined on $H$.
Following this idea, in order to define linear regression for an $X$-valued independent random variable, we consider such variable as $H$-valued, obtain the regression operator and then take the restriction of the operator on $X$. 
In this way, when $ H=L_2^\mu(\mathbb{R})$ and $X=\varphi_\mu(\mathcal{W}_2(\mathbb{R}))$, it is possible to exploit the classical FDA framework to perform 
all kinds of distribution on scalar/vector/etc... regression.
For brevity, we report only the definition with $V=\mathbb{R}$.

\begin{defi}\label{defi:dist_on_scalar}
Let $\Zvar$ an $X$-valued random variable, and $\Yvar$ a real valued one. 
Let $\Gamma_\beta:H\rightarrow \mathbb{R}$ be a functional linear regression model 
for such variables, with $\Zvar$ considered as $H$-valued and $\Gamma_\beta(v)= \langle \beta,v \rangle$.
A projected linear regression model for $(\Zvar,\Yvar)$ is given by $(\Gamma_\beta)_{|X}.$
\end{defi}

Now we turn to the cases which feature an $X$ valued independent variable and
a $Z$ valued dependent one, for $Z$ a generic Hilbert space.
Through the inclusion $X \hookrightarrow H$, we can consider a regression problem with $X$-valued dependent variable, as a problem with $H$-valued dependent variable.
Comparing this situation with the previous one, it is clear that we now 
face a \virgolette{dual} problem.
Indeed, while before we needed to restrict the domain from $H$ to $X$,
we now need to force the codomain of $\Gamma$ to lie inside $X$.
We would like to retain the same properties that make linear kernel operators appealing as regression operators between Hilbert spaces. 
A possibility could be considering a linear kernel operator $\Gamma$ with values in $H$ and restricting it to $\Gamma^{-1}(X)$.
However, this would imply that for any $z \not\in \Gamma^{-1}(X)$ no prediction
would be available.

We argue that a more reasonable approach consists in finding an operator
$\Gamma_{\mathrm{P}}: Z \rightarrow X$ as close as possible (in some sense that will be clear later) to the 
linear kernel operator $\Gamma$ aforementioned.
Hence, we relax the linearity assumption in favor of Lipschitzianity, and take as regression operator $\Pi_X\circ\Gamma$, whose image always lies in $X$.
Note that $\Gamma_{\mathrm{P}}$ inherits the interpretability of the kernel of $\Gamma$.

To motivate such choice, we give the following notion of a projected operator.
\begin{defi}
    Let $Z$ be a normed space and consider $\Zvar$ a $Z$-valued random variable.
    Let $\Gamma : Z \rightarrow H $ a generic Lipschitz operator between $Z$ and
    $H$. 
    A $(\Zvar, X)$-projection of $\Gamma$ is an operator 
    $\Gamma_{\mathrm{P}}: Z \rightarrow X$ such that:
       \[
           \Gamma_{\mathrm{P}}= \argmin_{T:Z \rightarrow X} \mathbb{E}_{\Zvar}[\|\Gamma(v)-T(v)\|^2]
       \]
\end{defi}

In other words, $\Gamma_{\mathrm{P}}$ provides the best pointwise approximation of the 
$H$-valued operator $\Gamma$, averaged w.r.t. the measure induced by $\Zvar$.
Hence, given a $\Zvar$ a $Z$-valued random variable and $\Yvar$ an $X$-valued random
variable and a linear regression model $\Gamma: Z \rightarrow H$ for $(\Zvar,\Yvar)$, 
the projected regression model induced by $\Gamma$ is $\Gamma_{\mathrm{P}}$. 

\begin{prop}\label{prop:projected_reg}
With the same notation as above, if $\mathbb E\left[ \|\Zvar\|^2 \right] < \infty$, then $\Gamma_{\mathrm{P}} = \Pi_X\circ\Gamma$. 
\begin{proof}
For any $T:Z \rightarrow X$, it holds: $\|\Gamma(z)-\Pi_X( \Gamma(z) )\|\leq \|\Gamma(v)-T(v)\|$. Moreover, $\Gamma$ and $\Pi_X \circ \Gamma$ are Lipschitz, and being $\Pi_X$ non-expansive, they share the same constant $L>0$:
\[
 \|\Gamma(v)-\Pi_X\circ\Gamma(v)\|^2\leq 2  L\|v\|^2 
\]
and thus $\mathbb{E}_{\Zvar}[\|\Gamma(z)-\Pi_X \circ \Gamma(z)\|^2]$ is bounded iff $\Zvar$ has finite second moment.
\end{proof}
\end{prop}

The only case left out from the treatment above is when both the 
independent and the dependent variables are X-valued.
This case, however, follows naturally by combining the two approaches
and we report the definition below.
\begin{defi}\label{def:proj_reg}
Let $\Zvar$ and $\Yvar$ two $X$-valued random variables. 
Let $\Gamma:H\rightarrow H$ be a functional linear regression model 
for the variables considered as $H$-valued.
A projected linear regression model for $(\Zvar,\Yvar)$ is given by $(\Pi_X \circ\Gamma)_{|X}.$
\end{defi}

\begin{rmk}\label{rmk:extension}
When considering a regression with $X$-valued independent variable, one may want to relax the restriction on $X$ in Definition~\ref{defi:dist_on_scalar} for various reasons; for instance one may have measurement errors, or by design the test set may consider points also outside $X$. In such cases it is worth considering the problem of how many continuous linear extensions of $\Gamma_{|X}$ are possible on the whole $H$. 
A sufficient condition for the uniqueness of such extension is the following:
there exist a sequence of linear subspaces of $H$, say $\{H_J\}_{J \geq 1}$, such that $\bigcup_J H_J$ is dense in $H$ and $X_J := H_J \cap X$ contains a basis of $H_J$ for every $J$. 
\end{rmk}

\begin{rmk}
When $ H=L_2^\mu(\mathbb{R})$ and $X=\varphi_\mu(\mathcal{W}_2(\mathbb{R}))$ the condition in Remark~\ref{rmk:extension} is verified, for instance, by Remark~\ref{rmk:basis} in Section~\ref{sec:splines}.
Moreover, observe that the uniqueness of the extension can also be proven thank to Jordan's representation of functions $f:\mathbb{R}\rightarrow\mathbb{R} $ with bounded variation (BV).
In fact any $f$ with BV can be written as the difference of monotone functions and thus $\Gamma(f)$ is fixed. Then by the density of BV functions in $H$, we define $\Gamma$ on the remaining elements of $H$.
\end{rmk}

\subsection{Comparison with intrinsic methods}\label{sec:vs_int}

\begin{figure}[t]
\centering
\includegraphics[width=0.5\linewidth]{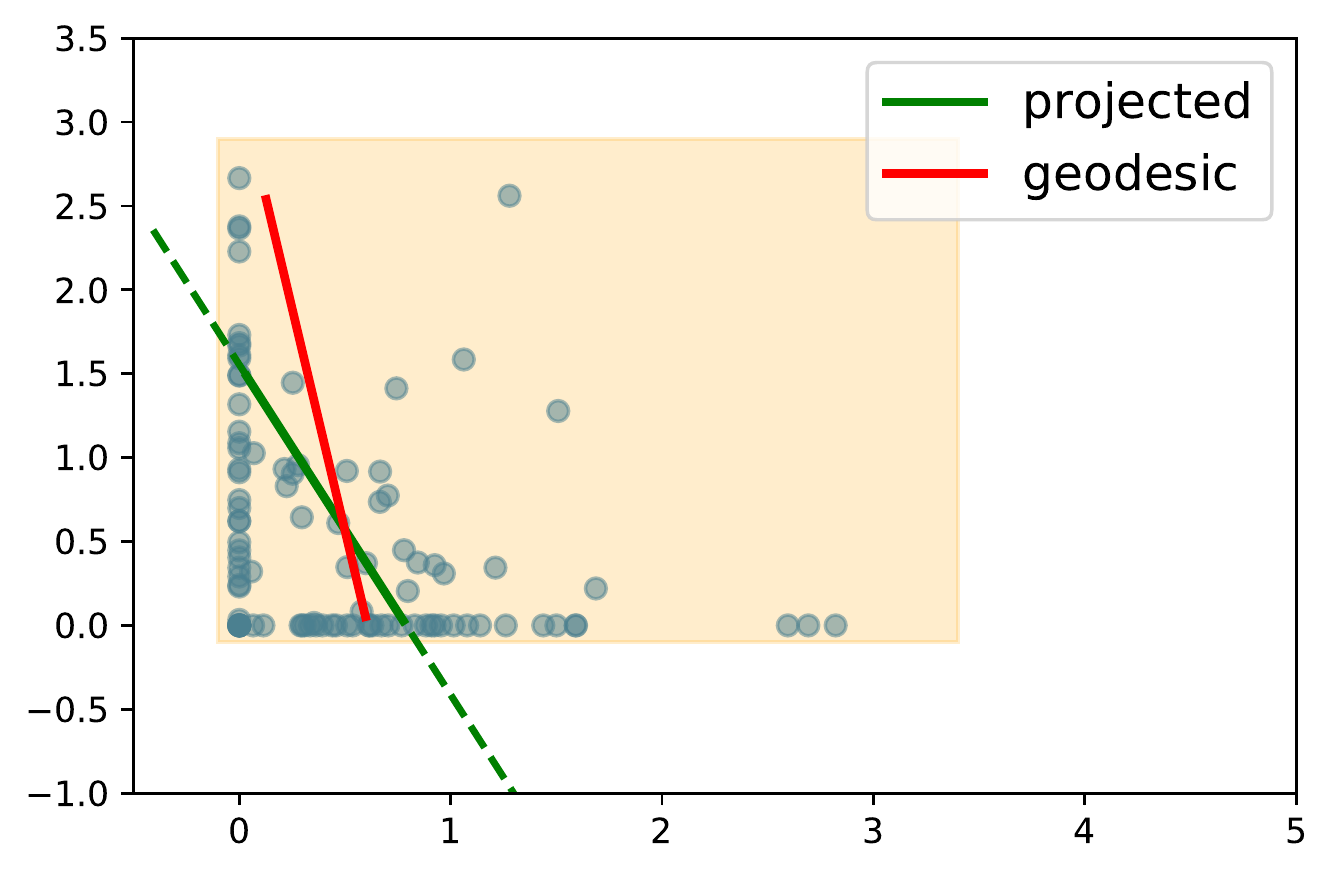}
\caption{Comparison of projected and geodesic PCA when $H=\mathbb{R}^2$ and $X$ is the shaded rectangle. The projected principal direction is rather different
from the geodesic one because most of the observations (blue dots) are concentrated around the borders}
\label{fig:geod_vs_proj}
\end{figure}

We now compare the projected methods defined earlier in this Section
and the intrinsic counterparts.
In particular, we focus on the \emph{geodesic} PCA defined in \cite{geodesic} and \cite{geod_vs_log} and on the distribution on distribution regression model in \cite{muller}.

\cite{geodesic} and \cite{geod_vs_log} define two different PCA, namely a 
global and a nested one; in particular the nested approach
presents analogies with other PCAs developed for manifold valued random variables \citep{jung2012analysis, huckemann2018backward, pennec2018barycentric};
we report the two definitions below.

\begin{defi}\label{defi:global}(Global geodesic PCA)
Let $\Xvar$ a random variable with values in $X$ with $\mathbb{E}[\Xvar] = x_0$.
A $(k, x_0)$-global geodesic PC is a set $C^*$ minimizing
$\mathbb{E}\left[d(\Xvar, C)^2\right]$ over the closed convex sets $C \subset X$ such that $x_0 \in C$ and $dim(C) \leq k$
\end{defi}

\begin{defi}\label{defi:nested}(Nested geodesic PCA)
Let $\Xvar$ a random variable with values in $X$ with $\mathbb{E}[\Xvar] = x_0$.
For $k=1$, a $(k, x_0)$-nested geodesic PC is a set $C^*_k$ such that $C^*_k$ is a minimizer of $\mathbb{E}\left[d(\Xvar, C)^2\right]$ over the closed  convex sets $C \subset X$ such that $x_0 \in C$ and $dim(C) \leq k$; for $k\geq 1$, a $(k, x_0)$-nested geodesic PC is a set $C^*_k$ such that $C^*_k$ is a minimizer of $\mathbb{E}\left[d(\Xvar, C)^2\right]$ over the closed  convex sets $C \subset X$ such that: $x_0 \in C$, $dim(C) \leq k$, and $C \supset C^*_{k-1}$, where $C^*_{k-1}$ is a  $(k-1, x_0)$-nested geodesic PC.
\end{defi}

The first key difference between the global and the nested geodesic PCA is that
the latter provides a notion of preferential directions in the principal 
component, while the first one does not.
In fact, the first nested principal component corresponds to the first
principal direction, and it is possible to find the remaining principal 
directions by imposing orthogonality constraints as we obtain nested PCs 
of higher dimensions.
Thus, the nested geodesic PCA is more suitable to explore and visualize
the variability in a data set, see also Section~\ref{sec:PCA_simulations}.
On the other hand, exactly because of the lack of such constraints, the
global PCA is in general more flexible and provides superior performance
in terms of \emph{reconstruction error}, cf. \ Section~\ref{sec:PCA_simulations}.

Comparing these definitions with the one of our projected PCA,
the key difference is that geodesic PCAs do not exploit 
the Hilbert structure of $H$. 
Thus, as we discuss in Section~\ref{sec:new_geod}, the numerical
routines needed to find such principal components rely on nonlinear
constrained optimization, which can be extremely demanding and nontrivial
to implement. This is in sharp contrast with our projected PCA
in Definition~\ref{def:proj_pca}, that, thanks to Lemma~\ref{lemma:pca_easy}
can be straightforwardly computed.
However, as a result, the projected PCA is in general less respectful of
the underlying metric structure. By investigating this issue in simpler
settings, for instance when $H = \mathbb{R}^d$ and $X$ is a convex polytope
in $\mathbb{R}^d$, we noticed that the differences between the 
projected principal directions and the nested geodesic ones become
appreciable only if the random variable $\Xvar$ gives significant probability
to values near the borders of $X$. See for instance Figure~\ref{fig:geod_vs_proj}.
While this intuition remains valid also in the more complex setting that we investigate in this paper, it is harder to imagine realizations of $\Xvar$
near the borders of $X$.

Note that the interpretability of the projected PCA is determined by 
the level of discrepancy between the definitions, as in Figure~\ref{fig:geod_vs_proj}, which depends on how much variability it is correctly captured by the component, that is how much of the variability captured by the projected component lies in $X$.
This intuition is formalized in Section~\ref{sec:reliability} where
two measures of \virgolette{reliability} of the projected PCA are proposed.

Turning to the regression context, \cite{muller} define a distribution on distribution linear regression model 
in the Wasserstein space.  Their approach considers two different tangent spaces
of $\mathcal{W}_2(\mathbb{R})$ (the first one centered in the 
barycenter of the independent variable and the second one centered
in the barycenter of the dependent variable) and map the observations
to the corresponding tangent spaces. They then use FDA tools to
estimate a functional linear model $\widehat{\Gamma}$ between those two spaces.
When the image of the regression operator
$\Gamma$ lies inside the image of the log map centered in the
dependent variable's barycenter, their distribution on distribution regression can be considered a properly intrinsic method.
This assumption is used to prove asymptotic properties of their methodology, but as the authors in \cite{muller} notice, is hardly verified in practice,
so that whenever the output of the regression operator is not a distribution,
they resort to squeezing such a value with some scalar multiplication,
namely \virgolette{boundary projection}, which in general is not a metric projection.
The boundary projection step gives an extrinsic nature to their model and we provide further comparisons with our methods in Section~\ref{sec:vs_extrinsic}.

\subsection{Comparison with other extrinsic methods}
\label{sec:vs_extrinsic}

\begin{figure}[t]
\centering
\includegraphics[width=0.8\linewidth]{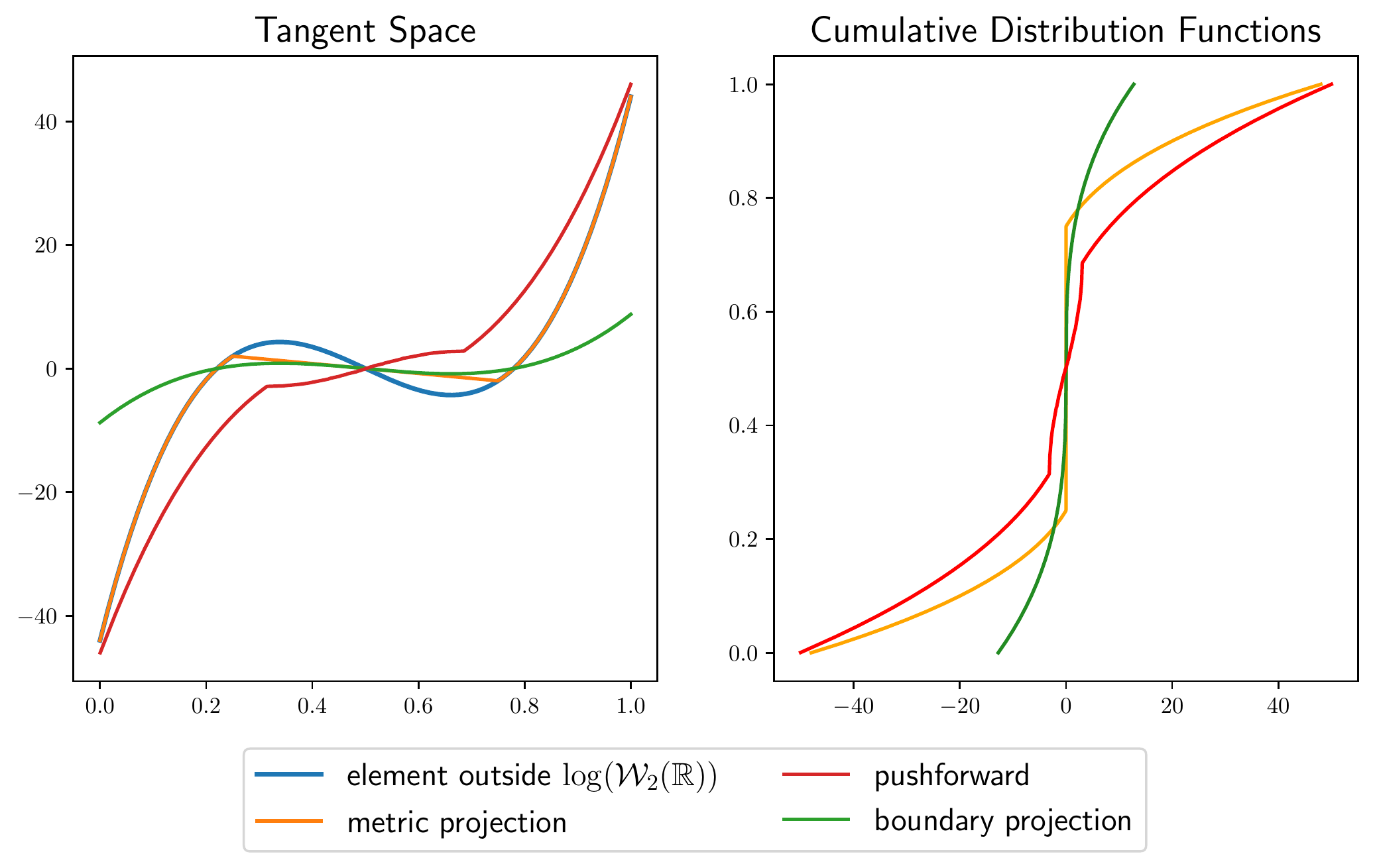}
\caption{
Comparison between different projections onto $X$ for a point $x \in H \setminus X$ (blue line) in the tangent space (left panel) and the associated cumulative distribution functions (right panel) when the base point $\mu$ is the uniform measure on $[0, 1]$.
The orange, green and red curves are obtained with metric projection, boundary projection and $\log_\mu \circ \exp_\mu$ respectively.}
\label{fig:metric_vs_boundary_vs_push}
\end{figure}

In this section, we offer a comparison of our projected methods with other extrinsic methods, namely the log PCA in \cite{geod_vs_log} and the distribution on distribution regression in \cite{muller}, which, as outlined in the previous section, may behave as an extrinsic method. Let us start with the former.

\cite{geod_vs_log} propose the definition of a log PCA as
an alternative to the geodesic PCAs in \cite{geodesic}.
Both the log and the projected PCA are extrinsic
methods: they proceed by carrying out the PCA in a linear space $H$ and 
then map back the results to the Wasserstein space, following an approach which had already been proposed by \cite{fletcher2004principal}.

For the log PCA, $H$ is the tangent space at $\mu$, for the projected $H$ is $L_2^\mu(\mathbb{R})$.
Given $U_k = \{w_1, \ldots, w_k\}$ the first $k$ $H$-principal components,
the log principal component in $\mathcal{W}_2(\mathbb{R})$ is $\exp_\mu(Sp(U_k))$ .
Analogously, by considering the convex cone $X =: \log_\mu(\mathcal{W}_2(\mathbb{R})) \subseteq H$, the principal component in $X$ is $\log_\mu\left(\exp_\mu(Sp(U_k))\right)$.

We notice two key differences between the log and projected PCA.
First, as pointed out in Remark~\ref{rmk:log_exp_prop}, 
$\text{log}_\mu \circ \text{exp}_\mu$ is not a metric projection in $L_2^\mu$
so that given a point $x \in H \setminus X$, $\log_\mu(\exp_\mu(x))$ might end
up being extremely different from $x$. See for instance Figure~\ref{fig:metric_vs_boundary_vs_push}
where for a point $x$ (blue line) that is close (in the $L_2^\mu$ norm) to $X$, $\log_\mu(\exp_\mu(x))$ turns out
to be quite far from $x$.
In the context of PCA, this means that as soon as the projection onto $Sp(U_k)$ of
observation lies outside of $X$, the log PCA quickly loses its interpretability.
Second, as discussed in Remark~\ref{rmk:convex_proj}, there is no guarantee
that $\log_\mu\left(\exp_\mu(Sp(U_k))\right)$ is contained in $Sp(U_k)$,
its dimension might increase and it might not even be convex.
For this same reason, in general, log PCA cannot define a set of (orthogonal) principal directions which span the principal component. Hence, it is not possible to work directly on the scores of the PCA. 

Combined, we believe that the above mentioned issues present a
major drawback of the log PCA when compared to the projected
PCA, as they prevent the possibility of doing proper dimensionality reduction and working on the scores of data points on the principal components. 
Finally, we also point out that approximating the $\exp_\mu$ map
is a nontrivial task, involving computing numerically the preimages 
of an arbitrary large number of sets and numerical differentiation, that 
can lead to numerical instability of the log PCA.

We end this disccussion with a comparison between the boundary projection in \cite{muller} and the metric projection. Their difference, for a possible regression output $x \in H \setminus X$
 is depicted in Figure~\ref{fig:metric_vs_boundary_vs_push}.
Note that, by construction, such a procedure shrinks the tails of the output.
Even when the regression output is slightly outside the image of the log map, the boundary projection result can be extremely far from the regression 
output and from the metric projection in terms of Wasserstein distance.
For example, in Figure~\ref{fig:metric_vs_boundary_vs_push}, the regression output 
and the projected method assign positive
probability to values in the range $[-45, 45]$, while the output of the
boundary projection assigns zero probability to values outside  $[-17, 17]$.
This underrepresentation of the variability might be a crucial issue
depending on the application considered. 

\section{Computing the metric projection through B-spline approximation}
\label{sec:metric}

The projected methods defined in Section~\ref{sec:proj_methods}
depend heavily on the availability of projection operators on the
closed convex cone $X = \varphi_\mu(\mathcal{W}_2(\mathbb{R}))$.
Being $X$ a cone inside a linear space, such operators are always
well defined, but their implementation might be nontrivial.
In this Section, we present a possible solution to this problem,
based on choosing a particular $\mu$ as base point and 
constructing a B-spline representation of the cone $X$.

\subsection{Choosing $\mu$ as the uniform distribution on $[0, 1]$}
\label{sec:uniform}

As already mentioned, our projected methods can be carried out
by choosing $\mu$ arbitrarily and there is no theoretical difference
between different choices of $\mu$, cf.\ Section~\ref{sec:weak_riemann}.
Nonetheless, in practice, a clever choice of $\mu$ can lead to
substantially easier and more numerically stable algorithms.
For instance, by choosing a measure $\mu$ with compact support $C$ in $
\mathbb{R}$, 
then the ambient space becomes $L_2^\mu(C)$ since we work up to zero-measure sets.
This greatly simplifies any numerical procedure since we could work with grids over bounded sets, and do not need to resort to any truncation procedure, which would be mandatory in case the support of $\mu$ was unbounded.
Moreover, note that evaluating the maps $\varphi_\mu$ in a certain measure
$\nu$ amounts to computing the transport map $T_\mu^\nu = F^-_\nu \circ F_\mu$,
hence it is clear that the choice of $F_\mu$ numerically influences the results.

For the aforementioned reasons, we argue that a reasonable choice is to center our analysis in $\mu=U([0,1])$.
In fact, in this case, $L_2^\mu(\mathbb{R})=L_2([0,1])$, and 
$F_\mu =id_{[0,1]}$ (the transport maps are simply given by quantile functions).

\subsection{Metric Projection}

Having chosen $\mu$ as Section~\ref{sec:uniform} leads to an explicit
characterization of the image of $\varphi_\mu$ as the set of square integrable a.e. non-decreasing functions on $[0, 1]$.
Hence, the operator $\Pi_X$ in Section~\ref{sec:proj_methods}
is the metric projection onto the cone of a.e. non-decreasing functions 
in $L_2([0,1])$.

Projection onto monotone functions has been widely studied in the field of
\emph{order restricted} inference, \citep{anevski2006general, dykstra2012advances}.
For instance, in \cite{anevski2011} an explicit characterization of such
a projection is given, which however does not lead to a closed form solution, while in \cite{PAVA} several numerical algorithms
to approximate the projection operator are proposed.
Those algorithms are based on approximating the function to be projected
with a step function defined on $n$ intervals and can be shown to have a
computational complexity that is linear in $n$ \citep{active_set}.

Despite the numerical convenience of the aforementioned  
approximations, we believe that they are not suited for distributional
data analysis.
First and foremost, suppose that observations are given as probability
density functions, so that one may want to interpret the results of a PCA,
for instance, in terms of pdfs and not of quantile functions.
If one were to estimate discontinuous principal directions through 
any of the algorithms in \cite{PAVA}, it would not be possible to 
do so, as the corresponding cdfs would not be differentiable.
In addition to that, the choice of the number of intervals $n$ is not obvious when quantile functions are not directly observed but obtained with transformation. If $n$ needs to be big to faithfully approximate the true quantile functions, this projection can be quite slow.

For these reasons, we propose to resort to a B-spline expansion, through which
we can derive an alternative approximation of the projection operator $\Pi_X$, without
incurring in the issues of the algorithms in \cite{PAVA}.
Moreover, we will also show in Section~\ref{sec:new_geod}  that the proposed B-spline expansion
also leads us to a simpler and faster reformulation of the geodesic PCA in \cite{geodesic}.

\subsection{Monotone B-splines representation}
\label{sec:splines}

In what follows, let $\mu = U([0, 1])$. Moreover, denote with $\bm x = [x_1, \ldots, x_k]^\prime \in \mathbb{R}^k$ a generic vector.

As already said, through the $\varphi_\mu$ map, we can identify $\mathcal W_2(\mathbb{R})$ with the space
\[
    L_2([0, 1])^{\uparrow} := \{F^{-} \in L_2([0, 1]) \text{ s.t. } F^{-} \text{ is monotonically nondecreasing}\}
\]
This leads us to consider a suitable B-spline basis for the space, to efficiently
evaluate all the computations needed in our algorithms and for a convenient way to express the constraints which define $L^2([0, 1])^{\uparrow}$.
In particular, we consider the basis of quadratic splines with equispaced knots in 
$[0, 1]$. The reason for this particular choice is two-folded.
First of all, splines of degree greater than one enjoy the nice property of uniform approximation
of all continuous functions as the maximum distance between knots goes to zero, in turn this
means that the closure of the linear space generated by the spline basis w.r.t the $L_2$ norm
coincides with $L_2([0,1])$.
Secondly, quadratic splines are particularly well suited to characterize monotonic functions
by looking at the coefficients of the (quadratic) B-spline expansion, as shown in the next
Proposition.

\begin{prop}\label{prop:spline_mono}
Let $\{\psi_j^k\}_{j=1}^J$ be a basis of B-splines of order $k$ defined over the knots $x_1,\ldots,x_{J+k+2}$.
Let $f(x) = \sum_{j=1}^J a_j \psi_j^k(x)$, then:
\begin{enumerate}
    \item If the coefficients $\{a_j\}$ are monotonically increasing (decreasing) $f$ is  monotonically increasing (decreasing)
    \item If $k=2$, then 1. holds with an \virgolette{if and only if} 
\end{enumerate}
\end{prop}

Before proceeding, let us fix some notation.
From now on, we omit the dimension index 
\virgolette{$k$} for the spline basis, writing $\psi_j$ for $\psi_j^2$,
moreover we will let $\{\psi_j\}_{j=1}^J$ with fixed $J > 0$ denote
a B-spline basis in $L_2([0, 1])$.

\begin{rmk}\label{rmk:RJ}
Let $\mathbb{R}^{J\uparrow}$ be the set of vectors $v \in \mathbb{R}^J$ with nondecreasing coefficients.
That is, letting $G = \{g_{ij}\}$ be the $J \times J$ binary matrix such that $\sum_j g_{ij} v_j = v_i-v_{i-1}$, for any element $\bm v \in \mathbb{R}^J$ it holds that $G \bm v\geq 0$. 
Using Proposition \ref{prop:spline_mono}, through the coordinates operator, the set $L_2([0, 1])^{\uparrow}\cap Span\{\psi_j\}^J_{j=1}$ is fully identifiable with 
$\mathbb{R}^{J\uparrow}$, endowed with the metric given by the symmetric 
positive definite matrix $E$ with entries 
\begin{equation}\label{eq:E}
E_{ij} = \langle \psi_i,\psi_j \rangle_{L_2([0, 1])}.
\end{equation}
The norm induced is therefore $\|\bm x\|_E^2 = \bm x^T E \bm x$.
\end{rmk}

\begin{rmk}\label{rmk:basis}
It is possible to find a basis for $\mathbb{R}^J$ with vectors lying in 
$\mathbb{R}^{J\uparrow}$ (and so in $X_J$), namely the vectors $(0,\ldots,0,1)$, $(0,\ldots,0,1,1)$ etc.
In other words, $Span(L_2([0, 1])^{\uparrow}\cap Span\{\psi_j\}^J_{j=1})=Span\{\psi_j\}^J_{j=1}$ for every $J>0$. This tells us that the convex cone of monotone splines is indeed quite big inside the spline space, and this a priori is beneficial for extrinsic methods, especially for PCA.  
\end{rmk}

From now on, to lighten the notation, we deliberately confuse the coefficients of the splines, living in $\mathbb{R}^J$ or $\mathbb{R}^{J\uparrow}$ (with the metric given by $E$), with the corresponding spline functions 
living in the subsets of $L_2([0,1])$ given by $L_2([0, 1])^{\uparrow}\cap Span\{\psi_j\}^J_{j=1}$ and $Span\{\psi_j\}^J_{j=1}$.

\begin{rmk}\label{rmk:polytope}
Lastly, we  point out that $\mathbb{R}^{J\uparrow}$ has the structure of a convex polytope, since the constraints given
by $G \bm v \geq 0$ (guaranteeing that $\bm v \in \mathbb{R}^{J\uparrow}$) are linear.
Such geometric property makes optimization on $\mathbb{R}^{J\uparrow}$ handy and is key for the empirical methods developed in the remaining of the paper.
\end{rmk}

As a consequence of Remark \ref{rmk:polytope}, the optimization problem 
given by the projection of a vector $\bm v\in \mathbb{R}^{J}$ onto $\mathbb{R}^{J\uparrow}$ can be formulated as follows:
\begin{equation}\label{eq:proj}
\Pi_{\mathbb{R}^{J\uparrow}}(\bm v) = \argmin_{G\bm w\geq 0} \| v-w \|_E.
\end{equation}
The computational complexity required to solve \eqref{eq:proj} is at most cubic in the number of basis elements $J$ \citep{POTRA2000281}.
 
Preliminary analysis showed that solving the optimization problem in~\eqref{eq:proj}
compares favorably with the Pool Adjacent Violators Algorithm (PAVA) in \cite{PAVA}.
In particular, computing PAVA with $n=100$ approximation intervals is roughly
eight times slower than~\eqref{eq:proj} with $J=20$ (a reasonable choice, leading to
negligible approximation error, in our examples, with a quadratic spline basis).
Increasing $n=1000$ for PAVA makes it 700 times slower than~\eqref{eq:proj}.

In addition to that, resorting to a discretized approximation of quantiles would also increase the cost of the projected PCA, due to the need of using some functional PCA implementation, as opposed to the low-dimensional multivariate model we are able to implement with the B-spline basis functions.

\section{Empirical Models with B-splines}
\label{sec:empirical}

In this Section, we present the empirical counterparts of the projected
PCA defined in Section~\ref{sec:proj_methods} and provide an illustrative
example of projected linear regression, namely when both the dependent
and independent variables are distributions.

Let $\{\psi_j\}_{j=1}^J$ be a fixed quadratic B-spline basis.
Upon approximating the observed quantile functions with their spline
expansion, thanks to Remark \ref{rmk:RJ}, we can develop our methodology in $\mathbb{R}^{J}$, considering
the metric induced by $E$ instead of the usual one.
Indeed, given a vector $\bm w \in \mathbb{R}^{J}$,
we can identify the corresponding function in $L_2$ by the map 
$\bm w \mapsto \sum_{j=}^J w_{j} \psi_j$.

For the projected PCA in Section~\ref{sec:emp_pca} and for the geodesic PCA in Section~\ref{sec:new_geod} we consider observations
$F^-_1, \ldots, F^-_n$, and let $F^-_0$ be the centering point of the 
PCA. In our examples, $F^-_0$ will always be the barycenter of the
observations. 
As a preprocessing step, we approximate each of these quantile functions
through a B-spline expansion and denote by 
$\bm a_i = \{a_{ij}\}_j$ and $\bm a_0 = \{a_{0j}\}_j$ the coefficients of the spline representation associated to 
$F^{-}_i$ and $F^{-}_0$ respectively, that is, $F^{-}_i \approx \sum_{j=}^J a_{ij} \psi_j$.
For the projected regression in Section~\ref{sec:emp_regression}, let observations $\{(F^-_{z}, F^-_{y})_i\}_{i=1}^n$, where the $F^{-}_{zi}$'s are realizations of the independent variable $\Zvar$
and the $F^{-}_{yi}$'s are realizations of the dependent variable $\Yvar$.
We apply the same preprocessing step and let $\bm a^{(z)}_i$ and $\bm a^{(y)}_i$
denote the coefficient of the spline approximation of $F^{-}_{zi}$ and $F^{-}_{yi}$ respectively.

\subsection{Empirical PCA}\label{sec:emp_pca}

Denote with $A$ the $(n \times J)$ matrix with rows $\bm a_1, \ldots, \bm a_n$. As in standard PCA, the first principal component centered in $\bm a_0$ is found by solving the optimization problem:
\begin{equation}\label{eq:pc_1}
\bm{w}^*_1 = \argmax_{\bm{w}: \|w\|_{E}=1}  \sum_i |\langle \bm a_i-\bm a_0, \bm{w} \rangle_E |^2 = \argmax_{\bm w: \|\bm w\|_{E}=1} \| A E \bm w \|^2 
\end{equation}
where $A$ is the matrix whose i--th row is given $\bm a_i- \bm a_0$.
The optimization problem~\eqref{eq:pc_1} can be solved similarly to a Rayleigh 
quotient: using Lagrange multipliers, \eqref{eq:pc_1} is equivalent to
\begin{equation}\label{eq:pc1_lagrange}
\mathcal L(\bm w) := \bm{w}^T (A \ E)^T \ A \ E \ \bm{w}-\lambda (\bm{w}^T \ E \ \bm{w}-1)
\end{equation}
Deriving~\eqref{eq:pc1_lagrange} w.r.t $\bm w$ and equating the derivative to zero
shows that the solutions to $d\mathcal L(\bm w) / d \bm w = 0$ are the eigenvectors of
the matrix $A^T A E$.
Hence, ordering the eigenvalues of $A^T A E$ in decreasing order, 
the first principal component $\bm w^*_1$ corresponds to the first eigenvector.
Using similar arguments it can be shown that $\bm w^*_2, \ldots \bm w^*_J$ correspond
to the remaining eigenvectors.

Once the first $k$ principal directions $\bm w^*_1, \ldots, \bm w^*_k$ are
found, the projection of a new observation $x^* = \sum_{j=1}^J a^*_j \psi_j$
onto  $U^{k, x_0}_X$ (see Definition~\ref{def:proj_pca}) is found exploiting Proposition~\ref{prop:metric_proj}.
In particular, the following optimization problem is to be solved:
\begin{equation}\label{eq:proj_emp}
\begin{aligned}
    \argmin_{\lambda_j\in\mathbb{R}} & \|(\langle \bm a^* - \bm a_0, \bm w^*_i \rangle_E - \lambda_i)_{i=1}^k \| \\
    \textrm{s.t. }  &G \Big(\sum_{i=1}^k \lambda_i \bm w^*_i + \bm a_0 \Big)\geq 0
\end{aligned}
\end{equation}
which is equivalent to the minimization of a norm inside a polytope,
that is a well-studied problem in $\mathbb{R}^J$
\citep[see][]{poli} and there exist a variety of fast numerical routines to solve it.

\subsection{Empirical Regression}\label{sec:emp_regression}

In this section, we provide the details of the estimation procedure for
a projected regression model where both the independent and the dependent
variables are distribution-valued.
It is straightforward to extend our methodology to cases when only one
of these variables is distribution-valued and the other one takes values
in $\mathbb{R}^q$.

First, we outline how to obtain an estimator for the linear operator $\Gamma$
in Definition~\ref{def:proj_reg}.
Following Section~\ref{sec:proj_reg} we first embed both $\Yvar$ and $\Zvar$
in $L_2([0, 1])$ through the inclusion operator $L_2([0, 1])^\uparrow \hookrightarrow L_2([0, 1])$, and assume the functional linear model presented in \cite{ramsay} and
\cite{regression_consistency}
\begin{equation}\label{eq:func_linear}
    \Yvar(t) = \alpha(t) + \int_0^1 \beta(t, s) \Zvar(s) ds + \varepsilon(t), \qquad t \in [0, 1]
\end{equation}
so that $\Gamma = \Gamma_{\alpha, \beta}$ is the operator $\Gamma_{\alpha, \beta}(v)(t) = \alpha(t) + \int_0^1 \beta(t, s) v(s) ds$.
The goal is then to estimate $\alpha \in L_2([0, 1])$ and $\beta\in L_2([0,1]^2)$. Further, we assume that $\varepsilon$ and $\Zvar$ are uncorrelated: $\mathbb{E}[\Zvar(s)  \varepsilon(t)]=0$  for every $t,s\in [0,1]$.

Consider now observations $\{(F^-_{z}, F^-_{y})_i\}_{i=1}^n$ and the corresponding spline coefficients. Further, we project $\alpha(t)$ on the
same spline basis, so that $\alpha \approx \sum_{j=1}^J \theta_{\alpha j} \psi(j)$ and $\beta(t, s)$ on the basis on 
$[0,1]^2$ with $J\times J$ elements, so that $\beta(t, s) \approx \sum_{i, j\prime=1}^{J} \Theta_{\beta i j} \psi_i(t) \psi_j(s)$.
Neglecting the spline approximation error, model \eqref{eq:func_linear} entails
\begin{equation}\label{eq:func_lin_spline}
    \bm a^{(y)}_i = \bm \theta_\alpha + \Theta_\beta E  \bm a^{(z)}_i + \bm a^{(\varepsilon)}_i, \qquad i=1, \ldots, n
\end{equation}
where $\bm a^{(\varepsilon)}_i$ denotes the spline expansion coefficients of the
unobserved error $\varepsilon_i(t)$.

We propose to estimate \eqref{eq:func_lin_spline} using the same approach
of \cite{regression_consistency}, but extending it to account for
spline approximations for both dependent and independent variables.
We focus only on the estimate $\widehat{\Theta}_\beta$ of $\Theta_\beta$ since once such estimate
is obtained, the estimate for $\bm a_\alpha$ can be straightforwardly derived, \citep[see][]{cai2006}
as:
\[
    \hat {\bm \theta}_\alpha = \overline{\bm a^{(y)}} - \widehat{\Theta}_\beta E \overline{\bm a^{(z)}}
\]
where $\overline{\bm a^{(y)}}$ and $\overline{\bm a^{(z)}}$ are the means of $\bm a^{(y)}$ and $\bm a^{(z)}$ respectively.

The estimator $\widehat{\Theta}_\beta$ is found by penalized least square minimization:
\begin{equation}\label{eq:reg_penalized}
    \widehat{\Theta}_\beta = \argmin_{\Theta} \frac{1}{n} \sum_{i=1}^n \| \left(\bm a^{(y)}_i - \overline{\bm a^{(y)}}\right) - \Theta E  \left(\bm a^{(z)}_i - \overline{\bm a^{(z)}} \right)  \|^2 + \rho \textrm{Pen}(1, \Theta)
\end{equation}
where $\rho > 0$ is a penalization parameter to be fixed (usually through
cross-validation) and  $\textrm{Pen}(1, \Theta)$ is a penalization term
defined in \cite{regression_consistency}.

Briefly, the term $\textrm{Pen}(1, \Theta)$ in~\eqref{eq:reg_penalized}
penalizes both the norm of $\beta(t, s)$ and its derivatives, thus 
favoring smoother solutions.
As shown in \cite{regression_consistency}, \eqref{eq:reg_penalized} has a closed
form solution.
Nonetheless, the form of our solution differs from the one presented in \cite{regression_consistency}, since they work directly on discretized functions
while we propose to estimate spline coefficients, and some care must be taken
since they can use (up to scaling) the usual inner product in the Euclidean
space of discretized functions, while we must consider the inner product 
induced by $E$.
However, the procedure for obtaining our result is identical to the one in \cite{regression_consistency}. Hence, we only report the expression for the estimate.

Let $\hat C$ be the matrix with entries
\[
    \hat C_{ks} = \langle \frac{1}{n} \sum_{i=1}^n \langle \bm a^{(z)}_i, b_k \rangle_E \ \bm a^{(z)}_i, b_s \rangle_E,
\]
where $b_k$ and $b_s$ are the $k$-th and $s$-th elements of the standard Euclidean basis in $\mathbb{R}^J$.
Further let $\hat D$ the matrix with entries
\[
    \hat D_{ks} = \langle \frac{1}{n} \sum_{i=1}^n \langle \bm a^{(z)}_i, b_k \rangle_E \ \bm a^{(y)}_i, b_s \rangle_E.
\]
Finally, let $E^\prime$ denote the matrix with entries $E^\prime_{ij} = <\psi_i^\prime, \psi_j^\prime>$ (where $\psi_i^\prime$ denotes the first derivative of the B-spline basis function $\psi_i$), $C_\rho = E^T \otimes (\hat C + \rho E^\prime)$, and $P = E^{\prime T} \otimes E + E^T \otimes E^\prime$, where $\otimes$ denotes the Kronecker product.
Then the solution of \eqref{eq:reg_penalized} can be expressed as
\[
    \text{vec}(\widehat \Theta_\beta) = (C_\rho + \rho P)^{-1} \text{vec}(\hat D)
\]
where $\text{vec}(\cdot)$ denotes the \emph{vectorization} of the matrix.

Finally, our projected regression model is the composition of the operator induced by $(\hat{\bm \theta}_\alpha, \hat{\Theta}_\beta)$ with the projection on $\mathbb{R}^{\uparrow J}$:
\[
   \mathbb{E}[\bm a^{(y)}_i \mid \bm a^{(z)}_i] = \Gamma_{\textrm{P}}(\bm a^{(z)}_i) = \Pi_{\mathbb{R}^{J\uparrow}} \left(\bm \hat{\bm \theta}_\alpha + \widehat{\Theta}_\beta E \bm a^{(z)}_i \right).
\]

\subsection{An alternative optimization routine for the geodesic PCA and a comment on the computational costs}
\label{sec:new_geod}

We now show how the framework in Section~\ref{sec:metric} can be 
employed also to derive faster numerical algorithms to find the
global and nested geodesic PCA as of Definition~\ref{defi:global} and Definition~\ref{defi:nested}.

\begin{prop}{(Global geodesic PCA)}\label{prop:geod}
A $k$ dimensional global geodesic PC centered in $\bm a_0$ 
is the subset of $\mathbb{R}^{J\uparrow}$
spanned by $\{\bm w_1,\cdots,\bm w_k\}$, linearly independent, which solve:
\begin{equation}\label{eq:glob}
\begin{aligned}
\argmin_{\{\bm \lambda_{i}\}_1^n, \{\bm w_j\}_1^k} \sum_{i=1}^n & \lvert \lvert \bm{a_i} - \bm{a_0} -\sum_{j=1}^k \lambda_{ij} \cdot \bm w_j \rvert \rvert^2_E
\\
\textrm{s.t. } & G \Big(\sum_j\lambda_{ij} \bm w_j + \bm a_0\Big)\geq 0
\end{aligned} 
\end{equation}
\end{prop}

\begin{prop}{(Nested geodesic PCA)}\label{prop:nest}
With the same notation as above, a $k$ dimensional nested geodesic PC, centered in $\bm a_0$ is the set spanned by $\{\bm w_1,\cdots,\bm w_k\}$ in $\mathbb{R}^{J\uparrow}$, where the $\bm w_i$s are found recursively from $\bm w_1$ to $\bm w_k$, such that $\bm w_h$ is a solution, for every $h$, of:
\begin{equation}\label{eq:geod}
\begin{aligned}
\argmin_{\{\bm \lambda_{i}\}_{i=1}^n, \bm w} & \sum_{i=1}^n \|\bm{a_i} - \bm{a_0} -\lambda_{i} \bm w\|^2_E \\
\textrm{s.t.   } & \langle \bm w_j,\bm w\rangle_E=0, \quad j=1, \ldots, h - 1 \\
               & G \Big(\lambda_{i} \bm w + \bm{a_0}\Big)\geq 0, \quad \|\bm w\|_E=1
\end{aligned}
\end{equation}
\end{prop}

To solve \eqref{eq:glob} and \eqref{eq:geod} we employ an interior point method
using the solver Ipopt \citep{ipopt}.
When comparing our implementation with $J=20$ spline basis and the one in \cite{geod_vs_log}, we notice a substantial performance improvement, by
a factor of 35 for a data set of $n=100$ distributions, due to the fact working
with spline approximations reduces greatly the number of parameters in the
optimization problem.

Further, note that \eqref{eq:glob} and \eqref{eq:pca_proj} seem extremely similar.
However, in \eqref{eq:pca_proj} the optimization is carried out having
fixed $\bm w^*_1, \ldots, \bm w^*_k$ and for a single observation, while
in \eqref{eq:glob} the optimization is done over a much larger set of parameters.
In fact, the number of parameters in \eqref{eq:glob} is $(n + k)J$, hence the
computational complexity needed to solve \eqref{eq:glob} is cubic in both
the number of bases and the number of observations.
On the other hand, the projected PCA requires a linear time in the number 
of observations (computation of $A^T A E$) and cubic time in the number of basis $J$ (eigendecomposition and projections of new observations).

\section{Asymptotic Properties}\label{sec:asympt}

In this section, we study the convergence of the proposed projected
empirical methods.
First of all, we show that as the number of spline basis $J$ increases,
the error due to the spline approximation vanishes
if the data is sufficiently regular.
Further, under a suitable set of assumptions, we establish consistency
results for the projected PCA and for the projected
distribution on distribution regression.

\subsection{Convergence of Quadratic B-splines}\label{sec:spline_conv}

In the following, denote with $W_k^r([0, 1])$ the space of functions 
whose weak derivatives up to order $k$ belong to $L_r([0, 1])$, further
denote with $D$ the (weak) derivative operator, so that $Df = f^\prime$, 
$D^2 f= f^{\prime\prime}$ and so on,  

\begin{prop}\label{prop:spline_conv}
Let $\mu$ a probability measure on $\mathbb{R}$, $F^{-}_\mu$ 
its quantile function such that $F^{-}_\mu \in W_3^\infty$.
For each $J$ let $\{\psi_j\}_{j=1}^J$ denote a quadratic B-spline basis on $J$ 
equispaced knots in $[0, 1]$.
Then there exist a sequence of spline functions 
$S_J = \sum_{j=1}^J \lambda^{(J)} \psi^{(J)}_j$, with $\lambda^{(J)}_j$ monotonically
non-decreasing in $j$ for every $J$, such that:
\[
    \| S_J - F^{-}_\mu \|_\infty \leq C \|D^2 f_\mu^-\|_\infty J^{-2}
\]
with $f^{-}_\mu =  D F^{-}_\mu$  and $C>0$ constant.
\end{prop}
Let us remark two important facts.

\begin{rmk}\label{rmk:inclusions}
Since the inclusion $L_\infty([0,1])\subset L_2([0,1])$ is 
continuous, thanks to H{\"o}lder inequality, 
the convergence rates hold also for the $L_2$ norm.
By default we will
use the $L_2$ norm if not stated differently.
\end{rmk}

\begin{rmk}\label{rmk:poincare}
By Poincar\'e inequality, if $\|D^3f\|_{\infty} < C$ then $f$ 
belongs to a sphere in $W_3^\infty([0,1])$ whose radius depends on 
$C$ and on the Poincar\'e constant of $[0,1]$; viceversa, all the 
elements in the sphere of radius $C$ in $W_3^\infty([0,1])$ 
clearly have (weak) derivatives bounded by $C$.
\end{rmk}

\subsection{Consistency}\label{sec:consistency}

In this Section we prove the consistency of the projected methods under some assumptions 
on the data-generating process.
In particular, we show that that there exists a number of basis functions
$J>0$ and a sample size $n$ such that the error committed by the 
empirical models in Section \ref{sec:empirical} is smaller than $
\varepsilon>0$, for any fixed $\varepsilon$.
 
\subsubsection{PCA}\label{sec:PCA_assumptions}

Consistency of spline-based PCA for functional data has been addressed,
among the first, by \cite{silverman1996smoothed} and \cite{PCA_consist}.
As one of the main building blocks of our projected PCA is the 
PCA in the ambient space, that is $L_2([0, 1])$, it is natural to
follow \cite{PCA_consist} in making the following assumptions.
Consider data $\mu_1, \ldots, \mu_n$, $F_1^{-}, \ldots, F_n^{-}$ the corresponding quantile functions, then:
\begin{itemize}
    \item[(P1)] The data generating process satisfies $F_1^{-}, \ldots, F_n^{-} \sim \mathcal{F}$ with the $F^{-}_i$ independent and $\mathbb{E}[\mathcal{F}] = 0$.
    \item[(P2)] $F_1^{-}, \ldots,  F_n^{-}$ can be approximated by functions in $W_3^\infty$ with 
uniformly bounded third derivative.
    \item[(P3)] $\mathbb{E}[\|F^-_i(t)\|^4]<\infty, \ i=1, \ldots, n$.
    \item[(P4)] The eigenvalues of the covariance operator of $\mathcal{F}$ have multiplicity $1$.
    \item[(P5)] The eigenfunctions of the covariance 
operator of $\mathcal{F}$ belong to some bounded set in $W^\infty_3([0,1])\subset W^2_3([0,1])$.
\end{itemize}

Before stating the main results, let us comment on assumptions (P1)-(P5).
First of all, (P2) is essential in order to apply 
Proposition~\ref{prop:spline_conv} and get uniform errors on the 
data set. Moreover, (P2) is satisfied, for instance, if the $F^{-}_i$'s 
lie in the $L_2$-closure of a ball of radius $M>0$ in $W_3^\infty$.
(P4) is a rather standard condition and is satisfied if $\mu_1, \ldots, \mu_n \in \mathcal{W}_4(\mathbb{R})$.
(P4) and (P5) imply the assumptions that in \cite{PCA_consist} are used for the consistency results. In particular, (P5) is stronger than the corresponding assumption in \cite{PCA_consist}, where the eigenfunctions are assumed to belong to $W_2^2([a,b])$. 
Similarly, in such work, there is no counterpart of assumption (P2); in fact we need these stronger regularity conditions to get uniform errors when using B-splines. Still some of the examples \cite{PCA_consist} provide of situations satisfying their assumptions, meet also our requirements.
Finally, the zero-mean assumption in (P1) might seem a little odd, since
we know that the quantile functions are monotonically nondecreasing.
However, observe that it is always possible to subtract the empirical mean
from the observations to satisfy (asymptotically) this assumption.

Let $J$ denote the dimension of a quadratic B-spline basis on $[0, 1]$ and 
let $\bm a_i^J$ the coefficients of the B-spline approximation of $F^{-}_i$.
In what follows, to lighten the notation, we refer to a set of 
spline coefficients both as elements of $\mathbb{R}^J$ with the $E
$-norm, or as functions in $L_2$, without making explicit 
reference to the coordinate operator and its inverse.

\begin{prop}
\label{prop:PCA_consist}
Under assumptions (P1)-(P5), for any $\varepsilon>0$  there exists a 
 sample size $n>0$ and a number of basis functions $J>0$ such that:
\[
\Big|\max_{\|w\|_{L_2}=1}\frac{1}{n} \sum_i \langle F^-_i, w \rangle_{L_2}^2-\max_{\|\bm w\|
_{E}=1} \frac{1}{n}\sum_i \langle\bm a_i^J, \bm w \rangle_E^2 \Big| < K\varepsilon
\]
for some constant $K>0$. 
\end{prop}
Proposition~\ref{prop:PCA_consist} ensures the consistency of the B-spline
approximation of the PCA for monotone functional data in 
$H$ which is equivalent to the consistent estimation of the projected principal directions. 

Suppose now to have computed $U^J_k=\{\bm w_h^{J*}\}_{h=1}^k$, that is 
the approximations of the principal directions 
$U_k =\{\bm w_h^{*}\}_{h=1}^k$ found with $J$ basis 
functions. We observe that 
$Sp(U^J_k)\cap L_2([0,1])^\uparrow = Sp(U^J_k)\cap \mathbb{R}^{J\uparrow}$. 
Since for any set of coefficients $\lambda_h$ we have the convergence 
$\sum \lambda_h \bm w_h^{J*} \rightarrow \sum \lambda_h w_h^{*}$,
we obtain that the projection of a point onto 
$Sp(U^J_k)\cap L_2([0,1])^\uparrow$ converges to the 
projection onto $Sp(U_k)\cap L_2([0,1])^\uparrow$. Thus we 
also have convergence of the projection onto the principal 
components.

\subsubsection{Regression}

We consider model \eqref{eq:func_linear} given samples $\{(F_{z}^-, F_y^-)_i\}_{i=1}^n$. 
We make the following assumptions:
\begin{itemize}
    \item[(R1)] The data generating process satisfies \eqref{eq:func_linear}  
    and $\mathbb{E}[\Zvar(s) \varepsilon(t)] = 0$ for every $t, s \in [0, 1]$.
    \item[(R2)] $\alpha \in L_2([0, 1])$ and $\beta \in L_2([0, 1] \times [0, 1])$.
    \item[(R3)] With probability $1$, each quantile function in the samples $\{(F_{z}^-, F_y^-)_i\}_{i=1}^n$ lies inside a sphere 
of radius $K>0$ in $W_\infty^3([0,1])$.
\end{itemize}

Without loss of generality, suppose that both the dependent and the independent
variables have been centered by subtracting their mean so that $\mathbb{E}[\Zvar] = \mathbb{E}[\Yvar] = 0$ and $\alpha = 0$.

The strategy to prove the consistency of the projected linear regression is 
the following.
First of all, we prove that the estimator $\widehat \Theta_J$ converges to 
the estimator $\widehat \Theta_{\textrm{PS}}$, defined in \cite{regression_consistency}, for large enough $n$ and $J$.
Second, we exploit the consistency of
the estimator in \cite{regression_consistency} 
combined with the approximation results of the metric projection, to establish consistency in terms of the prediction error of our projected regression operator. 

Briefly $\widehat \Theta_{\textrm{PS}}$ is obtained by minimizing an
objective function similar to the one in \eqref{eq:reg_penalized}, but
where the spline approximation is used only for $\Theta$, while
the $F_{zi}^-$'s and the $F_{yi}^-$'s are assumed fully observed, and not
approximated through splines.
Calling $B$ the vector of functions with entries $\psi_1, \ldots, \psi_J$,
$\widehat \Theta_{\textrm{PS}}$ is defined as:
\[
    \widehat{\Theta}_{\textrm{PS}} = \argmin_\Theta \frac{1}{n} \sum_i 
        \| F_{yi}^- - \langle F_{zi}^-,B^T \Theta B \rangle \|^2+\rho \text{Pen}(1,\Theta).
\]

Convergence of $\widehat \Theta_J$ to $\widehat{\Theta}_{\textrm{PS}}$ is 
shown in the next proposition

\begin{prop}\label{prop:reg_conv}
Under assumptions (R1)-(R3), if the number of samples is big enough $\widehat{\Theta}$ and $\widehat{\Theta}_J$ exists with probability close to $1$, and  there is $J>0$ such that  
$\|\widehat{\Theta}_{\textrm{PS}}-\widehat{\Theta}_J\|_{E\otimes E}<\varepsilon$. 
\end{prop}

Let $\widehat{\beta}_{\textrm{PS}}$ and $\widehat{\beta}_{J}$ be the kernels  $\widehat{\beta}_{\textrm{PS}}= B^T \widehat{\Theta}_{\textrm{PS}} B$ and $\widehat{\beta}_{J}= B^T \widehat{\Theta}_{J} B$. 
Since 
$\|\widehat{\beta}_{\textrm{PS}}(s,t)-\widehat{\beta}_J(s,t)\|_{L_2([0,1]^2)}=
\|\widehat{\Theta}_{\textrm{PS}}-\widehat{\Theta}_J\|_{E\otimes E}$, 
 we established strong convergence of our kernel to the estimator of \cite{regression_consistency}. 
This implies that the consistency results for the estimator $\widehat{\Theta}_{\textrm{PS}}$ holds also for $\widehat{\Theta}_{J}$,  with respect
 to the seminorm induced by the covariance operator of $\Zvar$.

Specifically, given $\Zvar$ $H$-valued random variable and its covariance
operator $\mathcal{C}_{\Zvar}$, for any $\varphi\in L_2([0,1]^2)$, we consider the semi-norm on $L_2([0,1]^2)$ given by:
\[
\|\varphi\|_{\Gamma_{\Zvar}} = 
\int_{[0,1]}\langle\mathcal{C}_{\Zvar}\varphi(\cdot,t),\varphi(\cdot,t) \rangle dt
\]

Thus, the following result is immediately implied since strong convergence implies
seminorm convergence (see Appendix~\ref{sec:proofs}).

\begin{cor}
For $J>0$ big enough $\mathbb{E}[\|\beta-\widehat{\beta}_J\|_{\mathcal{C}_{\Zvar}}]<\varepsilon$.
\begin{proof}
We use the seminorm triangle inequality:
\[
\|\beta-\widehat{\beta}_J\|_{\mathcal{C}_{\Zvar}}\leq 
\|\beta-\widehat{\beta}\|_{\mathcal{C}_{\Zvar}}+\|\widehat{\beta}-\widehat{\beta}_J\|_{\mathcal{C}_{\Zvar}}.
\]
The first term on the right hand side converges to zero thanks to Theorem 2 in \cite{regression_consistency}, while the second term converges to zero thanks to
Proposition~\ref{prop:reg_conv} and the previous observations.
\end{proof}
\end{cor}

Lastly, we need to take into account the projection step.
First, we notice that $\|\beta - \widehat{\beta}\|_{\Gamma_Z}$
corresponds to the expected prediction error, in fact, as in \cite{regression_consistency}:
\[
\|\beta-\widehat{\beta}_J\|_{\mathcal{C}_{\Zvar}}
= \int_{[0,1]} \mathbb{E}\left[\langle\Zvar,\beta(\cdot, t) - \widehat{\beta}_J(\cdot, t) \rangle^2 \mid \widehat{\beta}_J \right] dt,
\]
further, by H\"older's inequality $\mathbb{E}\left[ |\langle\Zvar,\beta - \widehat{\beta}_J \rangle | \, \big| \, \widehat{\beta}_J \right] \rightarrow 0$,
which straightforwardly yields $\mathbb{E}\left[\| \Gamma_\beta(z)-\Gamma_{\widehat{\beta}_J}(z)\| \, \big| \, \widehat{\beta}_J\right]\rightarrow 0$.

Thus, the following simple lemma ensures the consistency of the 
spline approximation of the projection on $X$ and
leads to the consistency of the projected 
regression in terms of prediction error.
Again, following Remark~\ref{rmk:RJ}, we can identify the space monotone $B$-splines with $J$ basis functions with $\mathbb{R}^{J\uparrow}$. Hence, to lighten the notation, we denote $\Pi_{\mathbb{R}^{J\uparrow}}$ the metric projection operator onto the space of monotone $B$-splines with $J$ basis functions.

\begin{lemma}\label{lemma:proj_conv}
Given $\beta_n\rightarrow \beta$ in $H$, for any 
$\varepsilon>0$ there exists $n,J>0$ such that 
$\|\Pi_{\mathbb{R}^{J\uparrow}}(\beta_n)- \Pi_{L_2([0, 1])^\uparrow}(\beta)\|\leq \varepsilon$.
\end{lemma}

\section{Numerical Illustrations for the PCA}\label{sec:PCA_simulations}

In this section we perform PCA on different simulated data sets and on a real data set of Covid-19 mortality data in the US. 
In particular, on the simulated data sets we compare the performance of our projected PCA (in terms of approximation error and interpretability of the directions) with the ones of intrinsic methods, showing that the projected PCA is a valid competitor in a diverse set of situations.
For the Covid-19 data set, we compare inference obtained using the projected, nested and log PCA, highlighting the practical benefits of the projected PCA over the log one.

For the projected, nested and global PCAs we need to fix a B-spline basis
to express the quantile functions. In particular, we fix an
equispaced quadratic B-spline basis with $J$ interior knots on $[0, 1]$.
Here, the number of basis $J$ is always fixed to $20$, which provided a negligible approximation error of the quantile functions. We did not observe any appreciable change when increasing it. 
In Appendix~\ref{sec:simulation2} we show further simulations where we perform sensitivity analysis as the number of basis increases for a fixed sample size, we provide empirical confirmation of the consistency results in Section~\ref{sec:asympt} and give practical guidance on how to choose $J$.

\subsection{Simulation studies}

We consider three different simulations to compare both the 
interpretability and the ability to compress information of different PCAs.

We compare our projected PCA with the nested and global geodesic PCAs \citep{geodesic, geod_vs_log} and the \emph{simplicial} PCA \citep{menafoglio}.

Briefly, the simplicial PCA applies a transformation that maps densities defined
on the same compact interval $I$ into functions in $L_2(I)$, called  \emph{centered log ratio}.
Then, a standard $L_2$ PCA is performed on the transformed pdfs and, by
the inverse of the centered log ratio transform, the results are
mapped back to the space of densities, called Bayes space
\citep[for a more accurate definition, see][]{egozcue2006hilbert}.
In particular, we remark that, to be well defined, the simplicial PCA requires
that all the pdfs have support equal to $I$, which is a strong assumption
in practice. Further details about simplicial PCA are given in Appendix~\ref{sec:simplicial}.

As for the projected PCA, to compute the simplicial PCA, we resort to a B-spline approximation, but this time of the transformed pdfs.
Hence, we need to select a B-spline basis on the support of
the pdfs $I$. 
In this case, we fix a cubic B-spline basis with
$$J^\prime = J = 20$$ interior knots on $I$, as this choice yielded a negligible approximation error for the transformed pdfs.

In the first scenario, we simulate data from
\begin{equation}\label{eq:simu1_dgp}
    \begin{aligned}
    p_i(x) &\propto \frac{1}{\sigma_i} \exp \left( (x - \mu_i)^2 \big/ (2 \sigma^2_i) \right) \mathbb{I}(x \in [-10, 10]), \quad i=1, \ldots 100 \\
    \mu_i &\sim 0.5 \mathcal{N}(-3, (0.2)^2) + 0.5 \mathcal{N}(3, (0.2)^2) \\
    \sigma_i &\sim \text{Uniform}([0.5, 2.0])
    \end{aligned}
 \end{equation}
Where \virgolette{proportional to} stands for the fact that we confine the density to the support $[-10, 10]$ and renormalize it so that it integrates to $1$.

\begin{figure}
    \centering
    \includegraphics[width=0.5\linewidth]{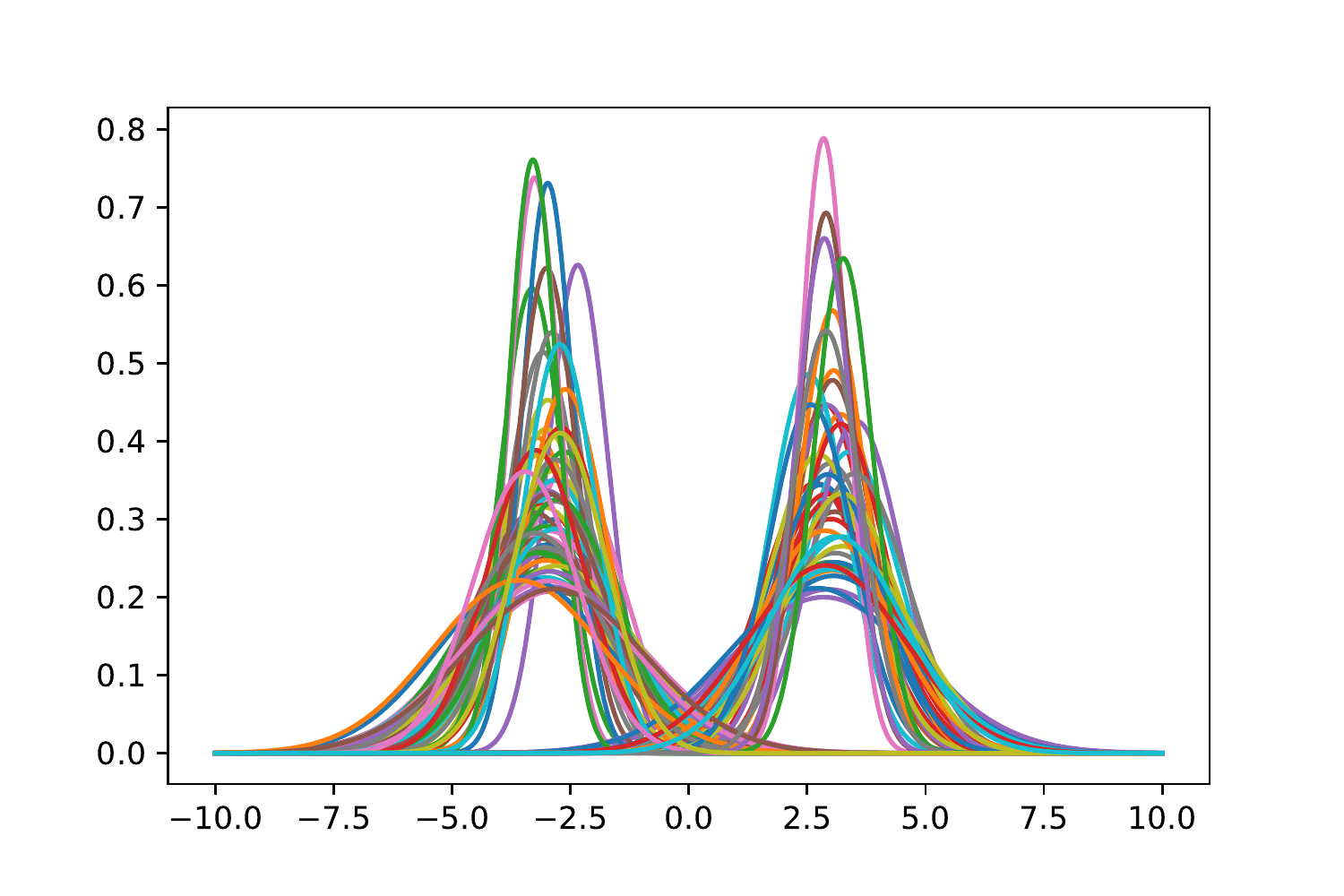}
    \caption{Data set of pdfs generated from~\eqref{eq:simu1_dgp}}
    \label{fig:simu1_data}
\end{figure}

Observe that there are two sources of variability across the pdfs from the data generating process~\eqref{eq:simu1_dgp}.
The first one is the location of the \emph{peak} $\mu_i$ and the second one is the 
\emph{width} of the distribution around the peak, controlled by $\sigma_i$.
See Figure~\ref{fig:simu1_data}.

\begin{figure}
    \centering
    \includegraphics[width=\linewidth]{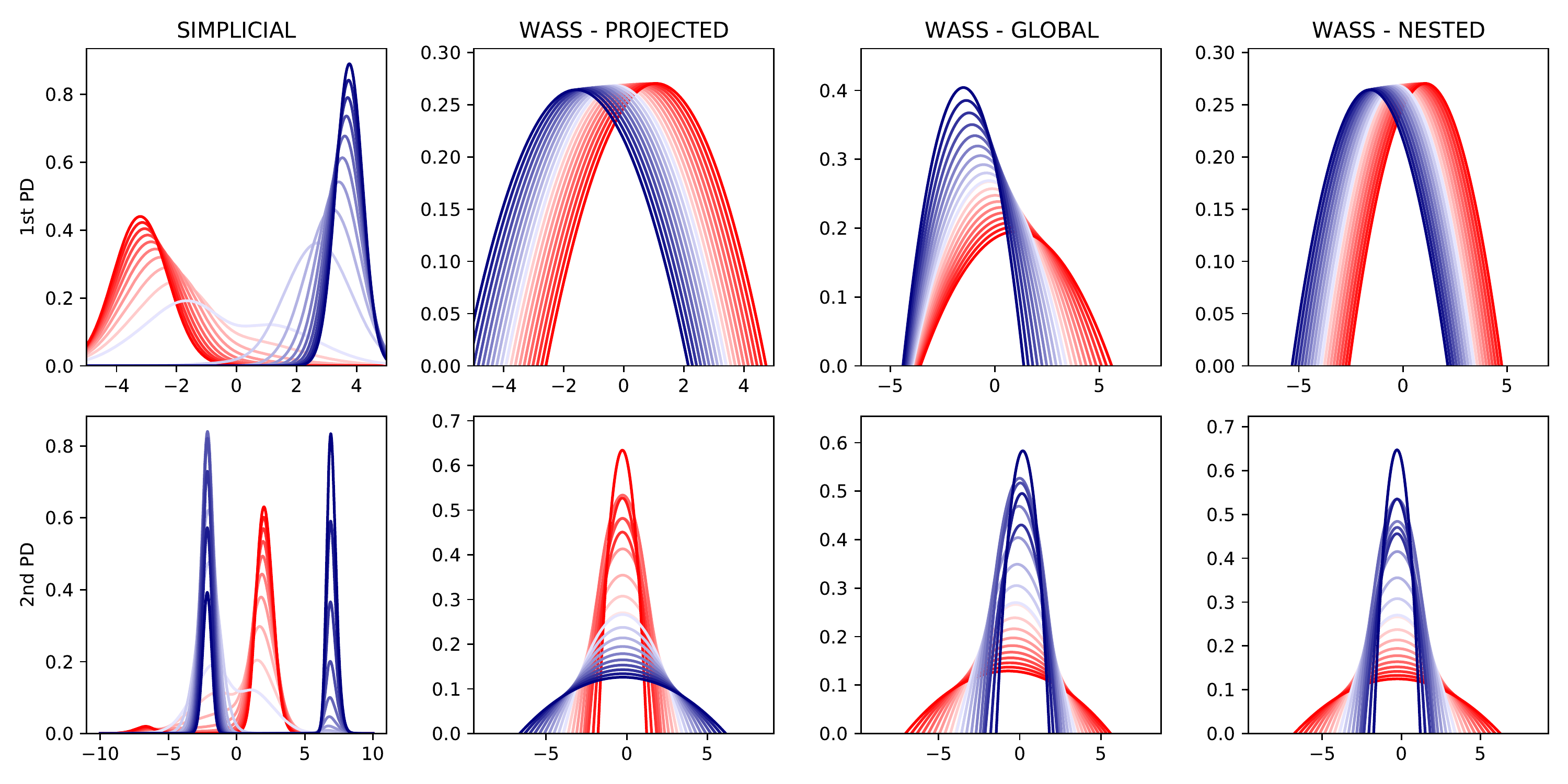}
    \caption{Top row: first principal direction. Bottom row: second principal direction.
    Each line represents the pdf associated to $\lambda \bm w_i$ where $\bm w_i$
    is the $i$--th principal direction ($i=1, 2$) and $\lambda$ is a score ranging
    from $-2$ (darkest blue) to $+2$ (darkest red).}
    \label{fig:simu1_results}
\end{figure}

Figure~\ref{fig:simu1_results} shows the first two principal directions obtained using
the different methods.
We can notice several differences between them. Focusing on the first principal direction,
we can see that the simplicial, projected and nested PCAs detect a change in the location
of the peak of the pdf.
In particular, the first direction for the Wasserstein PCAs represents a shift from 
left to right of this peak, while for the simplicial PCA the first direction is associated
to a peak in $3$ (blue lines, negative values of the scores) or to a peak in $-3$ (red
lines, positive value of the scores).
This also highlights the difference in the geometries underlying the Wasserstein and
Bayes spaces.
Looking at the second principal direction instead, we can see how in the Wasserstein
PCAs it clearly represents a change in the width of the distribution, while
for the simplicial PCA the interpretation is somewhat obscure.

The global geodesic PCA deserves a separate discussion. Indeed, from Definition~\ref{defi:global} it is clear that a global principal component is a convex set
without any notion of preferential directions, so that it is not possible to 
interpret separately the variation along the first and second direction found 
by the global PCA.

Now we present two additional simulations that quantify the amount of 
information that is \virgolette{lost} by performing the PCA.
As a metric, we consider the reconstruction error, that is, the quantity
\begin{equation}\label{eq:rec_error}
    RE_k = \frac{1}{n} \sum_{i=1}^n W_2(F^-_i, \widetilde{F}^-_i)
\end{equation}
where the $F^-_i$'s are the observed probability measures, $\widetilde{F}^-_i$ are 
the reconstructed ones and $k$ is the dimension of the principal component.
More in detail $\widetilde{F}^-_i$ is found by first projecting $(F^-_i - F^-_0)$ 
into $\mathbb{R}^k$ using the PCA and then applying the inverse transformation.
Informally, the reconstruction error is a measure of the quantity of information
lost by applying the PCA as a black-box dimensionality reduction.

As evident in Equation \eqref{eq:rec_error}, we measure the performance of 
PCAs just in terms of Wasserstein metric. This is likely to favor the
performance of the Wasserstein PCAs over the simplicial one.
Thus, the interesting performance comparison is the one between the geodesic PCAs and the projected PCA.
Nevertheless, we think that is worth reporting also the results for the simplicial PCA, which is an intrinsic method in the Bayes space, 
to show that the underlying metric structures are extremely different. 
This also helps to appreciate the results in Section \ref{sec:reg_simulations}.
Given the difference in the metric structure between Wasserstein and Bayes 
spaces, we believe that the choice between simplicial and Wasserstein frameworks
is not trivial and should be application-driven.

\begin{figure}
    \centering
    \begin{subfigure}{0.5\linewidth}
        \includegraphics[width=\linewidth]{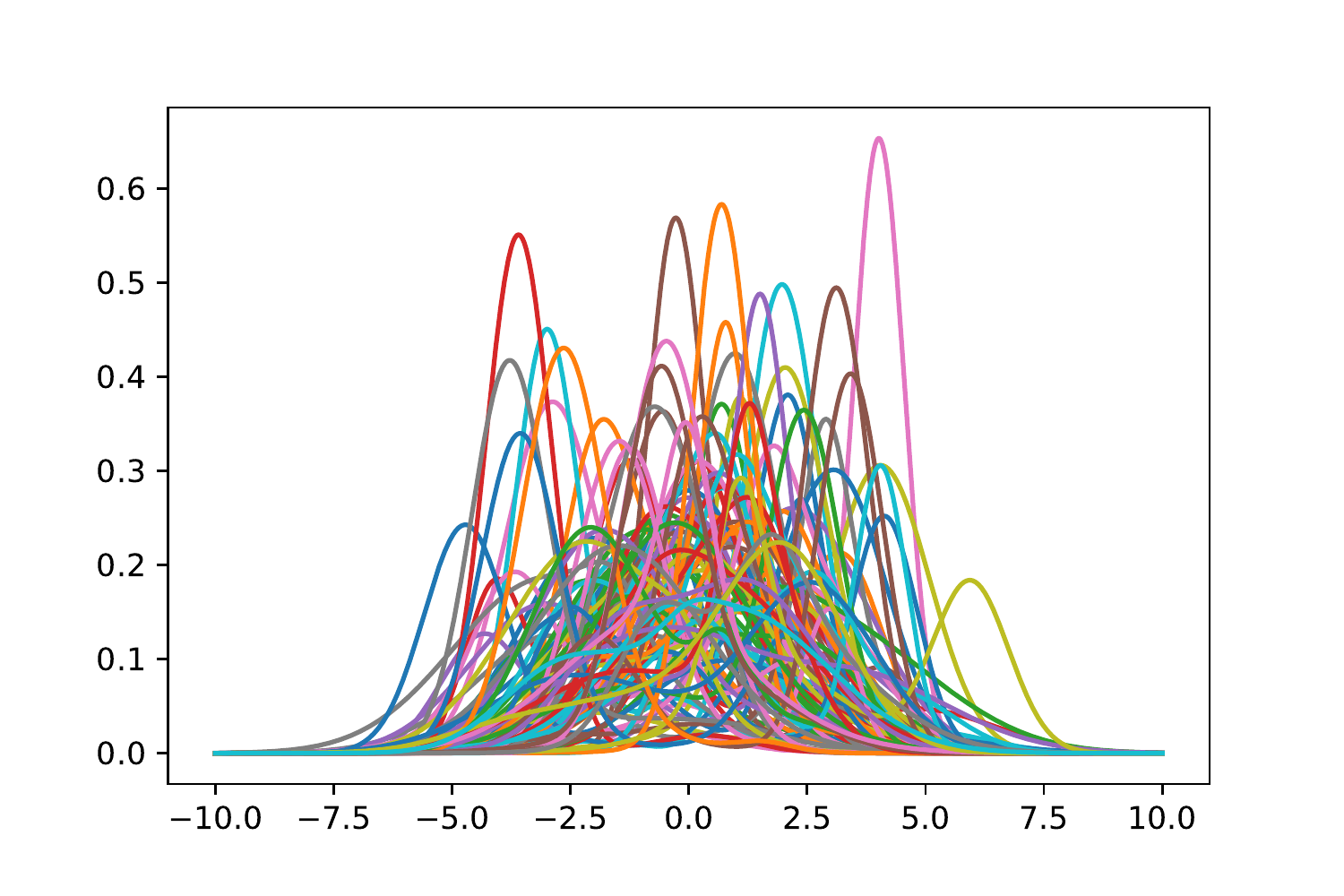}
        \caption{}
    \end{subfigure}%
    \begin{subfigure}{0.5\linewidth}
        \includegraphics[width=\linewidth]{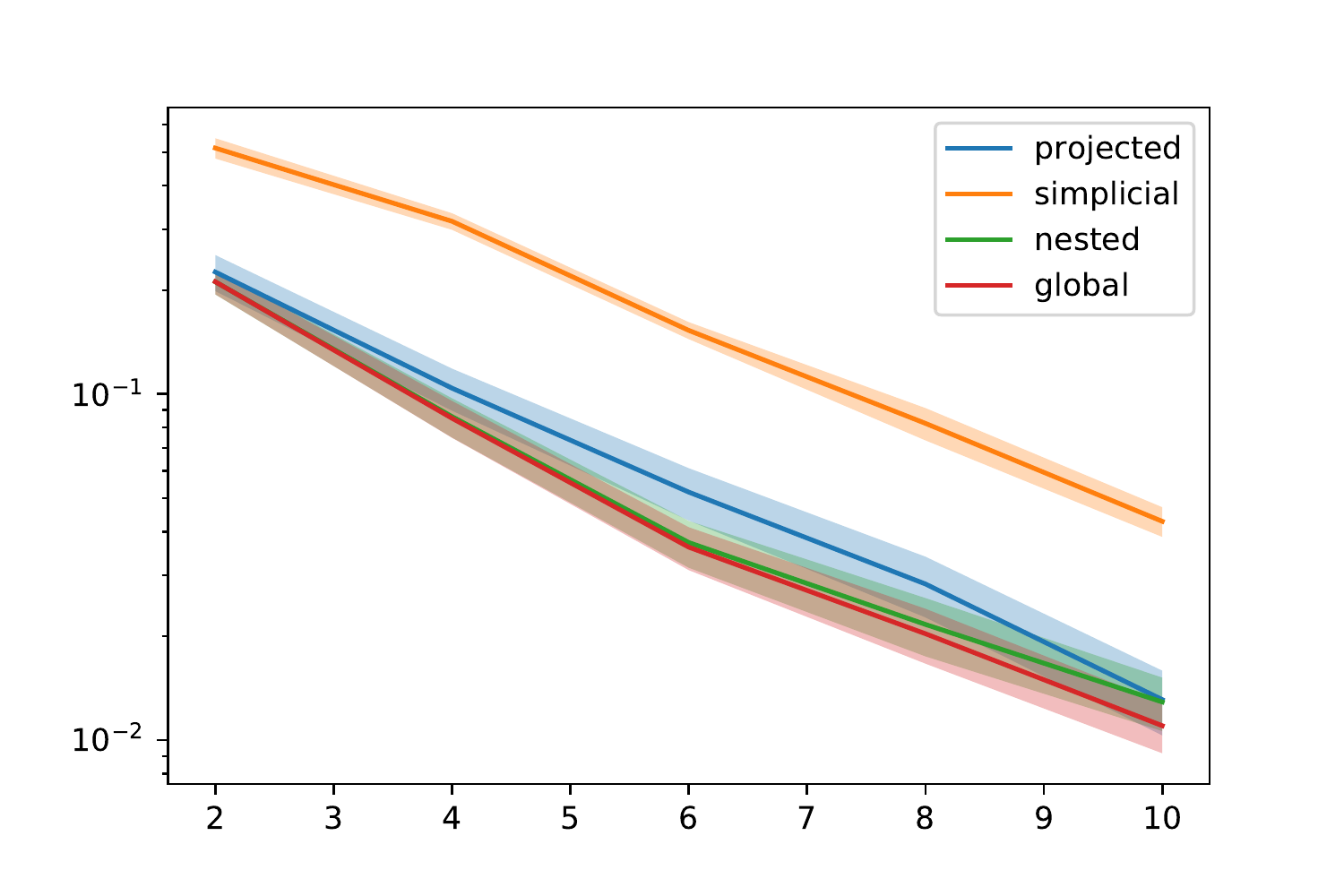}
        \caption{}
    \end{subfigure}
    \caption{Left panel: example of simulated data set for Scenario 2.
             Right panel: reconstruction error as a function of the dimension
             of the principal component employed for the different methods.
             The solid lines represent the mean of 10 independent runs on independent data sets from~\eqref{eq:simu2_dgp} and the shaded area represent $\pm$ one standard deviation.}
    \label{fig:simu2_data}
\end{figure}

To measure raw performance differences between geodesic and projected PCAs, we simulate data so that there is little recognizable structure
in them, unlike in the previous example.
The data generating process is as follows:
\begin{equation}\label{eq:simu2_dgp}
\begin{aligned}
    p_i(x) &\propto \sum_{j=1}^K w_{ij}  \frac{1}{\sigma_{ij}} \exp \left( (x - \mu_{ij})^2 \big/ (2 \sigma^2_{ij}) \right) \mathbb{I}(x \in [-10, 10]) + 10^{-5},  \quad i=1, \ldots 100 \\
    \bm w_i &\sim \text{Dirichlet}_K(1 / K) \\
    (\mu_{ij}, \sigma_{ij}) &\sim \mathcal{N}(d\mu_{ij}; 0, 2^2) \text{Uniform}(d\sigma_{ij}, 0.5, 2.0)
\end{aligned}
\end{equation}
Observe that~\eqref{eq:simu2_dgp} is a finite dimensional approximation of the
Dirichlet Process mixture model, a popular workhorse in Bayesian nonparametric
statistics, that is well known to be dense in the space of densities on $\mathbb{R}$,
see for instance \cite{ferguson1983bayesian}.
An example of the kind of pdfs generated from~\eqref{eq:simu2_dgp} is shown in
Figure~\ref{fig:simu2_data}(a).

To separate the effect of the B-spline smoothing procedure, in 
this scenario we evaluate the reconstruction error in \eqref{eq:rec_error} 
considering $\widetilde{\mu}_i$ to be the reconstructed quantile functions (for the Wasserstein PCAs) or pdfs (for the simplicial PCA) and $\mu_i$ to be
the probability measure represented by the B-spline approximation
of the quantile function or the (centered log ratio of) the pdf respectively.

Figure~\ref{fig:simu2_data}(b) shows the reconstruction error as a function
of the dimension of the principal component, that is, $RE_k$ as a function of $k$.
We can see how the three Wasserstein PCAs consistently outperform 
the simplicial one.
Moreover, as to be expected, the global geodesic PCA obtains the lowest
reconstruction error for all the choices of dimension $k$, with the 
nested geodesic PCA being a close runner-up.
However, the computational cost of finding the nested or global geodesic PCA
can become prohibitive as the sample size or the number of bases in the 
B-spline expansion or the dimension $k$ increases.
For comparison, finding the $10$-dimensional projected PCA is
around 1,000 times quicker than finding the corresponding global geodesic PCA
and 200 times quicker than finding the nested geodesic one.  

As an additional simulation, in Appendix~\ref{sec:simulation2} we investigate the effect of
the number of B-spline basis $J$. In particular, we conclude that,
for a fixed dimension $k$ the reconstruction error \eqref{eq:rec_error}
increases with the number of basis functions, both for the projected and the
simplicial PCA.
Furthermore, we also observe that the reconstruction error for the simplicial PCA exhibits a larger variance than the reconstruction error for the projected PCA.
Our insight is that this is due to the different degree of smoothness of the
pdfs and the quantile functions. Since the quantile functions are in general
smoother than the pdfs, their B-spline expansion should have lower variance.

\subsection{Assessing the reliability of the projected PCA}\label{sec:reliability}

A classical measure of performance of the standard Euclidean PCA,
also useful to determine the dimension of the principal component to
use, is the proportion of the explained variance.
For a $k$-dimensional Euclidean principal component, this quantity is 
easily computed as a ratio of eigenvalues: $\sum_{j=1}^k \lambda_j \big/ \sum_{j\geq 1} \lambda_j$.
Upon truncating the series at the denominator, the same quantity can also
be computed for PCA in infinite dimensional Hilbert spaces.

Due to the projection step involved in our definition of PCA, we argue that
the proportion of explained variance might not be a reliable indicator
of performance, nor should it be used to guide the choice of the dimension $k$.
Instead, we propose a fast alternative based on the Wasserstein distance that
we believe better represents the properties of the projected PCA,
that is, the normalized reconstruction error:
\[
    NRE_k = \frac{\frac{1}{n} \sum_{i=1}^n W_2(F^-_i, \widetilde F^-_i)}{\frac{1}{n} \sum_{i=1}^n W_2(F^-_i, F^-_0)},
\]
where the numerator corresponds to the reconstruction error in \eqref{eq:rec_error} and the denominator is the average distance between
the observed measures and their barycenter.
Observe that in Euclidean spaces, this quantity is closely related to the
proportion of explained variance, since in Euclidean spaces maximizing 
variance in a subspace, amounts to minimizing the average distance from the subspace to data points.

Given its extrinsic nature, for a fixed dimension, the projected PCA 
might sometimes fail to  capture the variability of some particular
data set and, in those situations, an intrinsic approach should be
preferred.
However, given the high computational cost associated to geodesic
PCAs, one would carry out such analysis only knowing that the results 
would be significantly better than the ones obtained by projected PCA.
This calls for discerning whether the poor performance of projected
PCA is due to its extrinsic nature or rather to the scarceness of 
structure in the data set under consideration: in the former situation it
is likely that a geodesic approach would yield better results,
in the latter instead, it is likely that results remain the same.
 
We propose now two empirical indicators of the \virgolette{reliability}
of the empirical projected PCA.
The first one measures, once a $k$-dimensional principal component is found, how reliable are the projected principal directions and the second one gives an idea of how different the projected PCA and the $L_2$ PCA
are. To assess the interpretability of the principal directions and the scores obtained with the projected PCA, we first compute for every principal direction $\bm w^*_h$ the quantities $\eta^{\min}_{h}$ and 
$\eta^{\max}_h$ such that
\[
    \eta^{\min}_h = \min_{\eta \in \mathbb{R}} \{\bm a_0 + \eta \bm w^{*}_h \in \mathbb{R}^{J\uparrow} \}
\]
where $\bm a_0$ is the spline coefficient vector associated with the barycenter
$F^-_0$. The scalar $\eta^{\max}_h$ is found analogously.
Hence $(\eta^{\min}_h \bm w^*_h, \eta^{\max}_h \bm w^*_h)$ is the segment spanned by
the principal direction living inside the convex cone $\mathbb{R}^{J\uparrow}$.
If the scores of all observations along this direction lie within the
range $(\eta^{\min}_h, \eta^{\max}_h)$, then the variability captured by (empirical) projected PCA
can be decomposed along the principal directions, whose scores are then highly interpretable.
Contrary, the PCA scores outside $(\eta^{\min}_h, \eta^{\max}_h)$ will be associated with functions which are not quantiles, and thus limiting the interpretability of the direction.
Hence, we propose the following \emph{interpretability score}
\begin{equation}\label{eq:is_score}
    IS_h = 1 - \frac{1}{n} \sum_{i=1}^n d\left(s_{ih},  [\eta^{\min}_h, \eta^{\max}_h] \right) \big/ |s_{ih}|,
\end{equation}
where $s_{ih}$ is the score of observation $i$ along direction $h$ 
according to the projected PCA.
A value of $IS_{h}$ equal to one corresponds to perfect interpretability,
that is, projected PCA behaves like a standard Euclidean PCA along direction $h$.
On the other hand, values of $IS_h$ closer to zero indicate that the decomposition of the variance along the principal directions lies outside $\mathbb{R}^{J\uparrow}$ for direction $h$.
The interpretability score can be fruitfully used also to evaluate the directions found with the nested PCA, 
upon replacing the $s_{ih}$'s in \eqref{eq:is_score} with the scores given by the nested PCA.

Note that the $IS_h$ score is useful to interpret the directions one at a time.
However, it can be the case that some scores along one direction $h^\prime$ lie outside
the $(\eta^{\min}_{h^\prime}, \eta^{\max}_{h^\prime})$ range but that the $L_2$ projection on the $h \geq h^\prime$ component still lies within the projected
component. For instance, this could imply that a projected PC could be similar to a nested one despite having very different directions. 
A discrepancy between the two can appear when the projections of some data points on the $L_2$ PCA lie outside $\mathbb{R}^{J\uparrow}$.
Using the terminology of Proposition~\ref{prop:metric_proj} this can be measured in terms of difference between the projections $\Pi_k(F^{-*} - F^-_0)$ and
$\Pi_{Sp(U_k)\cap (X-x_0)}(F^{-*} - F^-_0) = \Pi_{Sp(U_k)\cap (X-x_0)}(\Pi_k(F^{-*} - F^-_0))$, for a given observation
$F^{-*}$.
To quantify the loss of information at the level of the component (instead of
direction), we propose to measure the \virgolette{ghost variance} captured
by the $L_2$ PCA:
\[
    GV_k = \frac{1}{n} \sum_{i=1}^n \| \Pi_k(F^{-}_i - F^-_0) - \Pi_{U_X^{F^-_0, k}}(\Pi_k(F^{-}_i - F^-_0)) \|_2 \Big/ \|F^{-}_i - F^-_0\|_2,
\]
that is, the $GV_k$ score measures the quantity of information that is lost
due to the projection step or, in other words, the information that we trained our PCA on, but that does not appear in the Wasserstein Space.
If $GV_k=0$ then all the information captured by the $L_2$ PCA is inside the Wasserstein Space, then the projected PCA coincide with the nested one by definition.

Finally, although this situation never occurred in our experience,
it might happen that $GV_k$ is small but some $IS_k^\prime$ 
($k^\prime \leq k$) is large.
This means that the subspace identified by the projected PCA
is suitable for representing the data, but the single principal directions
are not interpretable. 
In this case, we suggest to take a hybrid approach: use the projected
PCA as a fast black-box dimensionality reduction step, thus reducing the
dimensionality of each observation from $J$ to $k$, and then use the
nested PCA, in dimension $k$, to estimate the directions,
the main advantage being the reduction in the computational cost to estimate
the nested PCA in this lower dimensional space.

\subsection{Analysis of the Covid-19 mortality data set}

\begin{figure}[t]
\centering
\begin{subfigure}{0.5\linewidth}
  \includegraphics[width=\linewidth]{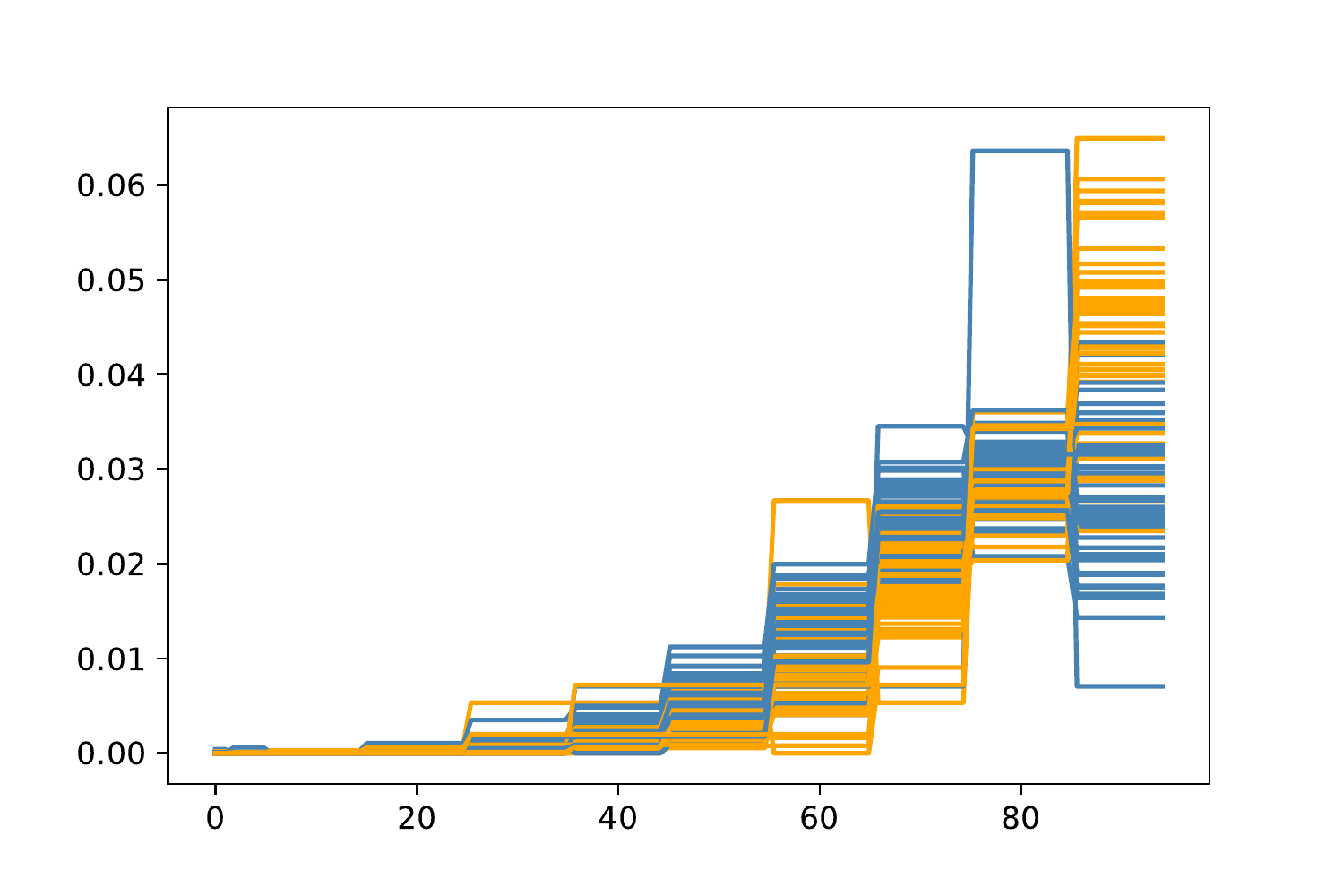}
  \caption{}
\end{subfigure}%
\begin{subfigure}{0.5\linewidth}
  \includegraphics[width=\linewidth]{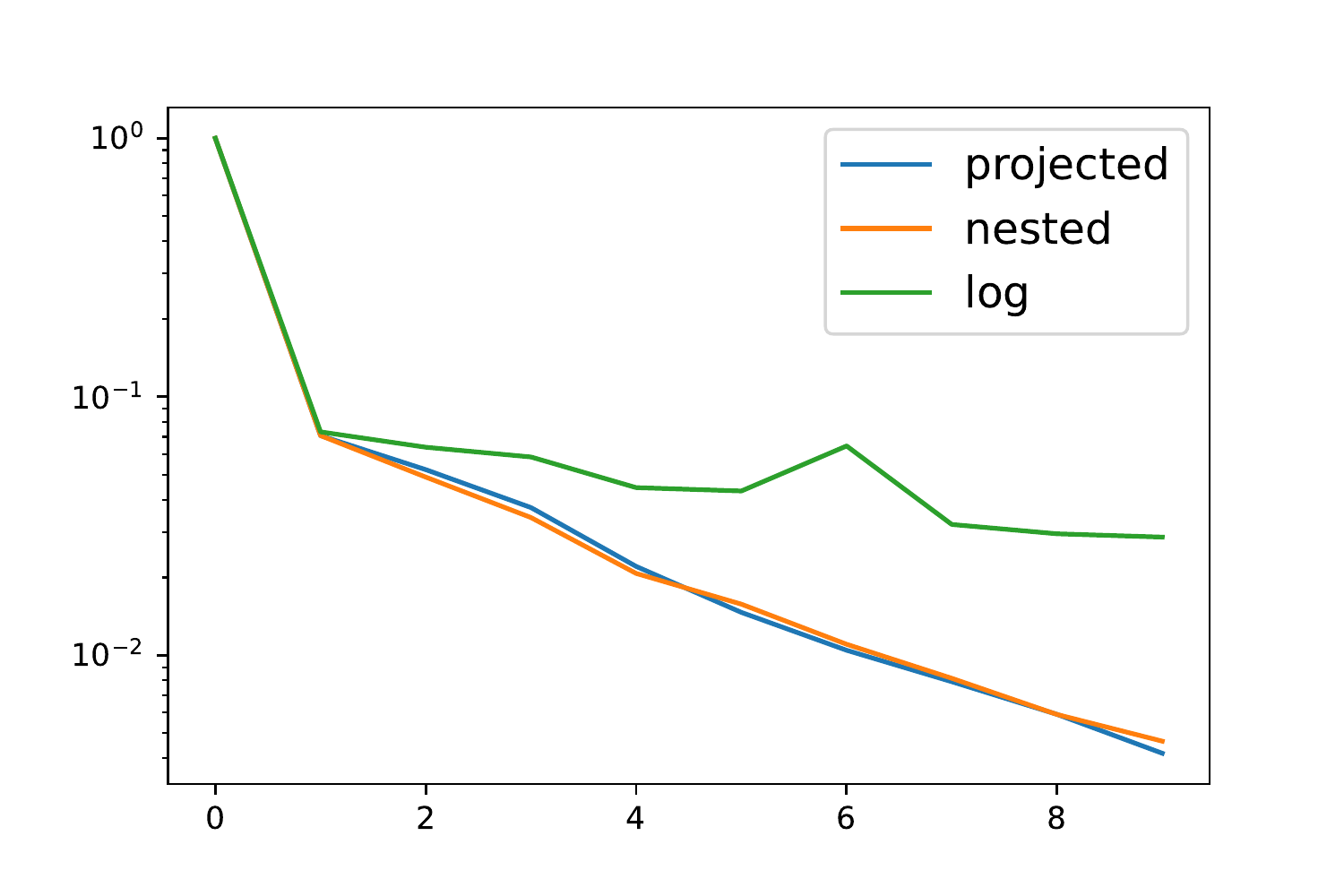}
  \caption{}
\end{subfigure}%
\caption{
Left panel: distributions of age at the time of death for Covid-19 patients divided by
sex: orange corresponds to female and blue to males.
Different lines correspond to different US states / inhabited territories.
Right panel: reconstruction error as a function of the dimension of the component for different PCAs.
The $0$-th principal component is the empirical mean.}

\label{fig:covid_data}
\end{figure}

We perform PCA analysis on the Covid-19 mortality data publicly available at \url{data.cdc.gov}
as of the first December 2020.
The data set collects the total number of deaths due to Covid 19 in the 
US from January 1st 2020 to the current date, data are subdivided by
state, sex, and age. In particular, the ages of the deceased are grouped
in eleven bins: $[0, 1), [1, 5), [5, 15), [15, 25), \allowbreak [25, 35), [35, 45), 
[45, 55), [55, 65), [75, 85), [85, +\infty)$ but we truncate the last 
bin to $95$ years for numerical convenience.
Further, we remove Puerto Rico from the analysis because it presented
too many missing values.
Our final data set, shown in Figure~\ref{fig:covid_data}(a), consists of 106 samples of the distribution of the
ages of patients deceased due to Covid-19, divided by sex and pertaining
53 between US states and inhabited territories.

We apply our usual B-spline approximation with $J=20$ basis to the
quantile functions obtained starting from the histograms in 
Figure~\ref{fig:covid_data}. This choice of $J$ yields an average
approximation error, in terms of Wasserstein distance, of $0.02$.
An error this low is
to be expected since the quantile functions are piecewise linear functions
defined on eleven intervals.

We use this real data set to make a hands-on comparison of the inference that can be obtained employing the projected, nested and log PCA.

We start by comparing the projected and nested PCAs.
The first direction found by the nested PCA is identical to the one found by the projected while the second is extremely close: the cosine between the two principal directions is approximately $0.99$.
In line with this, the interpretability scores equal $IS_1 = 1$ and $IS_2 \approx 0.89$, while $GV_2 = 0.05$.
Moreover, the two-dimensional projected principal component explains more than $90\%$ of the $L_2$ variability and $NRE_2 \approx 0.05$ for both projected and nested PCA. 
Given the reconstruction error and the $GV_2$ score, we can conclude that the two-dimensional projected principal component provides a very good fit to the data,
and that both selected principal directions are well behaved with respect to their scores, guaranteeing interpretable results.

\begin{figure}[t]
\centering
\includegraphics[width=\textwidth]{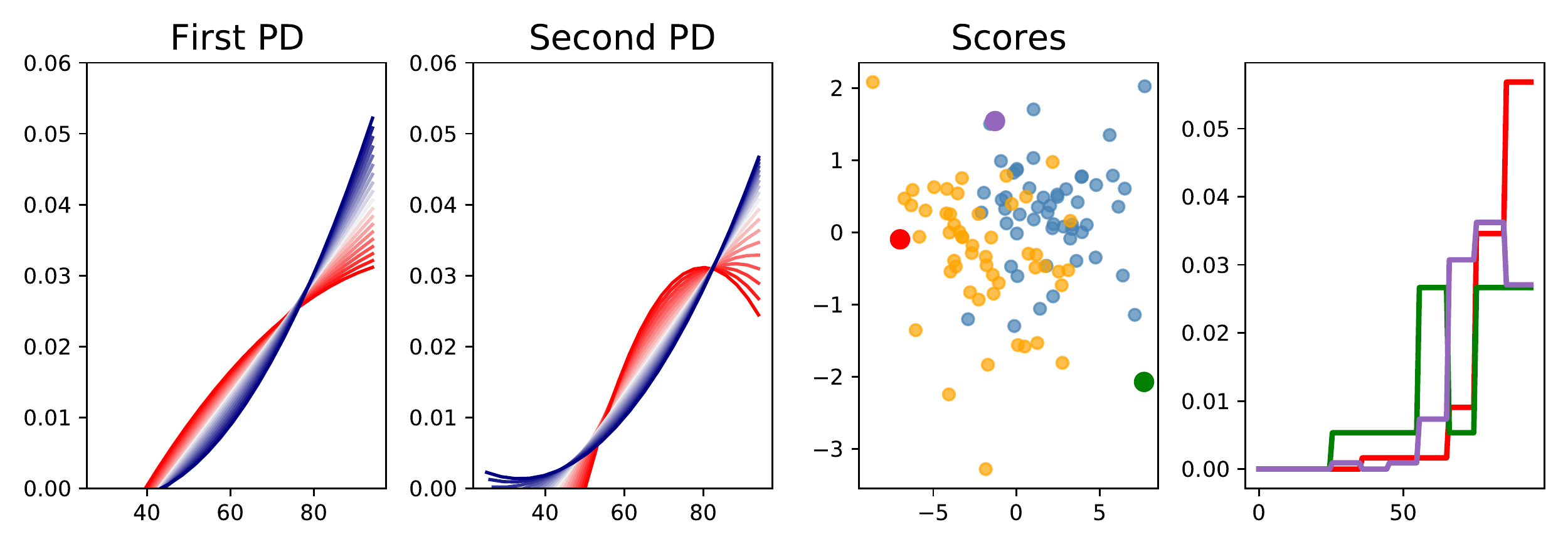}
\caption{
The first two panels show the variability along the first two principal directions (first and second panel), using the same visualization technique as in Figure~\ref{fig:simu1_results}.
The third panel reports the scores of the projections on the two dimensional principal component (orange for women and blue for men) and the fourth panel shows three particular distributions, also highlighted in the third panel.
In particular, the red distribution is the one of women in Vermont,
the green one are males in Alaska and the purple one are women in West Virginia.
}\label{fig:covid_res}
\end{figure}

Considering the discussion above and the fact that both the projected and nested PCA employ metric projection to map data points to the $k$-dimensional principal component, inference obtained with the nested PCA and with the projected one is almost identical in this case and we show results only for the projected PCA in Figure~\ref{fig:covid_res}. 
In particular, the first principal direction shows that the greatest variability
is due to the elders: low negative values along this direction correspond
to most of the mortality being concentrated among in the $80+$ range.
The red and the green distributions shown in the rightmost panel show two
antithetic behaviors which correspond to scores along the first principal
direction of roughly $-8.5$ and $7$ as shown in the third panel of Figure~\ref{fig:covid_res}.
In fact, the red distribution is concentrated almost exclusively on the last
two bins of the histogram, with the $85+$ bin weighting for more of
$60\%$ of the deaths.
At the opposite, the green distribution gives more weight to lower age values.
The second direction instead shows variability in the $40-80$ range.
The purple distribution, characterized by the highest score along this
direction, shows  that a significant percentage of deaths occurred
in the age range $60-75$.
Finally, the third panel of Figure~\ref{fig:covid_res} reports the scores along
the first two principal directions for the whole data set, blue dots representing
males and orange dots women.
We can appreciate how women tend to have lower scores on both directions.
This is in line with our understanding that Covid-19 is more severe
among the male population \citep[see for instance][]{nycovid}, which 
explains why males are more susceptible to death even at younger ages, while
deaths among women are more concentrated in the $70+$ age range, being the
elders in general more fragile.

The comparison with log PCA requires more attention.
First of all, note that the directions obtained with the projected and log PCA are the same by definition, since they are both obtained performing PCA in $L_2([0, 1])$, but 
the principal components may differ because different projection operators are employed when the orthogonal projection of a point onto the principal component lies outside of the image of $\varphi_\mu$, as discussed in Section~\ref{sec:vs_extrinsic}.
As expected from the comparison between the metric projection and the pushforward operator in Figure~\ref{fig:metric_vs_boundary_vs_push}, the fit to the data of the projected and log PCAs will be different. In particular, in this case we observe that the log PCA does a worse job in term of $NRE$, as shown in Figure~\ref{fig:covid_data}(b), especially when the dimension increases. This behavior can be also in part explained by the complexity of the numerical routines needed to approximate the pushforward operator (required by the log PCA) where it is natural to expect some numerical errors.

More in general, as discussed also in \cite{geod_vs_log}, we can conclude that the log PCA is not suited to study this particular data set because the $L_2$ PCA is different from the nested geodesic PCA (as testified by the $GV_2$ score). 
In fact, apart from the visual inspection of the $L_2$ principal directions -- which are not guaranteed to span the log-principal components -- not much can be obtained from the log PCA in this case, since it does not provide a consistent way of projecting data points on the principal component as pointed out in Section~\ref{sec:vs_extrinsic}.

\section{Numerical Illustrations for the Distribution on Distribution Regression}\label{sec:reg_simulations}

In this section, we propose a comparison between the Wasserstein projected and simplicial (see Appendix~\ref{sec:simplicial}) approaches when the task at hand is distribution on distribution regression, and show an application of the Wasserstein projected regression framework to a problem of wind speed forecasting.

\subsection{Simulation Study}\label{sec:reg_sim_study}

We consider two data generating processes as follows.
In the first setting, data are generating from the Wasserstein regression: independent variables $z_1, \ldots, z_n$ are generated by considering quantile functions $F^{-}_{z1}, \ldots, F^{-}_{zn}$ such that $F_{zi} = \sum_{h=1}^{30} a^{(z)}_{ih} \psi_j^{(3)}$ where $\psi_1^{(3)}, \ldots, \psi_{30}^{(3)}$ is a cubic spline basis over equispaced knots in $[0, 1]$ and $a^{(z)}_{i1} = 0$, $a^{(z)}_{i2} = \delta_{i1}$, $a^{(z)}_{ij} = a^{(z)}_{ij-1} + \delta_{ij-1}$, and $(\delta_{i2}, \ldots, \delta_{i 30}) \sim \text{Dirichlet}(1, \ldots, 1)$.
This data generating procedure ensures the $F^{-}_{zi}(0) = 0$, $F^{-}_{zi}(1) = 1$ and $F^{-}_{zi}$ is monotonically increasing, cf. Proposition~\ref{prop:spline_mono}.
The dependent variables $F^{-}_{y1}, \ldots, F^{-}_{yn}$ are generated using the same spline expansion of the dependent variables and letting 
$\bm a^{(y)}_{i} = B \bm a^{(z)}_{i}$. $B$ is a randomly generated matrix with rows $\bm b_1, \ldots, \bm b_{30}$, and each $\bm b_i$ is generated as follows:
$b_{i1} \sim \text{Uniform}(0, 0,5)$ $b_{ij} = b_{ij-1} + \tilde{b}_{ij}$ and $\tilde{b}_{ij} \sim \text{Uniform}(0, 0,5)$, so that the coefficients $a^{(y)}_{ij}$ are monotonically non decreasing for each $i$ and thus the $F^{-}_{yi}$'s can be considered quantile functions.

We compute the pushforward of the uniform distribution via numerical inversion and differentiation and obtain the pdf associated to each quantile function. Observe that this task is easier than approximating the pushforward of a generic $\mu$ through a generic $f$ (as \cite{geod_vs_log} do) since the quantile functions are monotonic and we have simple expressions for all the quantities related to $\mu$. Since the simplicial regression takes as input (a transformation of) the pdfs while the Wasserstein regression works directly on the quantile functions, and also due to the fact that numerical errors can be introduced in the data set during the inversion and differentiation, we consider as ground truth the pdfs and, for the Wasserstein approach, re-compute numerically the quantile functions.

In the second setting instead, we generate data from the simplicial regression model: 
independent variables $z_1, \ldots, z_n$ are generated by applying the inverse of the centered log ratio to a random spline expansion as follows.
For each $i = 1, \ldots, n$ let $\tilde p_{zi} = \sum_{j=1}^{30} a^{(z)}_{ij} \psi^{(3)}_j$ where
the $\psi^{(3)}_j$'s are the same B-spline basis as in the previous setting.
Here, the $a^{(z)}_{ij}$'s are generated iid from a Gaussian distribution with mean $0$ and standard deviation $0.2$.
The dependent variables are generated by letting $\tilde p_{yi} = \sum_{j=1}^{30} a^{(y)}_{ij} \psi^{(3)}_j$ and $\bm a^{(y)}_i = B \bm a^{(z)}_i$, where $B$ is a randomly generated $30 \times 30$ matrix with entries drawn iid from a standard normal distribution.
Finally the pdfs $p_{zi}$ ($p_{yi}$) are recovered by applying the inverse of the centered log ratio to $\tilde p_{zi}$ ($\tilde p_{yi}$), see Appendix~\ref{sec:simplicial} for more details.

Note that under the second data generating process, both the dependent and independent distributions have support in $[0, 1]$ by construction, whereas under the first data generating process the independent variables might have a larger support.
Thus, to fit the simplicial regression in the first scenario, as common practice (cf.\ Appendix~\ref{sec:simplicial}), we extend the support of all the distributions (both dependent and independent) to the smallest interval of the real line containing all the supports.
This is done by adding a small term to the pdfs (in our example, $10^{-12}$) and then renormalizing them.

\begin{table}
\centering
\begin{tabular}{c | c | c}
&  First scenario & Second scenario \\
\hline
Wasserstein & $(4 \times 10^{-7}, 7 \times 10 ^{-8})$ & $(5 \times 10^{-3}, 6 \times 10^{-3})$ \\
Simplicial & $(0.9, 2.66)$ & $(4 \times 10^{-4}, 5 \times 10^{-4})$
\end{tabular}
\caption{
Cross validation (leave one out) errors and standard deviations for the Wasserstein and Simplicial regression under the two simulated examples
}
\label{tab:regression}
\end{table}

For both examples, we simulated $100$ observations and compared the projected-Wasserstein and simplicial regression using leave-one-out cross-validation.
In particular, for both approaches we use $J=20$ quadratic spline basis
and choose the penalty term $\rho$ in \eqref{eq:reg_penalized} through grid
search.
Table~\ref{tab:regression} shows the pairs of mean squared error and standard deviation of the cross validation, the metric to compare the ground truth and the prediction is the $2$-Wasserstein distance.
As one might expect, the Wasserstein regression performs better in the first scenario while the simplicial regression performs better in the second scenario.
However, it is surprising how the Wasserstein geometry can capture (in terms of Wasserstein metric) dependence generated by a linear structure which we have shown to be very different from the Wasserstein one, making the projected regression a promising tool for such inferential problems

\subsection{Wind speed distribution forecasting from a set of experts}\label{sec:wind}

\begin{figure}[t]
\centering
\includegraphics[width=0.6\linewidth]{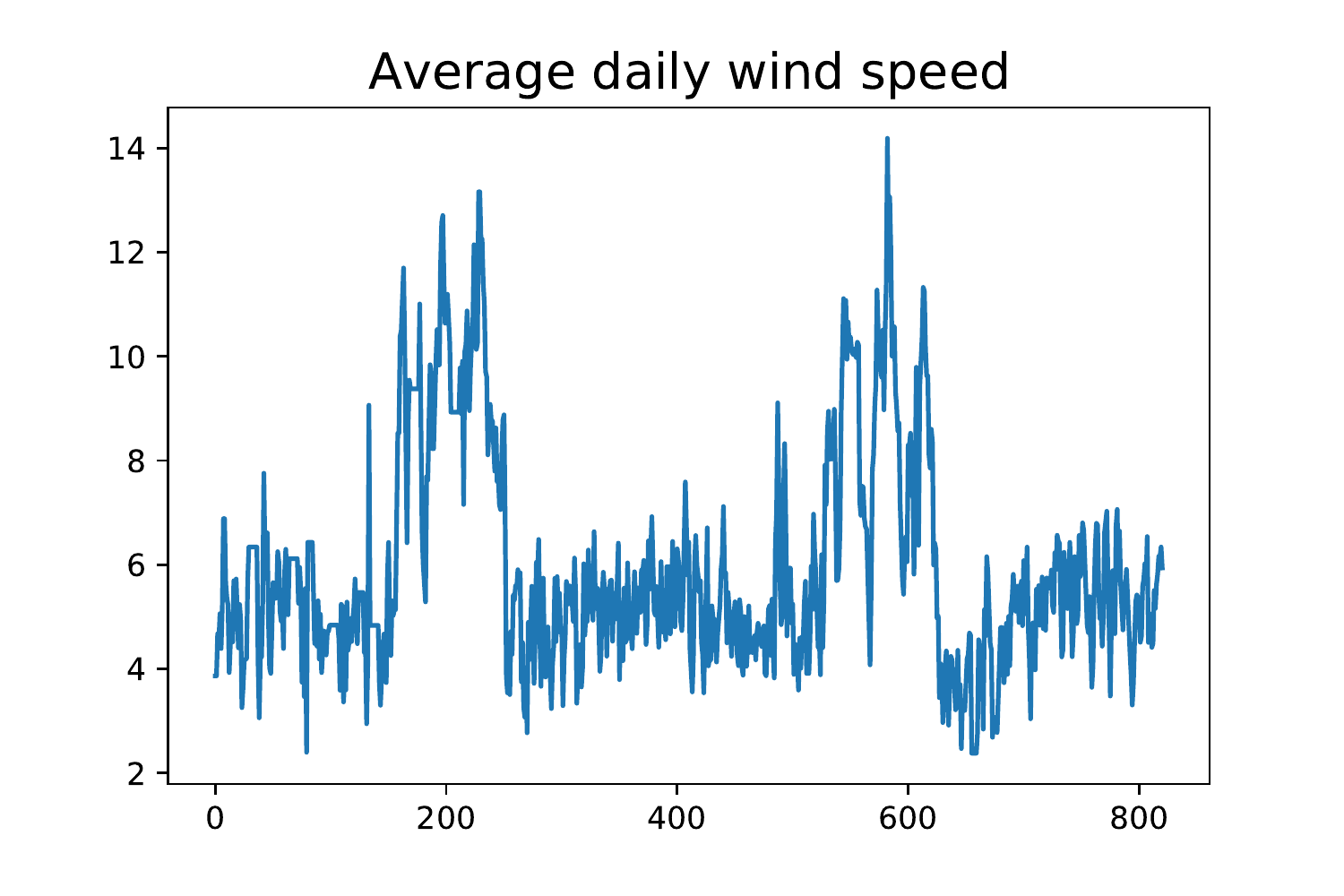}
\caption{Daily average wind speed}
\label{fig:wind_data}
\end{figure}

We consider the problem of forecasting the distribution of the wind speed
nearby a wind farm from a set of experts.
The data set is publicly available at \url{www.kaggle.com/theforcecoder/wind-power-forecasting}.
In particular, data consists of measurements of the wind speed collected
every ten minutes for a period of 821 days starting from the 31st December 2017.
The daily average wind speed is shown in Figure~\ref{fig:wind_data}.

We assume to have access to a set of \emph{experts}, that is
a set of trained models, that provide a probabilistic one-day-ahead forecast for the average wind speed.
Here, our goal is to combine this set of experts and provide a point estimate
of the wind speed distribution for the whole day, which can be helpful when
planning the maintenance of the wind mills for instance.

Formally, let $K$ denote the number of experts considered, $F^{-}_{zij}$ is the
quantile function associated to the probabilistic forecast of the average
wind speed for day $i$ given by expert $j = 1, \ldots, K$; $F^-_{yi}$ is the
empirical quantile function of the wind speed for day $i$.
In particular, we consider $K=4$ experts built from the \emph{Prophet}
model by Facebook \citep{taylor2018forecasting} as follows:
model $M1$ is the classical Prophet, without additional covariates or seasonality
trends; model $M2$ includes the ambient temperature as covariate but not seasonality;
model $M3$ includes a yearly seasonality and no covariates and model $M4$
includes both yearly seasonality and ambient temperature as covariate.
The models are estimated using variational inference on rolling samples of
$365$ days and produce one day ahead probabilistic forecasts for the average
wind speed. The final sample size corresponds to $n=456$.

We consider a trivial extension of the distribution on distribution 
regression model in Section~\ref{sec:emp_regression} as follows:
\begin{equation}\label{eq:reg_multi}
\mathbb{E}[F^-_{yi} \mid F^{-}_{zi1}, \ldots, F^{-}_{ziK}] = \Pi_{L_2([0, 1])^\uparrow} \Big( \alpha + \sum_{j=1}^K \int_0^1 \beta_j(t, s) F^{-}_{zij}(t) \ \mathrm{dt} \Big)
\end{equation}
Having approximated all the functions through a B-spline expansion, the model reads
\[
    \mathbb{E}[\bm a^{(y)}_i \mid \bm a^{(z)}_{i1}, \ldots, \bm a^{(z)}_{iJ}] = \Pi_{\mathbb{R}^{J\uparrow}}\Big(\bm \theta_\alpha + \sum_{j=1}^K \Theta_{\beta_j} E \bm a^{(z)}_{ij} \Big).
\]
The procedure for estimating $\bm \theta_\alpha$ and $\Theta_{\beta_1}, \ldots \Theta_{\beta_K}$ is analogous to the one outlined in Section~\ref{sec:emp_regression}.

We compare the prediction performance of five distribution on distribution regression models. Models $R1$ to $R4$ are obtained by fitting
model \eqref{eq:reg_multi} using only one of the four experts, $M1$ to $M4$,
while the fifth model ($RF$) is the \virgolette{full} model in \eqref{eq:reg_multi} considering all the four experts.
For this comparison, we perform a train-test split of the $456$ days for which
the experts produced the prediction, considering the last $100$ days as test.
We select hyperparameters (namely, the penalty coefficient $\rho$  in \eqref{eq:reg_penalized} and whether to include or not the intercept term $\alpha$) by a grid search cross validation on the training set, and compare
the mean square error on the held-out test set.
Results of the comparison are reported in Table~\ref{tab:reg_result}.
As expected, the model with the four predictors ($RF$) is the best performer.
Interestingly, all the other models $R1$-$R4$ perform similarly and  present a much higher mean 
square error when compared to $RF$, thus suggesting that the best performance
is achieved by combining the different experts together and no expert alone
can be a good predictor.
This is possibly explained by some experts being able to better forecast
one scenario (for instance, light winds) and other experts being able to better forecast other scenarios.

\begin{table}
\centering
\begin{tabular}{c| c c c c c}
& $R1$ & $R2$ & $R3$ & $R2$ & $RF$ \\
\hline
MSE & $(1.22 \pm 1.32)$ & $(1.19 \pm 1.26)$ & $(1.15 \pm 1.07)$ & $(1.24 \pm 1.23)$ & $(0.86 \pm 0.82)$
\end{tabular}
\caption{Mean square prediction error $\pm$ one standard deviation on the held-out test set.}
\label{tab:reg_result}
\end{table}

\begin{figure}[t]
\centering
\includegraphics[width=\linewidth]{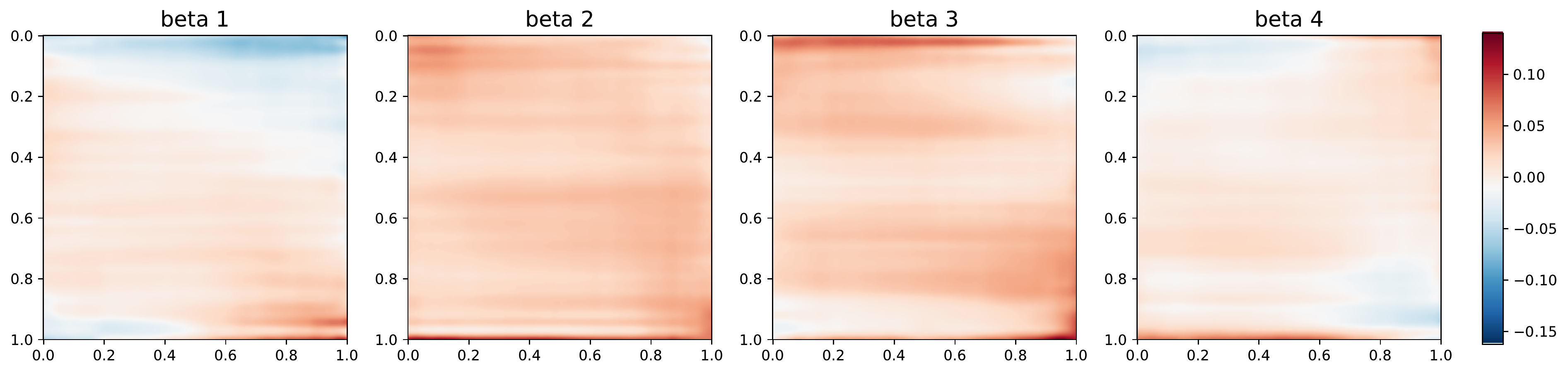}
\caption{Estimates of the $\beta_i(t, s)$'s evaluated on $[0, 1]^2$.
The variable $t$ runs across columns, and variable $s$ across rows}
\label{fig:beta}
\end{figure}

We conclude with some descriptive analysis.
Figure~\ref{fig:beta} shows the point estimates for
the coefficients $\beta_j$. We can interpret as highly influential for the regression the areas of the $\beta_j$'s with high absolute value, and as negligible area with values close to zero.

We can highlight some differences among the coefficients in Figure~\ref{fig:beta}.
In particular, model $M1$, seems
influent when predicting the tails of the distribution, in particular with negative weights for the left tail and positive weights for the right tail.
Model $M2$, seems to be affecting all the steps of the prediction and in particular to be model affecting the most the median of the distribution.
Model $M3$, appears to be, with $M2$, the most important model for the prediction: the absolute value in the corresponding regressor $\beta_3$ is often very high and with noticeable peaks corresponding to areas predicting the left tail and towards the right tail.
Finally, the regressor corresponding to $M4$ has very low values thus resulting in minor importance in terms of regression influence.

Interestingly, the experts providing the most precious inputs to our regression model are $M2$ and $M3$, that incorporate only the seasonality effect and the
temperature covariate respectively, while $M4$, which incorporates both,
seems to be less important.
Hence, the regression model in \eqref{eq:reg_multi} finds more effective combining experts trained on different covariates than correcting an expert already trained on all the covariates.
In particular, our insight is that $M2$ is responsible for centering the median of the output distribution. The tails of the distribution seem to need also the contribution of seasonality data, given by $M3$.
Finally, we also observe that the left tail of the wind distribution seems the most difficult to be predicted, needing very high positive and negative weights across different models, to be obtained.

\begin{figure}
\centering
\includegraphics[width=0.9\linewidth]{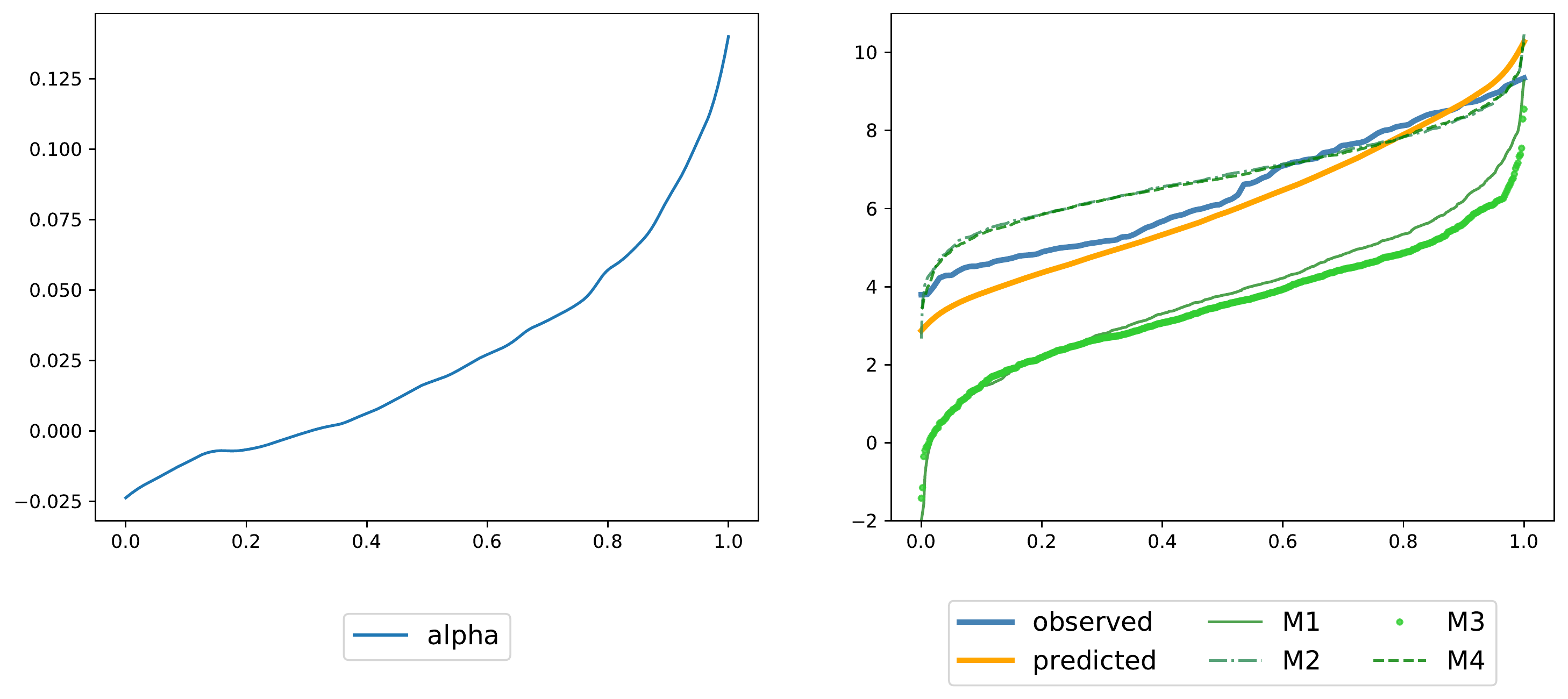}
\caption{Estimate of $\alpha$ (left) and prediction of one $F^{-}_y$ of the test
set (right).
In the right panel, the blue line corresponds to the empirical quantile function,
the orange one to the prediction from $RF$ and the green ones to the
average wind predictions obtained from the experts $M1$-$M4$.}
\end{figure}

\section{Discussion}\label{sec:discussion}

In this paper, we propose a novel class of \emph{projected} statistical
methods for distributional data on the real line, focusing in particular on 
the definition of a \emph{projected} PCA and a \emph{projected} linear regression.
By investigating the weak Riemannian structure of the Wasserstein space and the transport maps between probability measures, we represent the Wasserstein space as a closed convex cone inside an Hilbert space.

Similar to \emph{log} methods, our models exploit the possibility to map data into a linear space to perform statistics in an \emph{extrinsic} fashion.
However, instead of using operators like the \emph{exp} map or a some kind of boundary projection to return to the Wasserstein space, we rely on a metric projection operator that is more respectful of the underlying metric.

By choosing as base point the uniform measure on $[0,1]$, we are able to
efficiently approximate the metric projection operator so that our models 
combine the ease of implementation of \emph{extrinsic} methods
while retaining a performance similar to the one of \emph{intrinsic} methods.
Further, through a quadratic B-spline approximation, we can greatly reduce
the dimensionality of the optimization problems involved, resulting
in fast empirical methods. 
As a byproduct of this approach, we also derive faster numerical routines
for the \emph{geodesic} PCA in \cite{geodesic}.

We study asymptotic properties of the proposed methods, concluding that,
under reasonable regularity assumptions, our \emph{projected} models provide
consistent estimates and that the B-spline approximation error becomes 
negligible.
We showcase our approach in several simulation studies and using two real
world data sets, comparing our models to \emph{intrinsic} and \emph{extrinsic} ones and to the
\emph{simplicial} approach in \cite{menafoglio}, concluding that the \emph{projected} PCA and regression
constitute a valid candidate for performing inference on a data set of
distributions.

Although our \emph{projected} framework was proven to be viable in many practical situations,
some care must be taken when adopting it, especially when performing PCA. 
In fact, the \emph{extrinsic} nature of our method might not fit every data set, in which case a more computationally demanding \emph{intrinsic} PCA might be preferred, see for instance Appendix~\ref{sec:badex} for an example where the \emph{projected} principal directions are not interpretable.
On top of that, performing PCA in the Wasserstein space requires more attention than performing the usual Euclidean PCA: as pointed out in Appendix~\ref{sec:scores}, since principal components are not linear subspaces, decomposing the variance along the directions (i.e., looking at the scores) must be done carefully, and making sure that the directions are indeed interpretable.
To assist practitioners, in Section~\ref{sec:reliability} we have also proposed two scores that quantify the interpretability of the principal directions and the discrepancy between the \emph{nested} and \emph{projected} principal components.

Several extensions and modifications of our approach are possible.
One possibility is to extend our framework to encompass more models, such
as generalized linear models and independent component analysis.
Although this should be straightforward in theory, the numerical computations
could become more burdensome.
Furthermore, as an alternative to our approach based on B-splines approximation, one could use such B-spline expansion only to approximate
the metric projection operator.
Another interesting line of research would consist in building hybrid approaches (as anticipated in Section~\ref{sec:reliability}) to analyze distributions in the Wasserstein space, using both \emph{extrinsic} and \emph{intrinsic} methods to exploit the advantages of both worlds, while mitigating the disadvantages.
We also think that a deeper comparison between the Wasserstein and the simplicial geometries could help practitioners in choosing between them.

Finally, as pointed out by an anonymous referee, extensions to encompass measures supported on $\mathbb{R}^d$, $d>1$, are of great interest.
This is surely a very challenging problem, due to the geometric structure of $\mathcal{W}_2(\mathbb{R}^d)$. We identify three main obstacles in this sense.
First, the map onto the tangent space is not an isometry because the Wasserstein space is curved. Second, we lose the nice characterization of the tangent space and of the image of $\log_\mu$, so that the metric projection operator becomes harder to derive. Third, the computational cost would greatly increase due to the need of numerically approximating the transport maps needed to compute the distances.


\acks{We thank two anonymous referees for their detailed and helpful reviews, which helped us improving the quality and the clarity of this work.
We also thank Riccardo Scimone for helpful feedback and comments on an earlier draft of this paper and Federico Bassetti, Alessandra Guglielmi and Piercesare Secchi for helpful discussions. 
}


\appendix

\section{Proofs}\label{sec:proofs}

\noindent
\underline{\textit{Assumptions on $x_0$}.}

\smallskip\noindent    
Let $B_\varepsilon(x_0)=\{x \in H  \, \big| \, ||x-x_0||<\varepsilon\}$, a ball of radius $\varepsilon$ in $H$. 
Given a set $C$, we refer to $\text{aff}(C)$ as the smallest affine subset containing $C$, found as the intersection of all affine subspaces containing $C$. Similarly $\mathcal{H}(C)$ is the convex hull of $C$, the smallest convex subset of $H$ containing it. 
The relative interior of a set $C$ is defined as its interior considering as ambient space $\text{aff}(C)$: $\text{relint}(C)=\{x\in C \, \big| \, \exists B_\varepsilon(x_0) \text{ such that }  B_\varepsilon(x_0)\cap \text{aff}(C)\subset C \}$.

Throughout our paper we assume that the random variable $\Xvar$ is such that 
(i) there exists $x_0 = \mathbb{E}[\Xvar]$ and  (ii) $x_0\in \text{relint}(\mathcal{H}(\text{supp}(\Xvar)))$ where $\text{supp}(\Xvar)$ is the support of $\Xvar$.
These assumptions are indeed quite natural and require that the distribution of $\Xvar$ has a well defined barycenter, which is not in a \virgolette{degenerate} position with respect to the convex hull of its support, which may happen in infinite dimensional Hilbert Spaces. 
See, for instance, \cite{berezin2019barycenters} for an example of distributions not verifying this second assumption.

\medskip
\noindent
\underline{\textit{Proof of} Lemma \ref{lemma:pca_easy}.}

\smallskip\noindent
The proof is divided in two steps. First, we prove that $(x_0 + Sp(U_k)) \cap X$ has dimension $k$. Then, we show that $U_X^{x_0, k} = (x_0 + Sp(U_k)) \cap X$. 
Without loss of generality, for ease of notation, we perform an affine change of variable so that $x_0 = 0$, but, with a slight abuse of notation, we keep denoting with $\Xvar$ and $X$ the transformed random variable and the convex cone respectively.

To prove the first part, let $\mathcal{H}(\Xvar)$ be the convex hull of the support of $\Xvar$ and $\text{aff}(\mathcal{H}(\Xvar))=K$ be the smallest affine subset of $H$ containing $\mathcal{H}(\Xvar)$.
We know by assumption that there is an open ball in $K$ which contains $x_0 = 0$ and is contained in $\mathcal{H}(\Xvar)$. 
Moreover, for every $k \leq dim(K)$, $Sp(U_k)\subset K$. 
Note that we can clearly suppose $k \leq dim(K)$, otherwise principal components analysis is useless. With this assumption, since $x_0 = 0$ is in the relative intern of $\mathcal{H}(\Xvar)$, we have 
$k=dim(Sp(U_k) \cap \mathcal{H}(\Xvar)) \leq dim(Sp(U_k) \cap X)\leq k$.

Now we prove that a $(k, 0)$-projected principal component is given by $Sp(U_k)\cap X$. 
To prove this, let $C^*$ be a $(k, 0)$-projected principal component and $A^* =  A \cap  X$, with $A=Sp(U_k)$: we know  (i) $x_0 = 0 \in A^*$, (ii) $dim(A^*) = k$ by definition and (iii) $A^* \subseteq \Pi_X(A)$, so we have $A^*\subset C^*$.

Since $dim(C^*)=k$ there is $C$ linear subspace of dimension $k$ such that $C^*\subset C$. Consider $C'= C \cap  X$: clearly $C^*\subset C'$,  so that  $A^* \subset C^* \subset C'$. 
Moreover, $A^*\subset C'$, which implies $A\cap X \subset C\cap  X$ 
and thus $Sp(A \cap X)\subset Sp(C \cap X)$.
The proof is concluded if $dim(Sp(A \cap X)) = dim(Sp(C \cap X))=k$. In fact, in this case
$A=Sp(A \cap X)$ and $C=Sp(C \cap X)$ which means that $A\subset C$ and since $dim(A)=dim(C)=k$, $A$ and $C$ coincide, proving $A^*=C^*$.

To prove this final claim, observe that $dim(Sp(A \cap X)) < k$ implies $dim(A \cap X) < k$, which contradicts  the proof of the first part of this Lemma. Similarly, $dim(Sp(C \cap X))=k$ since $dim(C^*) = k$ by hypothesis.





\hfill\BlackBox

\medskip
\noindent
\underline{\textit{Proof of} Proposition \ref{prop:err_convergence}.}

\smallskip\noindent
The fact that  $\|\Pi_{U_X^{x_0,k}}(x)-x\| \geq \|\Pi_{U_X^{x_0,k+1}}(x)-x\|$ follows easily by noticing that $U_X^{x_0,k}\subset U_X^{x_0,k+1}$.

Now, to prove that $\|\Pi_{U_X^{x_0,k}}(x)-x\| \rightarrow 0$ as $k$ increases, we first notice that, by the properties of the principal components in $H$ we have $\Pi_{Sp(U_{k})}(x-x_0)\xrightarrow{k} x-x_0$ for every $x\in X$, which implies $\|\Pi_{Sp(U_k)+x_0}(x)-x\|\rightarrow 0$. 
Then, denote $x_1=\Pi_{U_X^{x_0,1}}(x)$ and let $r_k$ be the line between
$x_1$ and $x$. Let:
\[
x_k=\argmin_{x'\in r_k\cap Sp(U_k)+x_0}\|x'-x\|
\] 
We clearly have have $x_k\rightarrow x$. Finally, by convexity we know $x_k\in U_X^{x_0,k}$, which implies
$\|\Pi_{U_X^{x_0,k}}(x)-x\|\leq \|x_k-x\|\rightarrow 0$.

\hfill\BlackBox

\medskip
\noindent
\underline{\textit{Proof of} Proposition \ref{prop:metric_proj}.}

\smallskip\noindent
Again, without loss of generality, for ease of notation, we perform an affine change of variable so that $x_0 = 0$, but, with a slight abuse of notation, we keep denoting with $\Xvar$ and $X$ the transformed random variable and convex cone respectively.

We start by noticing that being $\Pi_{k}$ the orthogonal projection onto 
a subspace, $x-\Pi_k(x) \bot Span(U_k)$ and thus for $v \in Span(U_k)$:
\[
\|x^* - v\|^2=\|x^* -\Pi_k(x^*)\|^2+\|\Pi_k(x^*)-v\|^2
\]
Then
\[
    \argmin_{v\in U_X^{0,k}} \| x^* - v \| =  
        \argmin_{v \in Sp(U_k)\cap X} \|\Pi_k(x^*)-v\|
\]
and the result follows. 
\hfill\BlackBox

\medskip\noindent
\underline{\textit{Proof of} Proposition \ref{prop:spline_mono}.}
\begin{enumerate}
\item As shown in the supplementary of \cite{shape_costr_add_models} by standard B-spline formulas we obtain that given $f(x) = \sum_{j=1}^J a_j \psi^k_j(x)$, then $f'(x) = \sum_{j=1}^J (a_j- a_{j-1})\cdot \psi^{k-1}_j(x)$. Being the B-spline basis function nonnegative by definition, we obtain the result.
\item With $k=2$, $f'(x)$ on the interval $[x_{j+1},x_j]$ has the following expression:
\[
\frac{x-x_j}{x_{j+1}-x_j}\cdot (\alpha_j-\alpha_{j-1})+\frac{x_{j+1}-x}{x_{j+1}-x_j}\cdot (\alpha_{j-1}-\alpha_{j-2})
\]

so:
\[
lim_{x\rightarrow x_{j+1}^-} f'(x) = \alpha_j-\alpha_{j-1}
\]
and the result follows. \hfill\BlackBox
\end{enumerate}

\medskip\noindent
\underline{\textit{Proof of} Proposition \ref{prop:geod} and \ref{prop:nest}.}

\smallskip\noindent
We report here Propositions 3.3 and 3.4 of \cite{geodesic}, with the notation adapted to our manuscript. In the following $H$ is a separable Hilbert space, $X$ is a closed convex subset of $H$, $\Xvar$ is an $X$-valued square integrable random variable, $x_0$ a point in $X$ and $k\geq 1$ an integer.

\begin{prop}\label{prop:bigot_geod}
Let $U^*=\{u^*_1,..,u^*_k\}$ be a minimizer over orthonormal sets $U$ of $H$ of cardinality $k$, of $D^{x_0}_X(\Xvar,U) := \mathbb{E}d^2(\Xvar,(x_0+ Sp(U))\cap X)$, then $U^{x_0}_X := (x_0+ Sp(U))\cap X$ is a $(k,x_0)-$global principal component of $\Xvar$.
\end{prop}

\begin{prop}\label{prop:bigot_nest}
Let $U^*=\{u^*_1,..,u^*_k\}$ be an orthonormal set such that $U^*_i = \{u^*_1,..,u^*_i\}$ is a minimimizer of $D^{x_0}_X(\Xvar,U)$ over the orthonormal sets of cardinality \virgolette{$i$} such that $U\supset U^*_{i-1}$; then $U^{*x_0}_X$
is a $(k,x_0)-$nested principal convex component of $\Xvar$.
\end{prop}

Applying Propositions \ref{prop:bigot_geod} and \ref{prop:bigot_nest}  we can obtain equivalent definitions of geodesic and nested PCA as optimization problems in $L_2([0,1])$. If we fix $J \in \mathbb{N}>0$ and a quadratic B-spline basis $\{\psi_j\}_{j=1}^J$, we can use Propositions \ref{prop:bigot_geod} and \ref{prop:bigot_nest} with $X=L_2([0,1])^{J\uparrow}$ and $H=L_2([0,1])^J$.  Thanks to Remark \ref{rmk:RJ} we obtain the results.

\hfill\BlackBox

\medskip\noindent
\underline{\textit{Proof of} Proposition \ref{prop:spline_conv}.}

\smallskip\noindent
    Let $S_J = \sum_{j=1}^J \lambda^{(J)}_j \psi^{(J)}_j$ and its derivative
    $s_J = \sum_j (\lambda^{(J)}_j - \lambda^{(J)}_{j-1}) \widetilde \psi^{(J)}_j$ where
    $\widetilde \psi^{(J)}_j$  denotes the linear spline basis on the same equispaced grid in $[0, 1]$.
    
    Let $f^{-}_\mu = (F^{-}_\mu)^\prime$, of course it can be seen that $f^{-}_\mu$ is
    non-negative. Moreover, it is obvious that $f^{-}_\mu \in W_2^\infty([0, 1])$.
    Then, from \cite{deboor1974splines} we get that there exist $s_J$ such that
    $\|s_J -  f^{-}_\mu\|_\infty \leq C \|D^2f^{-}_\mu\|_\infty J^{-2}$, where $C$ is a constant depending
    on the interval $[0,1]$ but not on $n$.
    
    Hence, we can determine the coefficients $\{\lambda_j^{(J)}\}$, starting from
    the spline $s_J$, up to a translation factor.

    We fix a particular set of coefficients
    by letting $S_J(0) = \lambda^{(J)}_1 =F^{-}_\mu(0)$ for each~$J$.
    So that:
    \[
        S_J(x) - F^{-}_\mu(x) = \int_{0}^x s_J(t)dt - \int_{0}^x f^{-}_\mu(t) dt - S_J(0) + F^{-}_\mu (0) = \int_{0}^x s_J(t) - f^{-}_\mu(t) dt
    \]
    By using the previous result, the integral we have that
    $S_J(x) - F^{-}_\mu (x) \leq C J^{-2}$ for all $x$ which proves the proposition. \hfill\BlackBox

\medskip\noindent
\underline{\textit{Proof of} Proposition \ref{prop:PCA_consist}.} 

\smallskip\noindent
By the Assumptions in Section \ref{sec:PCA_assumptions} and Remark \ref{rmk:inclusions} there exists a ball $B_K$ in 
$W_3^\infty([0,1])$ of radius $K$ for some $K>0$, such that each $F^-_i$ can be $\varepsilon$-approximated by  
$\widetilde{F}^-_i \in W_3^\infty([0,1])$ 
with $\widetilde{F}^-\in B_K$. We can suppose 
that also the eigenvectors of the covariance operator of the 
generating process belong to such sphere, otherwise we just 
increase its radius of some finite amount.

By Proposition~\ref{prop:spline_conv} we can choose a spline basis (that is, a number 
of elements $J>0$), such that we get a $\varepsilon$-uniformly good approximation of $B_K$ (and thus we can $2\varepsilon$-approximate its $L_2$ closure).
To lighten notation, thanks to Remark~\ref{rmk:RJ} we deliberately
confuse $\mathbb{R}^{J\uparrow}$ and the space monotone $B$-splines with $J$ basis functions, the inner product we are referring to will always be clear by
looking at its entries.

Now consider the following inequalities, with $\bm a_i^J$ obtained as $2\varepsilon$ approximations of $F^-_i$, $\bm w^J \in \mathbb{R^J}$, $w \in L_2([0, 1])$:
\begin{multline*}
\Big|\frac{1}{n}\sum_i \langle F^-_i,w \rangle^2-\frac{1}{n}\sum_i \langle \bm a_i^J, \bm w^J \rangle^2 \Big|\leq  \\
\frac{1}{n} \Big|\sum_i \langle F^-_i,w \rangle^2-\sum_i \langle \bm a_i^J, w \rangle^2+
\sum_i \langle \bm a_i^J, w \rangle^2-\sum_i \langle \bm a_i^J, \bm w \rangle^2 \Big|,
\end{multline*}
where the inner product $\langle \bm a_i^J, w \rangle$ is to be intended as the $L_2$ inner product between the spline function with coefficients $\bm a_i^J$ and the $L_2$ function $w$.
Consider now:
\begin{align*}
\frac{1}{n}\sum_i (\langle F^-_i, w \rangle^2 &- \langle \bm a_i^J,w \rangle^2) = \\
& \frac{1}{n}\sum_i (\langle F^-_i,w \rangle -\langle \bm a_i^J,w\rangle) (\langle F^-_i,w \rangle + \langle \bm a_i^J,w \rangle) = \\
& \frac{1}{n}\sum_i \langle F^-_i - \bm a^J_i, w \rangle  \langle F^-_i + \bm a_i^J, w \rangle \leq \\
& \frac{1}{n}\sum_i \Big| \langle F^-_i - \bm a^J_i,w \rangle \Big| \cdot \Big|\langle F^-_i + \bm a_i^J,w \rangle \Big|\leq \\
& \frac{1}{n} \sum_i 2\varepsilon \|w\|^2 2K = 
 4\varepsilon K\|w\|^2
\end{align*}
Similarly:
\[
\Big| \frac{1}{n}\sum_i (\langle \bm a_i^J,w \rangle^2- \langle \bm a_i^J, \bm w^J \rangle^2) \Big| \leq \|\bm a_i^J\|^2\cdot\|w- \bm w^J\|\cdot (\|w\|+\|\bm w^J\|)  
\]

We know that a solution to the problem 
$\max_{\|w\|_{L_2}=1}\frac{1}{n} \sum_i \langle F^-_i, w \rangle^2$ is given by the 
first eigenfunction $\widehat{w}$ of the covariance operator of the 
empirical process. Now we are in the condition to apply results in  
\cite{Dauxois1982AsymptoticTF}, or in \cite{PCA_consist} (with 
$\alpha\rightarrow 0$) to conclude that  $\widehat{w}$ converges to 
the first eigenfunction $\bar{w}$ of the covariance operator of 
the process that generates $F^-_i$. By hypothesis, such eigenfunction 
$\bar{w}$ lies in $B_K$ and thus can be approximated with our 
fixed spline basis. Thus for high enough $n$, also $\widehat{w}$ can 
be approximated up to $2\varepsilon$.

Let $\bm a_{\widehat{w}}$ be the coefficients of the spline expansion of 
$\widehat{w}$ spline approximation, that is, $\|w- \bm a_w\|\leq 2\varepsilon$.  
Observe that 
$\Big| \| \widehat{w} \|_2 - \| \bm a_{\widehat{w}}\|_E \Big| \leq 2 \varepsilon$, just as
 $\|\bm a^i_J\| \leq K+2\varepsilon$. Thus, up to adding another 
 $\varepsilon$ to the approximation error 
 $\| \widehat{w}- \bm a_{\widehat{w}}\|$, we can suppose $\|\bm a_{\widehat{w}}\|_2=1$.
Hence:
\[
\Big|\frac{1}{n}\sum_i (\langle \bm a_i^J,\widehat{w} \rangle^2- \langle \bm a_i^J, \bm a_{\widehat{w}} \rangle^2) \Big| \leq
 (K+2\varepsilon)\cdot 3\varepsilon\cdot 2
\]

Which leads to:
\[
\Big|\max_{\|w\|_{L_2}=1}\sum_i \langle \bm a_i^J,w \rangle^2 - \max_{\|\bm w^J\|_E=1}\sum_i
 \langle \bm a_i^J, \bm w^J \rangle^2 \Big| \leq (K+2\varepsilon)\cdot 3\varepsilon\cdot 2
\]

Finally, combining the above results and the fact that 
$|\max f-\max g|\leq \max|f-g|$ for any pair of real valued 
functions $f$ and $g$, we obtain:
\begin{multline*}
\Big| \max_{\|w\|_{L_2}=1}\frac{1}{n} \sum_i \langle f_i,w \rangle^2-
\max_{\|\bm w^J\|_{E}=1} \frac{1}{n}\sum_i \langle \bm a_i^J, \bm w^J \rangle^2 \Big|\leq \\
\max_{\|w\|_{L_2}=1} 4\varepsilon K\|w\| + (K+2\varepsilon)\cdot
 6\varepsilon \leq 6\varepsilon K(1 + 2\varepsilon)
\end{multline*}

Thus for instance if we ask that $\varepsilon<1$, we obtain the 
desired result with $D = 18\cdot K$. Consistency follows since 
$\|\bm a_{\widehat{w}}-\bar{w}\|\leq \|\bm a_{\widehat{w}}-\widehat{w}\|+\|\widehat{w}-
\bar{w}\|$. \hfill\BlackBox

\bigskip\noindent
\underline{\textit{Proof of} Lemma \ref{lemma:proj_conv}.} 

\smallskip\noindent
Since for any $x \in X$ we have 
$\Pi_{\mathbb{R}^{J\uparrow}}(x)\rightarrow x$, for any $v \in H$:
\[
\|v-\Pi_{\mathbb{R}^{J\uparrow}}(v)\|\leq \| v-
\Pi_{\mathbb{R}^{J\uparrow}}
(\Pi_X( v))\|\leq \| v-\Pi_X( v)\|+
\|\Pi_X( v)-\Pi_{\mathbb{R}^{J\uparrow}}(\Pi_X( v))\|  
\]
which implies $\Pi_{\mathbb{R}^{J\uparrow}}( v)\rightarrow \Pi_X( v)$.
Consider now $\beta_n\rightarrow \beta$ in $H$; we have the inequality:
\[
\|\Pi_{\mathbb{R}^{J\uparrow}}(\beta_n)-\Pi(\beta)\|\leq 
\|\Pi_{\mathbb{R}^{J\uparrow}}(\beta_n)-\Pi_X(\beta_n)\|+
\|\Pi_X(\beta_n)-\Pi_X(\beta)\|
\]
the first term of the right hand side of the inequality can be 
sent to $0$ by increasing $J$, the other by increasing $n$. \hfill\BlackBox

\bigskip\noindent
\underline{\textit{Proof of} Proposition \ref{prop:reg_conv}.}

\smallskip\noindent
We call $a_i$ the spline coefficients associated to $x_i$ and $b_i$ the ones associated to $y_i$. Again we deliberately confuse the spaces where the coefficients live to lighten the notation.
Since the penalty term does not depend on the data, we have:
\begin{align*}
& \frac{1}{n} \Big| \sum_i \|y_i  - \langle x_i,B^TAB\rangle \|^2 - \sum_i \|b_i-\langle a_i,B^TAB \rangle _{L_2([0,1])}\|^2| = \\
& \qquad \frac{1}{n}| \sum_i (\|y_i-\langle x_i,B^TAB \rangle \|^2 - \|b_i-\langle a_i,B^TAB \rangle _{L_2([0,1])}\|^2)|\leq \\
 & \qquad \frac{1}{n}  \sum_i \|y_i-\langle x_i,B^TAB \rangle \|^2- \|b_i-\langle a_i,B^TAB \rangle _{L_2([0,1])}\|^2 |
\end{align*}
Now, since
\begin{align*}
& \Big| \|y_i - \langle x_i,B^TAB \rangle\|^2- \|b_i-\langle a_i,B^TAB\rangle_{L_2([0,1])}\|^2 \Big| = \\
& \qquad \Big|(\|y_i-\langle x_i,B^TAB \rangle \|-\|b_i-\langle a_i,B^TAB \rangle \|)\times \\
& \qquad \qquad (\|y_i-\langle x_i,B^TAB \rangle \|+\|b_i-\langle a_i,B^TAB \rangle \|) \Big|
\end{align*}

Then for some constant $K$ depending on the bounds in the Assumptions, we get:
\begin{align*}
&\Big| \|y_i-\langle x_i,B^TAB \rangle \|^2 - \|b_i-\langle a_i,B^TAB \rangle _{L_2([0,1])}\|^2 \Big| \leq \\
& \qquad  \|y_i-\langle x_i,B^TAB \rangle -b_i + \langle a_i,B^TAB \rangle\|2K = \\
& \qquad  \left(\|y_i-b_i\| + \langle a_i-x_i,B^TAB \rangle \right)2K
\end{align*}

Thus, if $J$ is such that we have $\varepsilon$-approximations of the data, by Cauchy-Schwartz we obtain:
\[
\frac{1}{n} \Big| \sum_i \|y_i-\langle x_i,B^TAB \rangle\|^2-
\sum_i \|b_i-\langle a_i,B^TAB \rangle_{L_2([0,1])}\|^2 \Big| \leq K'\cdot\varepsilon 
\] 
for some $K'$ constant.

Thanks to the results in \cite{regression_consistency}, for any 
$\varepsilon>0$, if the number of samples is big, 
$\widehat{\Theta}$ and $\widehat{\Theta}_J$ exist with probability 
$1-\varepsilon$ and are unique. Since the value of the 
minimization problem the solve are arbitrarily close, then the 
minimizers converge in $\mathbb{R}^{J\times J}$ with the metric given by the spline basis.  \hfill\BlackBox

\bigskip\noindent
\underline{\textit{Strong convergence implies semi-norm convergence}.}

\smallskip\noindent
Let $\Zvar$ be an $H$-valued random variable and $\mathcal{C}_{\Zvar}$ the covariance operator associated to 
 $\Zvar$, that is:
\[ 
 (\mathcal{C}_{\Zvar}f)(s)= 
 \int_{[0,1]} cov(\bm x(s),\bm x(t))f(t) dt.
 \]

In the following, we denote with $\|\cdot\|_{L_2}$ the $L_2([0, 1]^2)$ norm. Further, recall that  $\|cov(\Zvar(s),\Zvar(t))\|_{L_2([0,1]^2)}= 
 \mathbb{E}[\|\Zvar\|^2]$.
We want to look at the behavior of $\|\widehat{\beta}_{\textrm{PS}}-\widehat{\beta}_J\|_{\mathcal{C}_{\Zvar}}$.
\begin{align*}
& \int_{[0,1]} \langle \mathcal{C}_{\Zvar}(\widehat{\beta}_{\textrm{PS}}(s,t)-
\widehat{\beta}_J(s,t)),
\widehat{\beta}_{\textrm{PS}}(s,t)-\widehat{\beta}_J(s,t) \rangle dt \leq \\
& \qquad \| \mathcal{C}_{\Zvar}(\widehat{\beta}_{\textrm{PS}}(s,t)-
\widehat{\beta}_J(s,t))\|_{L_2}\cdot 
\|\widehat{\beta}_{\textrm{PS}}(s,t)-\widehat{\beta}_J(s,t)\|_{L_2}
\leq \\
& \qquad \mathbb{E}[\|\bm x\|^2]\cdot\|\widehat{\beta}_{\textrm{PS}}(s,t)-
\widehat{\beta}_J(s,t)\|_{L_2}\cdot 
\|\widehat{\beta}_{\textrm{PS}}(s,t)-\widehat{\beta}_J(s,t)\|_{L_2}.
\end{align*}

So $\|\widehat{\beta}_{\textrm{PS}}-\widehat{\beta}_J\|_{\mathcal{C}_{\Zvar}}\leq M\cdot \|\widehat{\beta}_{\textrm{PS}}-\widehat{\beta}_J\|_{L_2}^2$ for some constant $M$. Thus $\|\cdot\|_{L_2}$ convergence implies 
$\|\cdot\|_{\mathcal{C}_{\Zvar}}$ convergence.

\section{The simplicial approach}\label{sec:simplicial}

The simplicial approach to distributional data analysis is based on 
the definition of Bayes space $\mathcal{B}^2(I)$ \citep{egozcue2006hilbert}.
Formally, let $I \subset \mathbb{R}$ a closed interval, the Bayes spaces
$\mathcal{B}^2(I)$ is defined the equivalence class of probability densities
$p(x)$ on $I$ (that is $p(x) \geq 0$ and $\int_I p(x) dx = 1$) with square
integrable logarithm.

The Bayes space is endowed with a linear space starting from the definition
of the perturbation and powering operators, that are analogous to the 
sum and multiplication times a scalar, and inner product.
Moreover \cite{menafoglio2014kriging} defines an isometric isomorphism between
$\mathcal{B}^2(I)$ and $L_2([0, 1])$ through the so-called centered log ratio (clr) map defined as 
\begin{equation}\label{eq:clr}
\widetilde{p}(x) := \text{clr}(p)(x) = \log(p (x)) - \frac{1}{b - a} \int_a^b \log p(t) dt
\end{equation}
for every $p \in \mathcal{B}^2(I)$. The inverse map is defined as
\[
    p(x) = \text{clr}^{-1}(\widetilde{p})(x) = \frac{\exp(\widetilde{p}(x))}{\int_I \exp(\widetilde{p}(x)) dx}
\]

Thus, it is possible to define a \emph{simplicial} PCA and \emph{simplicial}
regression on the Bayes space starting from the clr map.
In particular, let $p_1, \ldots, p_n$ be observed densities on the interval $I$ and let $\widetilde{p}_i = \text{clr}(p_i)$.
Denote with $\widetilde{w}_1, \ldots, \widetilde{w}_k$ the first $k$ principal
directions estimated from the $\widetilde{p}_i$'s, then a $k$ dimensional
simplicial principal component is the span of $\{w_i = \text{clr}^{-1}(\widetilde{w}_i) \}_{i=1}^k$ in $\mathcal{B}^2(I)$.

Similarly, for pdfs $\{(p_z, p_y)_i\}_{i=1}^n$ a simplicial regression
model is defined starting from the clr transformed variables.
Let $\widetilde \Gamma$ denote a functional regression model in $L_2$ for variables $\{(\widetilde p_z, \widetilde p_y)_i\}_{i=1}^n$, then the simplicial
regression states:
\[
    \mathbb{E}[p_{yi} \mid p_{zi}] = \text{clr}^{-1}\left( \widetilde \Gamma (\widetilde p_{zi}) \right).
\]

Apart from the different geometries of the Wasserstein and Bayes space, which are discussed in Sections \ref{sec:PCA_simulations} and \ref{sec:reg_simulations}, we can highlight one particular drawback from the simplicial approach, which we believe poses a significant
limit to its usefulness.
In fact, the main assumption is that all the pdfs $p_i$ share the same support,
which might not be the case (for instance, it is not the case for our example
in Section~\ref{sec:wind}).
In practice, one may circumvent this need by either \virgolette{padding}
all the pdfs to the same support, i.e considering
\begin{equation}\label{eq:pad}
\overline{p}_i(x) \propto p_i(x) + \varepsilon \mathbb{I}[x \in I],
\end{equation}
where $\mathbb{I}[\cdot]$ denotes the indicator function, and the proportionality is due to the need of re-normalizing the $\overline{p}_i$'s so that they integrate to 1. 
Another approach could consist in considering $I$ as the intersection of all
the supports of the different $p_i$'s let truncate all the pdfs to the shared
interval $I$.

Both approaches present undesired side effects that can greatly alter the results.
The second approach might end up with a very small interval $I$, so that a lot
of information is lost due to this pre-processing step.
The drawback of the first approach instead is due to numerical instability.
In fact, one would like $\varepsilon$ in \eqref{eq:pad} to be small in order
not to corrupt the true signal, given by $p_i$. However, considering
the transformation in \eqref{eq:clr} having a small $\varepsilon$ would cause
the $\widetilde{p}_i$ to present some extreme values (negative) in correspondence
to $\varepsilon$.
Performing PCA on a data set processed in this way would greatly alter the results,
as most of the variability of the $\widetilde{p}_i$'s would be masked by
a difference in their support.

\section{Additional Simulations}\label{sec:simulation2}

\subsection{Sensitivity Analysis to the Number of Basis Functions}

\begin{figure}
    \centering
    \includegraphics[width=0.5\linewidth]{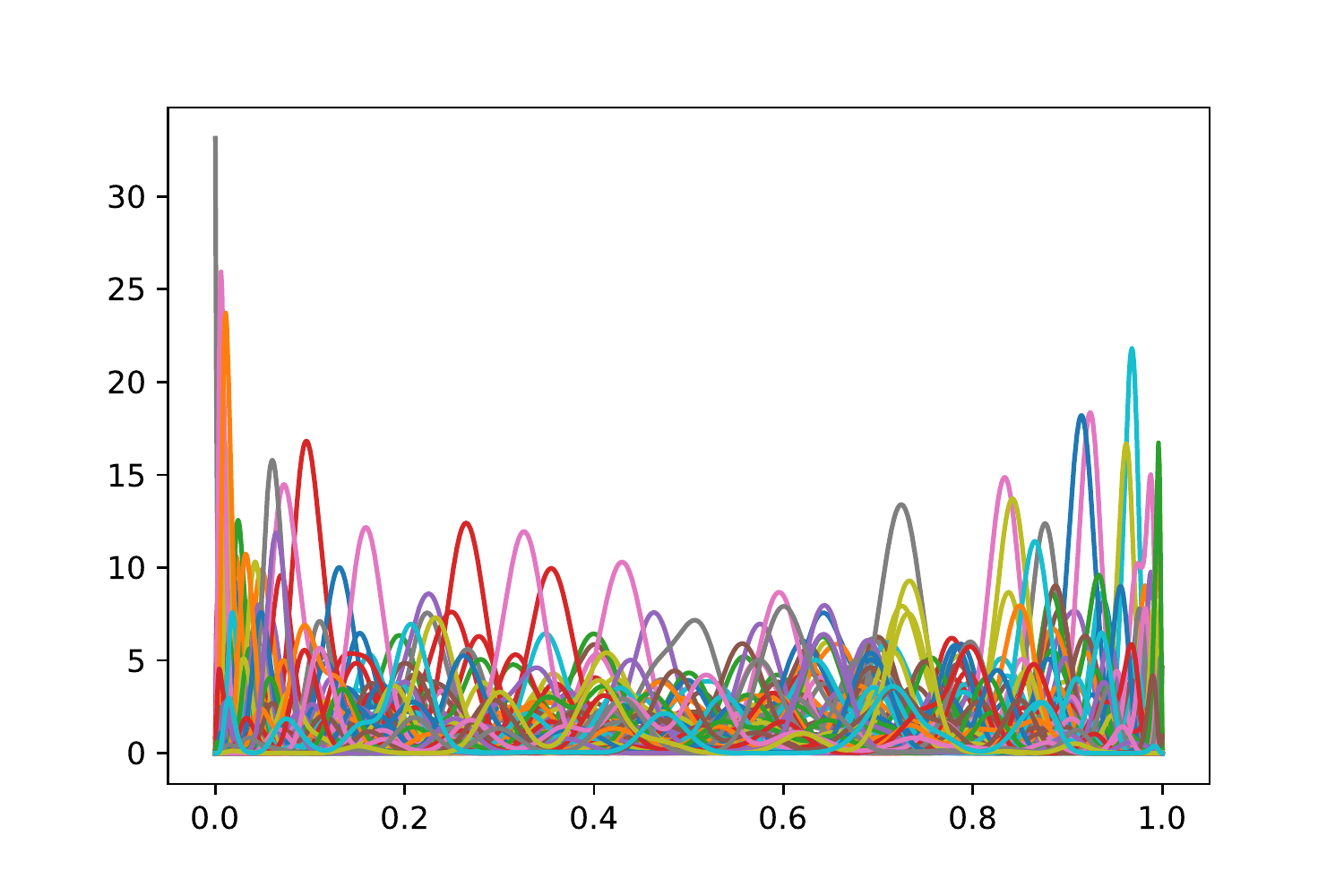}
    \caption{Example of data set from~\eqref{eq:simu3_dgp}}
    \label{fig:simu3_data}
\end{figure}

In this simulation, we show how the number of B-spline basis functions affects the inference
in our projected PCA and in the simplicial one.
In this Scenario, the probability measures are simulated as mixture of beta densities, also
known as Bernstein polynomials, as follows:
\begin{equation}\label{eq:simu3_dgp}
\begin{aligned}
    p_i(x) &= \sum_{j=1}^K w_{ij} \beta(x; j, K-j) \\
    \bm w_i & \sim \text{Dirichlet}_K(0.01)
\end{aligned}
\end{equation}
Where $\beta(x; a, b)$ denotes the density of a beta distributed random
variable with parameters $(a, b)$ evaluated in $x$.
By definition, the $p_i$s generated from~\eqref{eq:simu3_dgp} have a fixed
support $I=[0, 1]$. See Figure~\ref{fig:simu3_data}.

In this setting instead, we let $\mu_i$ in~\eqref{eq:rec_error}
be the probability measure
associated to $p_i$ and not its smoothed version.
Hence, in addition to the amount of information lost during the PCA another
factor comes into play: the amount of information that is lost due to
the B-spline representation.

\begin{figure}
    \centering
    \includegraphics[width=\linewidth]{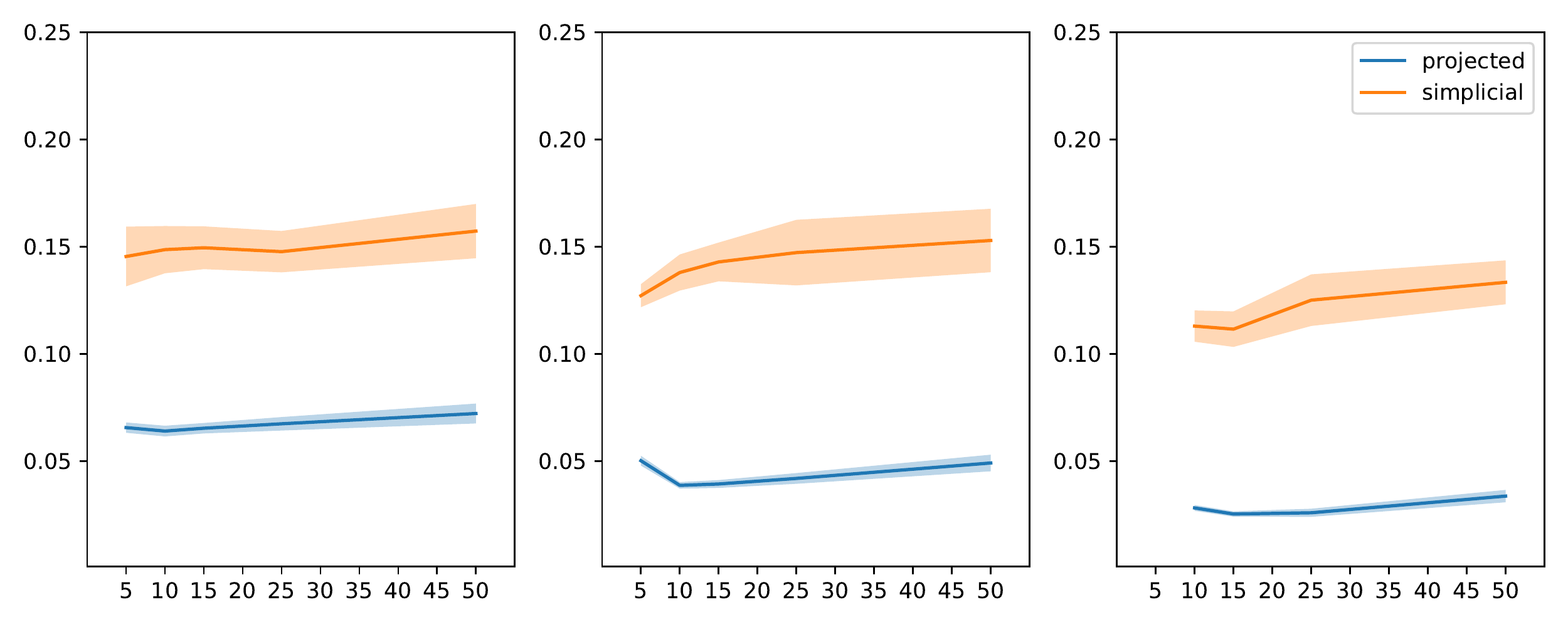}
    \caption{Results for the third scenario.
             All the panels show the reconstruction error as a function of the number 
             of the spline basis functions.
             From left to right the results are obtained using the $2$, $5$ and $10$ dimensional PCA.
             The solid lines represent the mean of 10 independent runs on independent data sets from~\eqref{eq:simu3_dgp} and the shaded area represent $\pm$ one standard deviation.}
    \label{fig:simu3_res}
\end{figure}

Figure~\ref{fig:simu3_res} shows the results. We can see that the reconstruction
errors decrease when the dimension of the principal component increases both for the
simplicial and projected PCA.
Moreover, as the number of B-spline basis increase, the performance tend to get
a little bit worse for both the approaches.
We believe that this is due to an increased variance in the B-spline estimation
of the quantile functions and (clr of) pdfs.
In fact, computing the spline approximation for a single function amounts
to solving a linear regression problem and increasing the dimension of the 
B-spline basis corresponds to increasing the number of regressors.
Hence, letting $B$ the matrix with columns $\psi_1, \ldots, \psi_J$ (evaluated
on a grid), the variance of the OLS estimate of the coefficients $\bm a$ is
proportional to $(B^T B)^{-1}$.
When increasing the number of B-splines, the entries in $B^T B$ become
closer to zero, since the support of each of the spline basis becomes smaller.
This leads to smaller precision (and higher variance) in the estimator for
$\bm a$.

Another interesting thing to notice is that the simplicial PCA exhibits a much
larger variance in the reconstruction error.
This is possibly due to the different degree of smoothness of the quantile
functions and of the pdfs. As the quantile functions are smoother than
the pdfs, their B-spline basis expansion should have lower variance and
be more similar to the true quantiles.

\subsection{Empirical Verification of Consistency Results and Choosing $J$}

\begin{figure}[t]
    \centering
    \begin{subfigure}{0.8\textwidth}
        \centering
        \includegraphics[width=\linewidth]{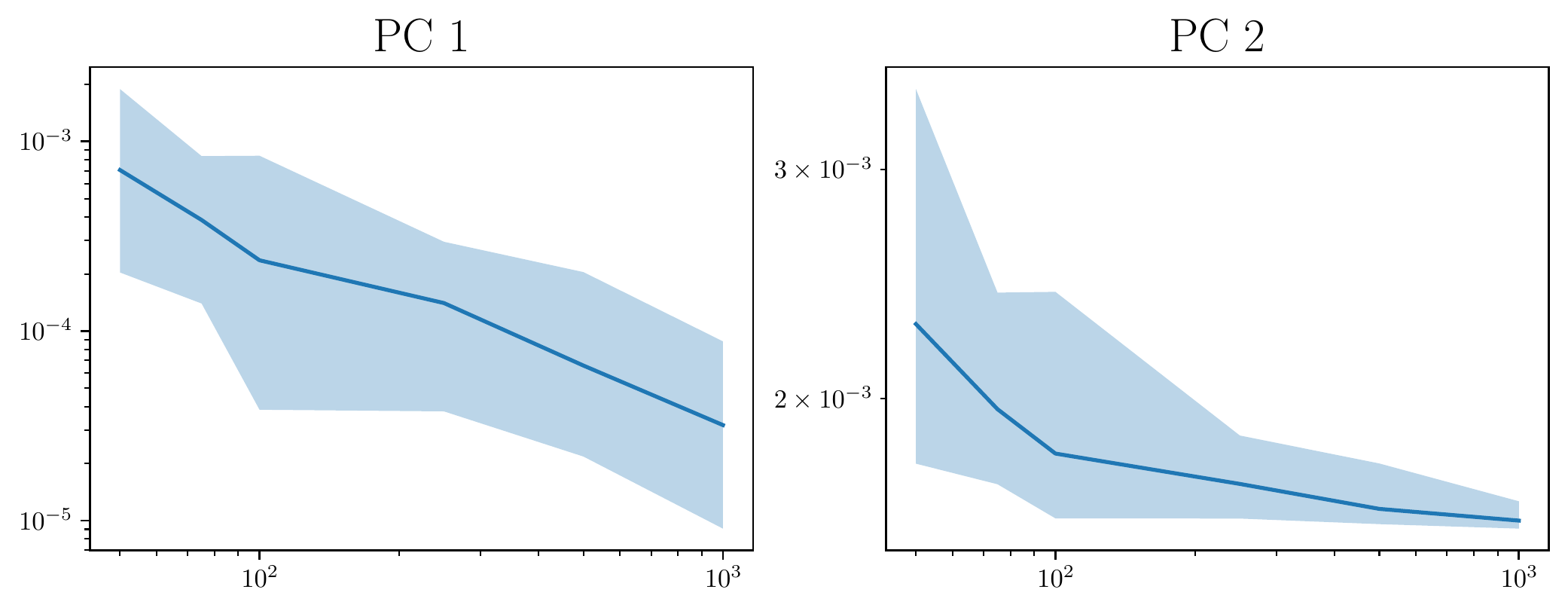}
        \caption{Gaussian data}
    \end{subfigure}
    \begin{subfigure}{0.8\textwidth}
        \centering
        \includegraphics[width=\linewidth]{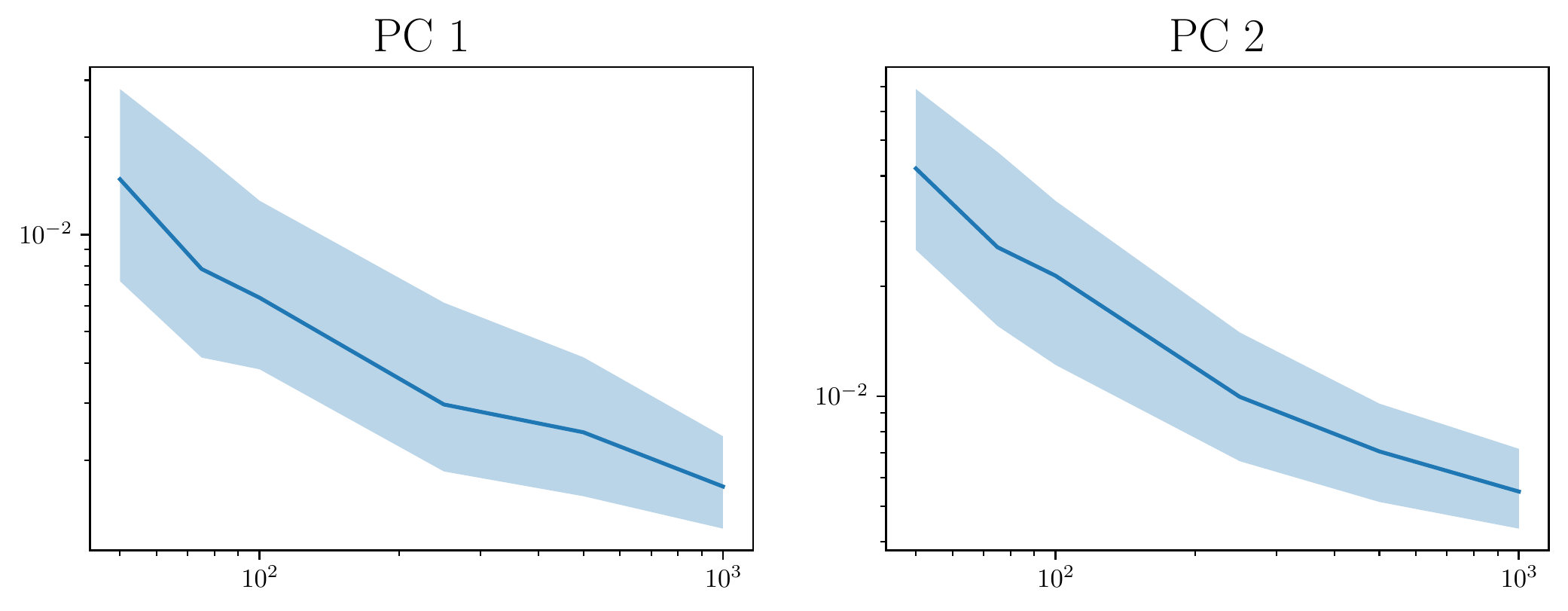}
        \caption{DPM data}
    \end{subfigure}
    \caption{$L_2$ distance between estimated and true principal directions when $J=20$ as a function of $n$. Solid line represents the median and the shaded area to a $90\%$ confidence interval estimated from 100 independent repetition.}
    \label{fig:consist_fixJ}
\end{figure}

\begin{figure}[t]
    \centering
    \begin{subfigure}{0.9\textwidth}
        \centering
        \includegraphics[width=\linewidth]{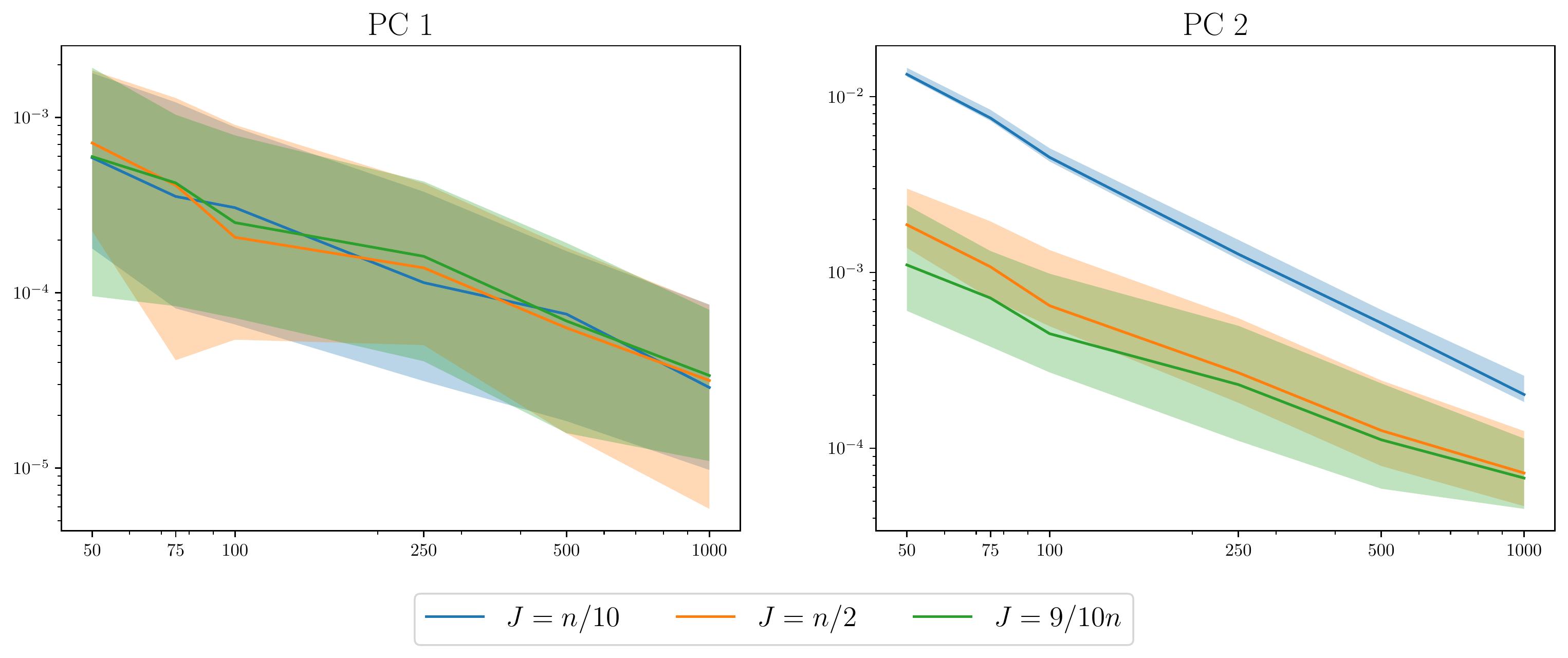}
        \caption{Gaussian data}
    \end{subfigure}

    \begin{subfigure}{0.9\textwidth}
        \centering
        \includegraphics[width=\linewidth]{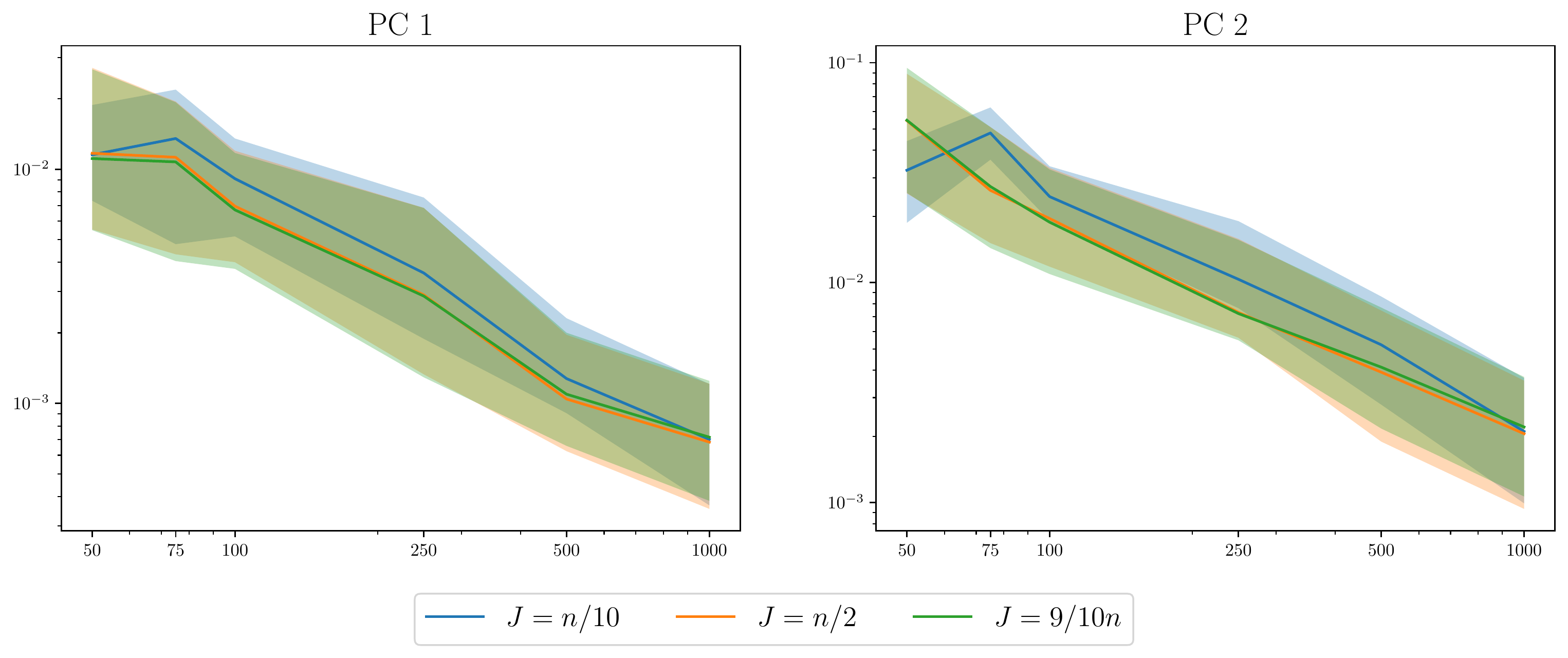}
        \caption{DPM data}
    \end{subfigure}
    \caption{$L_2$ distance between estimated and true principal directions as a function of $n$ for different choices of $J$. Solid line represents the median and the shaded area to a $90\%$ confidence interval estimated from 100 independent repetition.}
    \label{fig:consist_incJ}
\end{figure}

In this section, we provide additional simulations to verify the consistency results established in Section~\ref{sec:asympt}.

For the PCA, we consider the two data generating processes in equations \eqref{eq:simu1_dgp} (Gaussian) and \eqref{eq:simu2_dgp} (DPM). First, first we fix $J=20$ spline basis (as we do throughout Section~\ref{sec:PCA_simulations}) and let $n$ increase. Then, we also let $J$ increase linearly with $n$. 
We estimate the \virgolette{true} principal directions by simulating $10^5$ observations and using $2500$ elements in the B-spline basis. Then, for any choice of $n$ and $J$ we generate another data set and compute the corresponding first two principal directions via the projected PCA and compute the $L_2$ norm between the \virgolette{true} directions and the estimated ones.

Figure~\ref{fig:consist_fixJ} shows the case of fixed $J$ for both data generation strategies. It is clear that in both cases the error quickly decreases to zero (observe that both the $x$ and $y$ axes are in log scale), but the convergence speed is surely sub-exponential when looking, for instance, at the second principal direction.

When increasing the number of basis elements with $n$, we consider three strategies letting $J = n/10$, $n/2$ and $9/10 n$ respectively (rounded to the closest integer). 
Figure~\ref{fig:consist_incJ} shows the errors between the true and estimated principal directions in this case. Note that the convergence rate looks exponential for both data generating processes for every choice of $J = J(n)$ (increasing with $n$).
In the case of Gaussian data, we observe smaller errors (as low as $10^{-5}$ for the first direction and $10^{-4}$ for the second direction) than in the case of the more challenging DPM data set, see Figure~\ref{fig:consist_incJ}.
For the former data set, using a large number of basis functions such as $9/10 n$ or $n/2$ provides a much better fit than using $n/10$ basis functions on the second principal direction. 
For DPM data, the errors are in general two orders of magnitude higher than with Gaussian data. This is likely due to the different data generating process, which results in a more challenging
problem. Interestingly, the errors are almost equal for all values of $J$ (when fixing $n$).

Let us now analyze the projected regression. The independent variable are generated similarly to Section~\ref{sec:reg_simulations}, by discretizing the interval [0, 1] in 1,000 equispaced intervals, the value of the quantile function $F^-_{z_i}$ in the $j$-th interval equals $\sum_{k=1}^j \delta_{ik}$ and $(\delta_{i1}, \ldots, \delta_{i1000}) \sim \text{Dirichlet}(0.01, \ldots, 0.01) + \mathcal{U}([0, 5])$.
We fix the kernel $\beta^\star(t, s)$ (details are given below) and let quantile functions $F^{yi} = \Pi_{L_2([0, 1]^\uparrow)} \circ \Gamma_{\beta^\star}(F^-_{zi}) + \mathcal{N}(0, (0.1)^2)$.

We consider two different choices of $\beta^\star$: a smooth function $\beta_1^\star(t, s) = (t - 1/2)^3 + (s - 1/2)^3$, for which we expect that a small number of spline basis will give a low error, and a rougher function $\beta_2^\star(t, s)$ defined as
\[
  \beta_2^\star(t, s) = \sum_{k, h=1}^{10} \beta_1^\star(0.1 k, 0.1 h) \mathbb{I}[(t, s) \in [0.1 (k-1), 0.1 k) \times [0.1 (h-1), 0.1 h )]
\] 
that is, $\beta_2^\star$ corresponds to an approximation of $\beta_1^\star$ on a $10 \times 10$ grid.
As in the case of PCA, we present two simulations for each choice of $\beta^\star_i$, i=1,2, where we first fix the number of spline basis $J=20$ while increasing the sample size $n$ and second compare the performance for various values of $J$. We do not adopt the same strategy of setting $J$ as a fraction of the number of $n$ since the number of parameters to estimates grows quadratically with $J$ which makes the computational cost substantial when $J \geq 100$.
We measure both the seminorm error $\|\widehat{\beta} - \beta^\star\|_{\mathcal{C}_{\mathcal Z}}$ and the mean square prediction error on an unseen \virgolette{test} set of $1,000$ samples.

\begin{figure}[ht]
    \centering
    \begin{subfigure}{0.9\textwidth}
        \centering
        \includegraphics[width=\linewidth]{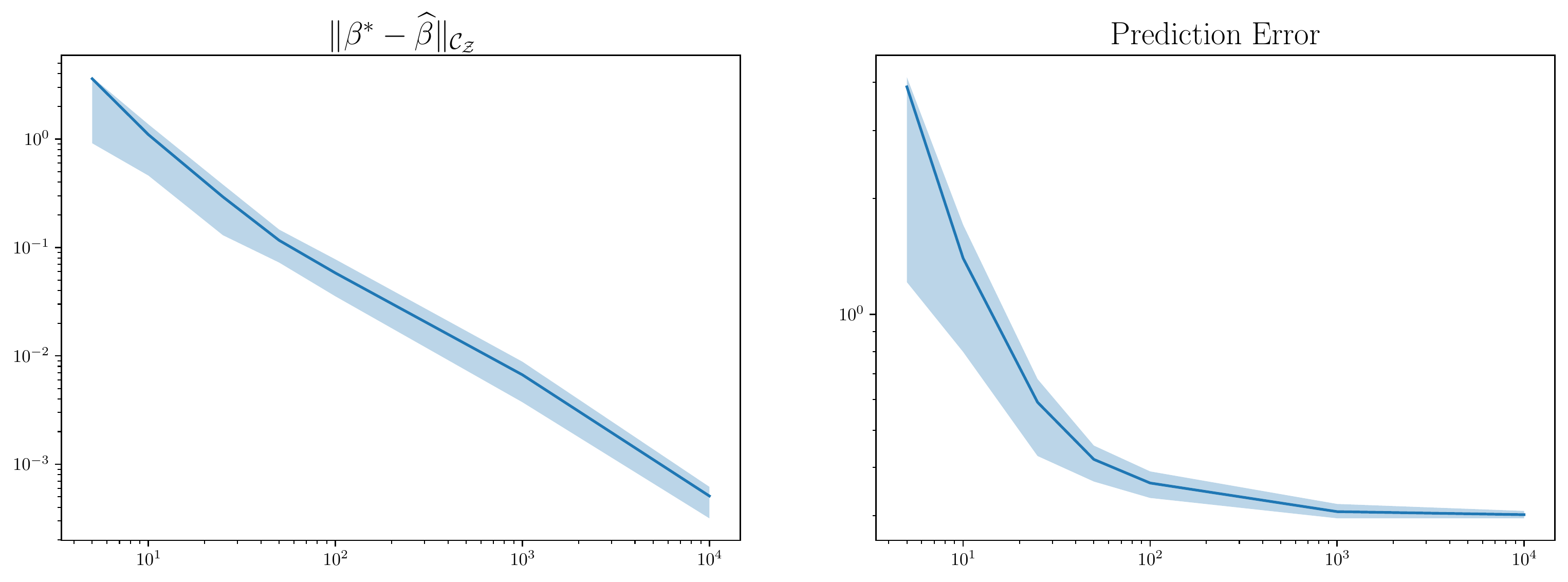}
        \caption{$\beta^\star_1$}
    \end{subfigure}

    \begin{subfigure}{0.9\textwidth}
        \centering
        \includegraphics[width=\linewidth]{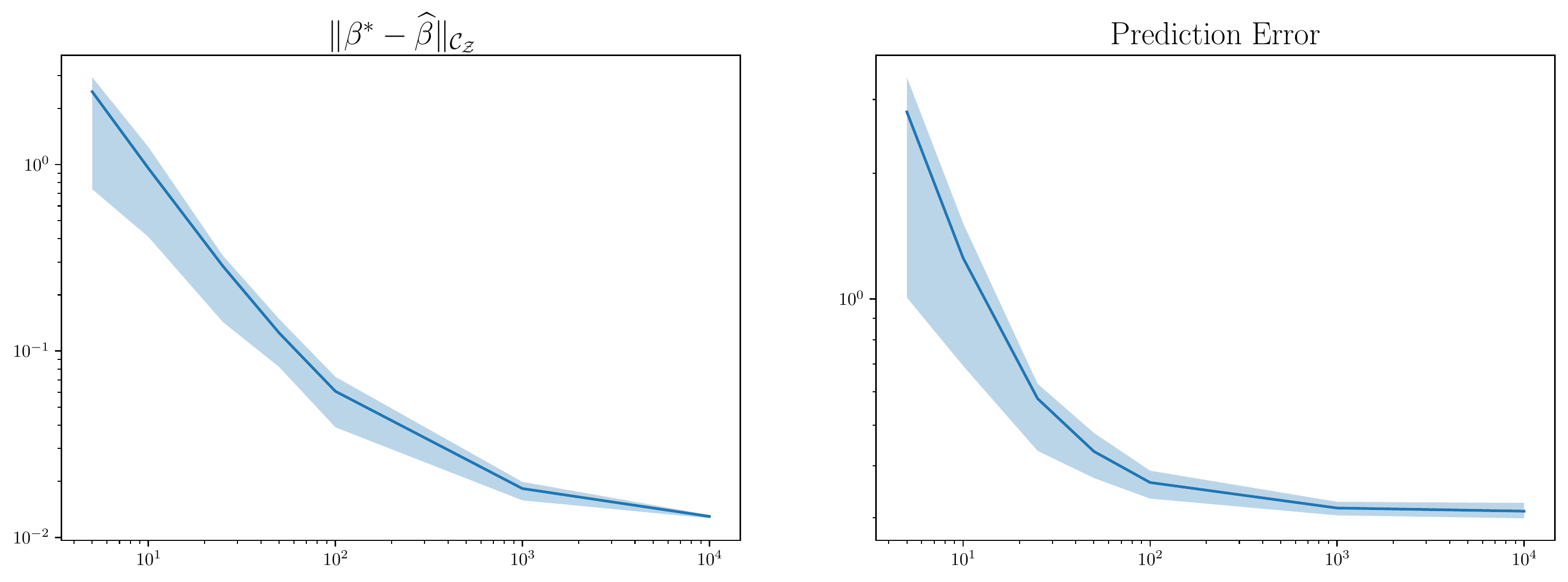}
        \caption{$\beta^\star_2$}
    \end{subfigure}
    \caption{Seminorm error (left) and mean square prediction error (right) for different choices of the kernel used to generate data, when $J=20$ as a function of $n$. Solid line represents the median and the shaded area to a $90\%$ confidence interval estimated from 100 independent repetition.}
    \label{fig:reg_fixJ}
\end{figure}

\begin{figure}[ht]
    \centering
    \begin{subfigure}{0.9\textwidth}
        \centering
        \includegraphics[width=\linewidth]{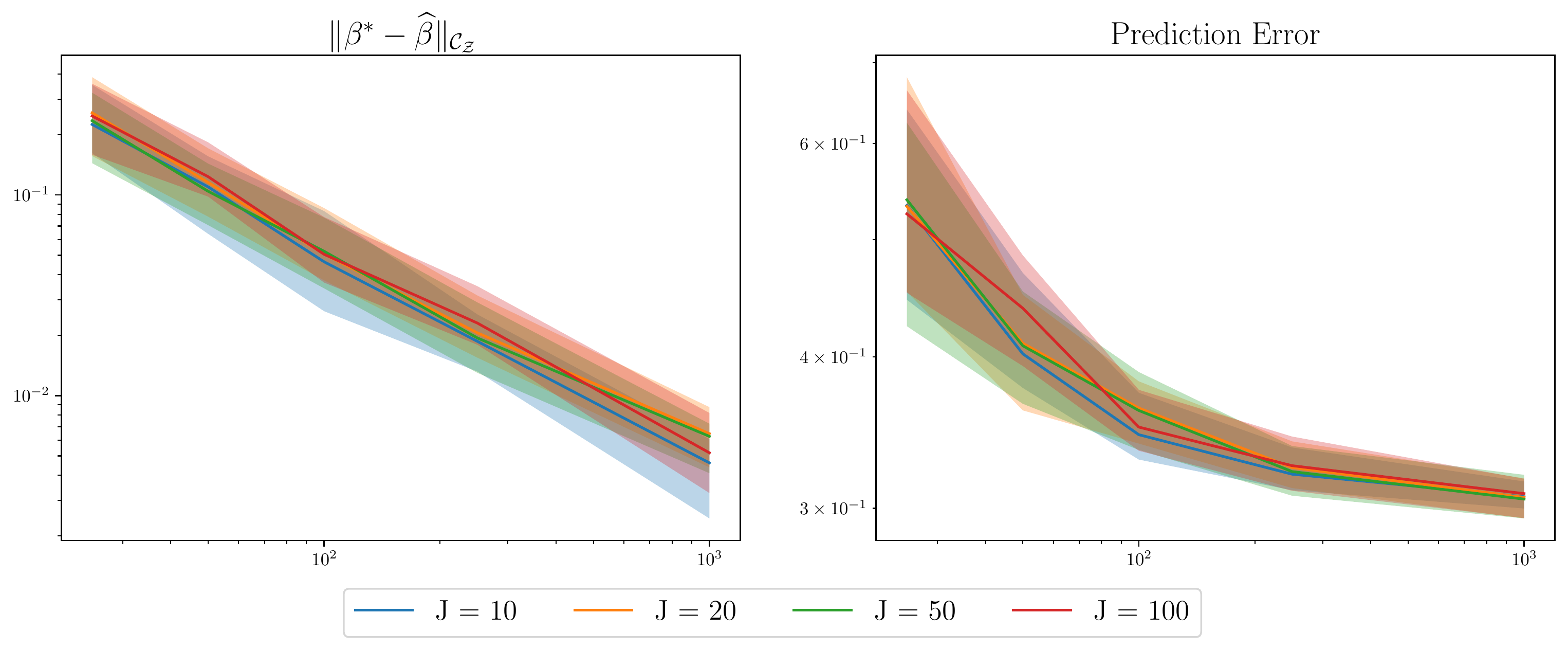}
        \caption{$\beta^\star_1$}
    \end{subfigure}

    \begin{subfigure}{0.9\textwidth}
        \centering
        \includegraphics[width=\linewidth]{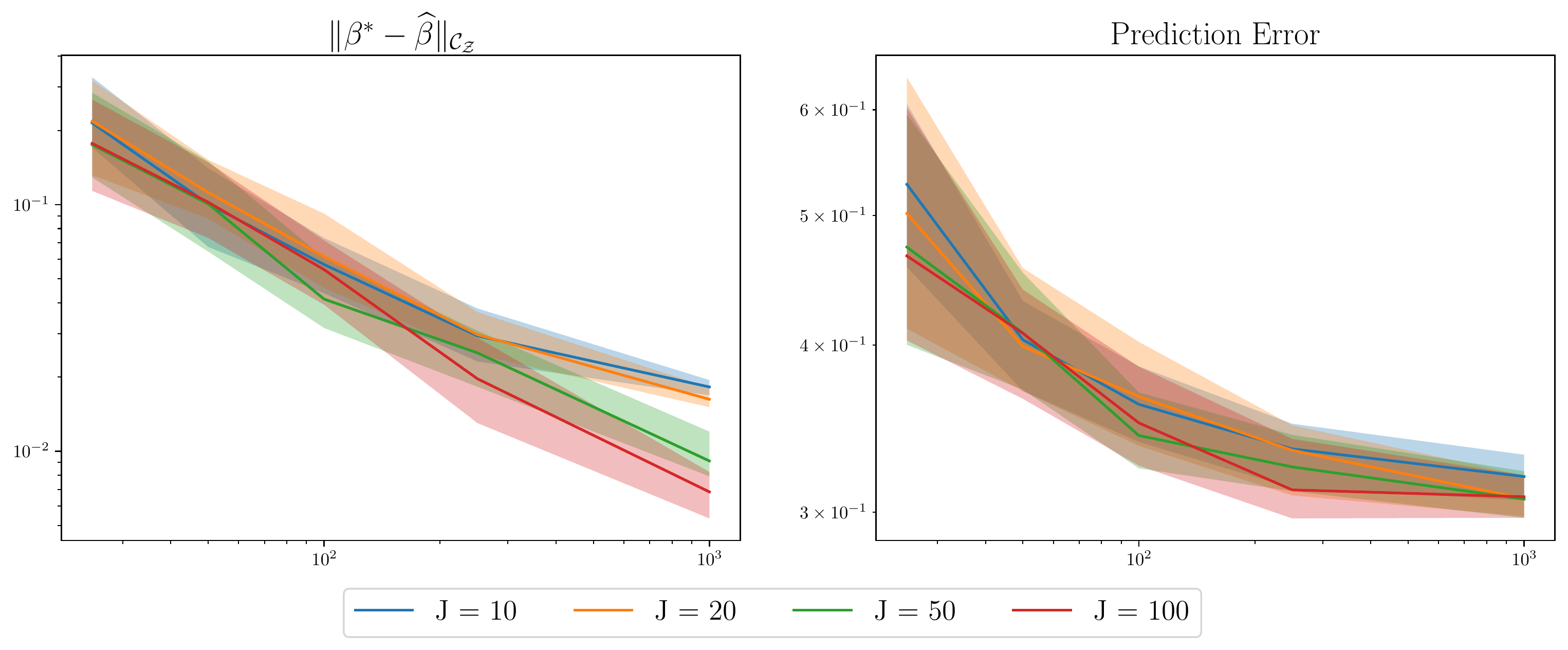}
        \caption{$\beta^\star_2$}
    \end{subfigure}
    \caption{Seminorm error (left) and mean square prediction error (right) for different choices of the kernel used to generate data, as a function of $n$ for different values of $J$. Solid line represents the median and the shaded area to a $90\%$ confidence interval estimated from 100 independent repetition.}
    \label{fig:reg_incJ}
\end{figure}

Figure~\ref{fig:reg_fixJ} shows the seminorm error and the prediction error when $J=20$ as $n$ increases, while in Figure~\ref{fig:reg_incJ} various values of $J$ are also considered.
When data are generated from $\beta^\star_1$, $J=20$ spline basis is more than enough (and actually $J=10$ would suffice) and the seminorm error in Figure~\ref{fig:reg_fixJ}(a) and Figure~\ref{fig:reg_incJ}(a) decays exponentially while the prediction error reaches the irreducible error with $n=10^3$ samples.
When data are generated from $\beta^\star_2$ the seminorm error does not show the same exponential decay when $J=20$ (see Figure~\ref{fig:reg_fixJ}(b)), but it does for larger values of $J$, in particular it seems that the error obtained with $J=50$ is the same obtained when $J=100$, see Figure~\ref{fig:reg_incJ}(b).
Hence, it is clear that the choice of $J$ is crucial to obtain a fast decay of the error: when the kernel to be approximated is not very smooth, a larger values of spline basis elements are needed, as one would expect.

We conclude this discussion by giving a practical advice on how to select $J$ for a given data set. Our suggestion is to let $J$ to be the smallest value that allows for a reconstruction error smaller than a given threshold, which may depend on the specific inferential task. 
For instance, if the problem is PCA and the goal is to provide a descriptive analysis of the variability, a (relative) approximation error below $0.05$ will typically give satisfactory results. If instead the goal is only to perform dimensionality reduction and working on the scores of a PCA as Euclidean data, one should aim for a lower approximation error, possibly of the order of $10^{-4}$.
A similar reasoning can be applied to the regression: if the goal is mainly to interpret the estimate $\widehat{\beta}$ a larger reconstruction error can be allowed. If instead one is interested in obtaining very accurate predictions, a lower error is preferred. For instance, when $\beta^\star_1$ is used to generate the data, the reconstruction error for both dependent and independent variables is below $10^{-4}$ for $J \geq 20$, while to get to the same error when $\beta^\star_2$ is used one must use $J=100$ basis.

\section{Limitations of the projected framework}\label{sec:limitations}

\subsection{When the projected PCA performs poorly}\label{sec:badex}

\begin{figure}[t]
    \centering
    \includegraphics[width=0.6\linewidth]{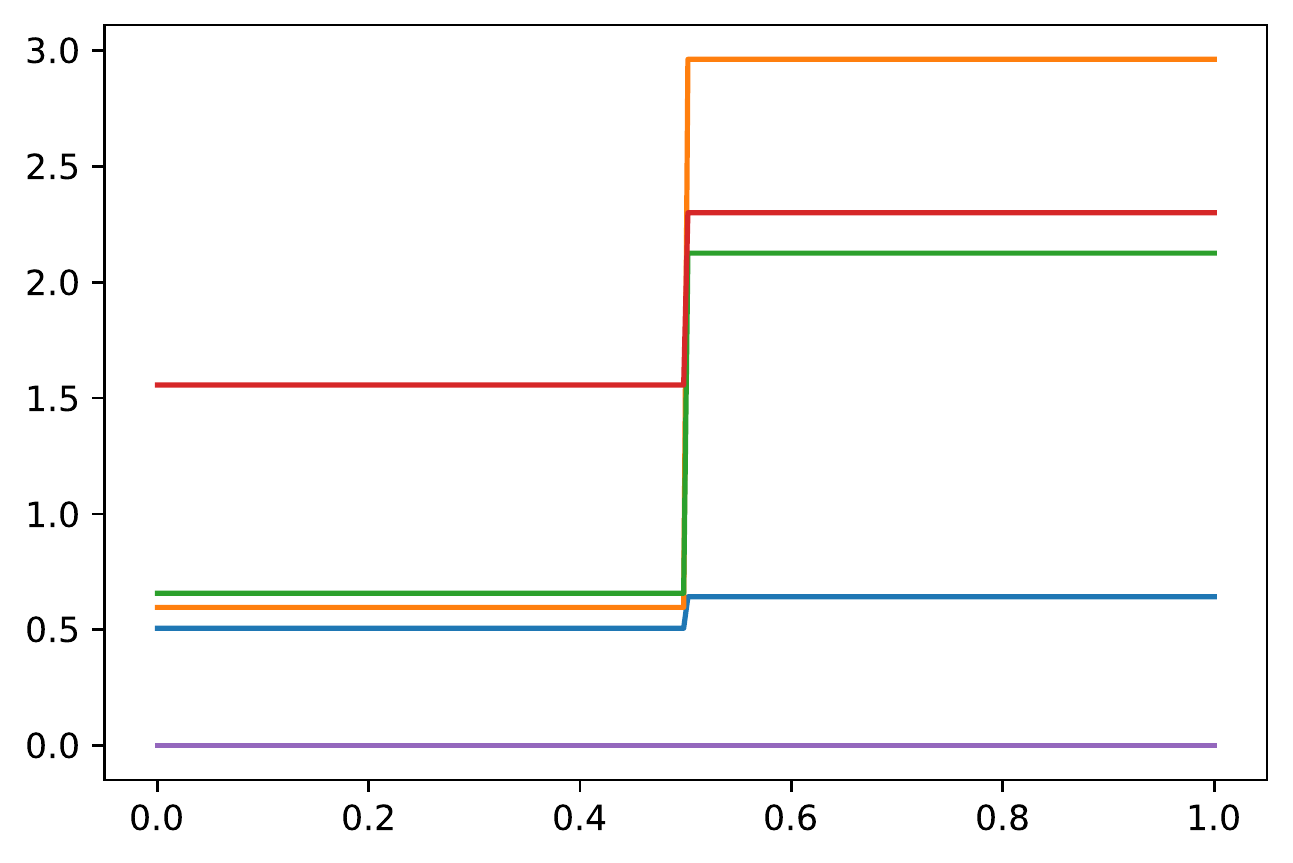}
    \caption{Five quantile functions from the data generating process considered in Appendix~\ref{sec:badex}}
    \label{fig:badex}
\end{figure}

Here, we show an example to highlight the limitations of the proposed framework, specifically of the projected PCA. 
The main idea behind this example is that the projected principal directions will be different from the nested geodesic ones when data are concentrated around the \virgolette{borders} of $X$, as in the trivial example shown in Figure~\ref{fig:geod_vs_proj}.
In the Wasserstein case, $X$ is the space of quantile functions so that the border composed of functions that are constant on a subset of $[0, 1]$. 

Hence, we consider the following data generating process, modeling directly the quantile functions
\[
    F^-_i(t) = \begin{cases}
        v_{i1}, \quad & \text{if } t < 0.5 \\
        v_{i1} + v_{i2}, \quad &\text{if } t > 0.5
    \end{cases}
\]
where $v_{ij} \sim \max\{0, \mathcal{N}(0, 1)\}$ independently. See Figure~\ref{fig:badex} for a random sample from this data generating process.

In this case, computing the projected PCA results in an interpretability score $IS_k$ equal to one for $k=1, 2$ and equal to zero for $k=3, 4, \ldots$. Hence, from the third principal direction onward, the projected PCA does not give any reliable information and, if those directions are needed, in this case a nested PCA could be preferred.
Despite the poor interpretability scores from the third direction onward, the reconstruction errors are always good as $NRE_1 = 0.26$ and $NRE_k \approx 10^{-6}$ for $k\geq 2$. Moreover, the ghost variances $GV_k$ are smaller than $10^{-10}$ for all values of $k$, so that this particular data set would be a good candidate for the hybrid methods mentioned in Section~\ref{sec:reliability}.

In summary, in our experience, the performance of the projected PCA can suffer when considering the interpretability of the directions associated to lower variability, but usually (at least always in our examples) gives a reasonable reconstruction error and ghost variance.

\subsection{Inconsistent scores when increasing dimensions}\label{sec:scores}

Here, we highlight a feature which is shared by both projected and nested PCA, 
that is, the scores of the projection onto a projected principal component are dependent on the dimension of the principal component, as already noted in Section \ref{sec:proj_PCA}. 

This can be considered a limitation to those frameworks, because it contributes to the complexity of the analysis: one has always to fix the dimension of the chosen principal component and use the scores accordingly obtained. 
For instance, the scores, both for nested and projected PCAs, coincide with the $L_2$ scores when the dimension of the principal components is equal to the cardinality of the spline basis $J$. This happens because the principal components are not linear subspaces.  As a consequence also the interpretability score of a direction is dimension-dependent. 

Hence, the choice of the dimension $k$ must be carried out balancing (i) a parsimonious representation, (ii) a low reconstruction error, so that the projections on the principal components yield good approximations of the data, and (iii) the intepretability score of the directions. 

Thus, opposed to standard Euclidean PCA, where the $k+1$-th direction does not change the behavior of the data along the previous $k$ directions (i.e., the scores),
when doing (any) PCA in Wasserstein space the whole picture must always be taken into account, both for nested and projected PCA to assess the interpretability of the results.

Finally, note that such interpretability might be low for both intrinsic and extrinsic methods, but this means that the Wasserstein metric may not be the most adequate to capture and explain the variability of the data set.

\FloatBarrier

\vskip 0.2in
\bibliography{references}

\end{document}